\documentclass[traditabstract]{aa}
\pdfoutput=1
\usepackage{lmodern}    \usepackage[T1]{fontenc}
\usepackage{pdflscape}  

\usepackage{graphicx}  
\usepackage{aalongtable} 
\usepackage{soul}      
\usepackage{amsmath, amssymb, stmaryrd}     
\usepackage[perpage,symbol*,hang]{footmisc} 
\usepackage[title,titletoc]{appendix}       

\usepackage{natbib}
\bibpunct{(}{)}{;}{a}{}{,} 

\usepackage[pdftex,bookmarksopen,bookmarksnumbered]{hyperref}

\defcitealias{2006A&A...459..423B}{B06}

\newcommand{\spr}{\textit{s}-process}
\newcommand{\rpr}{\textit{r}-process}
\newcommand{\alfe}{\ensuremath{\alpha}-elements}

\newcommand{\teff}{\ensuremath{T_{\mathrm{eff}}}}
\newcommand{\vmic}{\ensuremath{v_{\mathrm{t}}}}
\newcommand{\vrad}{\ensuremath{V_{\mathrm{rad}}}}
\newcommand{\fei}{Fe\,{\sc i}}
\newcommand{\feii}{Fe\,{\sc ii}}
\newcommand{\tii}{Ti\,{\sc i}}
\newcommand{\tiii}{Ti\,{\sc ii}}
\newcommand{\gf}{\textit{\sffamily{g}\rmfamily{f}}}
\newcommand{\logg}{\ensuremath{\log g}}
\newcommand{\kiex}{\ensuremath{\chi_{\mathrm{ex}}}}
\newcommand{\msol}{\ensuremath{\mathcal{M}_{\odot}}}
\newcommand{\kxb}{\ensuremath{\mathrm{slope}_{\chi}}}
\newcommand{\ewb}{\ensuremath{\mathrm{slope}_{W}}}
\newcommand{\kxbs}{\ensuremath{\sigma(\kxb)}}
\newcommand{\ewbs}{\ensuremath{\sigma(\ewb)}}

\begin{document}
\title{A high resolution VLT/FLAMES study of individual stars in the
centre of the Fornax dwarf spheroidal galaxy\thanks{Based on FLAMES 
observations collected at the European Southern Observatory, 
proposal number 171.B-0588}}

\author{Bruno~Letarte\inst{1,2,3}
\and V.~Hill\inst{4,5}
\and E.~Tolstoy\inst{1}
\and P.~Jablonka\inst{4,6}
\and M.~Shetrone\inst{7}
\and K.A.~Venn\inst{8}
\and M.~Spite\inst{4}
\and M.J.~Irwin\inst{9}
\and G.~Battaglia\inst{10}
\and A.~Helmi\inst{1}
\and F.~Primas\inst{10}
\and P~Fran\c cois\inst{4}
\and A.~Kaufer\inst{11}
\and T.~Szeifert\inst{11}
\and N.~Arimoto\inst{12}
\and K.~Sadakane\inst{13}
}

\institute{Kapteyn Astronomical Institute, University of Groningen, PO Box 800, 9700AV Groningen, the Netherlands
\and California Institute of Technology, 1200E. California Blvd, MC105-24, Pasadena, CA 91125 USA 
\and South African Astronomical Observatory, P.O. Box 9, Observatory 7935, South Africa 
\and GEPI, Observatoire de Paris, CNRS, Universit\'e Paris Diderot,  F-92125, Meudon, Cedex, France
\and Universit\'e de Nice Sophia-Antipolis, CNRS, Observatoire de C\^{o}te d'Azur, Laboratoire Cassiop\'ee, F-06304 Nice Cedex 4, France
\and Observatoire de Gen\`eve, University of Geneva, CH-1290, Sauverny, Switzerland
\and McDonald Observatory, University of Texas, Fort Davis, TX 79734, USA
\and Dept. of Physics \& Astronomy, University of Victoria, 3800 Finerty Road, Victoria, BC V8P 1A1, Canada
\and Institute of Astronomy, University of Cambridge, Madingley Road, Cambridge CB3 0HA, UK
\and European Southern Observatory, Karl-Schwarzschild-str. 2, D-85748, Garching bei M\"{u}nchen, Germany
\and European Southern Observatory, Alonso de Cordova 3107, Santiago, Chile
\and National Astronomical Observatory of Japan, 2-21-1 Osawa, Mitaka, Tokyo 181-8588, Japan
\and Astronomical Institute, Osaka Kyoiku University, Asahigaoka, Kashiwara, Osaka 582-8582, Japan
}

\date{Received 6th October 2009 / Accepted 6th July 2010}

\abstract{
For the first time we show the detailed late-stage chemical evolution
history of small nearby dwarf spheroidal galaxy in the Local Group.  We
present the results of a high resolution (R$\sim$20000) FLAMES/GIRAFFE
abundance study at ESO/VLT of 81 photometrically selected red giant
branch stars in the central 25$'$ of the Fornax dwarf spheroidal galaxy.
We also carried out a detailed comparison of the effects of recent
developments in abundance analysis (e.g., spherical models vs.
plane-parallel) and the automation that is required to efficiently deal
with such large data sets.  We present abundances of \alfe\ (Mg, Si, Ca
and Ti), iron-peak elements (Fe, Ni and Cr) and heavy elements (Y, Ba,
La, Nd and Eu).  Our sample was randomly selected, and is clearly
dominated by the younger and more metal rich component of Fornax which
represents the major fraction of stars in the central region.  This
means that the majority of our stars  are 1$-$4~Gyr old, and thus
represent the end phase of chemical evolution in this system. Our sample
of stars has unusually low [$\alpha$/Fe], [Ni/Fe] and [Na/Fe] compared
to the Milky Way stellar populations at the same [Fe/H]. The
particularly important role of stellar winds from low metallicity AGB
stars in the creation of \spr\ elements is clearly seen from the high
[Ba/Y].  Furthermore, we present evidence for an \spr origin of Eu.  }


\authorrunning{Letarte et al.}
\titlerunning{VLT/FLAMES HR abundances study of Fornax dSph}
\maketitle

\section{Introduction}

Local Group dwarf spheroidal galaxies (dSph) offer the opportunity to
study the star formation history (SFH) and chemical evolution of small
systems. They are also believed to provide an insight into the galaxy
assembly process. We have undertaken a large observational programme
\citep[DART, Dwarf Abundances and Radial velocity Team,
][]{2006Msngr.123...33T} using high resolution spectroscopy to study the
detailed nucleosynthesis properties of large samples of individual Red
Giant Branch (RGB) stars in 4 nearby dSph galaxies: Fornax, Sculptor,
Sextans \& Carina. Here we present the analysis of a sample of 81
individual stars in the Fornax dSph.

The Fornax dSph galaxy is a relatively isolated and luminous, dark
matter dominated dwarf galaxy with a total mass of $\sim 10^9 \msol$
\citep[e.g.,][]{2007ApJ...667L..53W, Battaglia07phd}, at a distance of
135 kpc \citep[e.g.,][]{2007MNRAS.380.1255R}.  It is well resolved into
individual stars, and colour-magnitude diagram (CMD) analyses have been
made going down to the oldest main sequence turn-offs in the central
region \citep[e.g.,][]{1998PASP..110..533S, 1998ApJ...501L..33B,
2000A&A...355...56S, 2005nfcd.conf...25G}.  Fornax also contains a
significant carbon star population \citep[e.g.,][]{Azzopardi99,
Aaronson80}.  Near Infra-Red CMDs have also been made of the RGB stars
\citep[][]{2007A&A...467.1025G}.  There have also been extensive
variable star searches, resulting in $\sim$500 RR~Lyr variables, 17
anomalous Cepheids, 6 population II Cepheids and 85 short period
$\delta$-Scuti and SX Phoenicis stars \citep[][]{Bersier02, Poretti08}.
These works have all shown that Fornax has had a complex SFH where the
majority of stars have formed at intermediate ages 2~$-$~6~Gyr ago with
a peak $\sim 5$~Gyr ago.  Fornax also contains a young stellar
population (few 100~Myr old) as well as an ancient one ($>$10 Gyr old),
as in all dwarf spheroidal galaxies studied to date (provided the
observations are deep enough to reach the oldest turnoff).

In common with most other dSph, Fornax has no obvious HI associated to
it at present, down to a density limit of $4 \times 10^{18} \rm cm^{-2}$
in the centre (or $\rm 0.037\,M_{\odot} pc^{-2}$) and $10^{19} \rm
cm^{-2}$ at the tidal radius \citep[][]{1999AJ....117.1758Y}.  Unusually
for dwarf galaxies Fornax contains five globular clusters
\citep[][]{1961AJ.....66...83H}.  There is evidence that Fornax
contains stellar substructure, in shell-like features and hence can
be considered evidence of recent merger activity
\citep[][(B06)]{2004AJ....127..832C, 2006AJ....131..912O,
2005PASA...22..162C, 2006A&A...459..423B}.

Low resolution spectroscopic studies of the Ca~II triplet lines have
been carried out for increasing numbers of individual RGB stars in
Fornax, to determine kinematics and [Fe/H] estimates for samples of
$\sim$30 stars \citep[][]{2001MNRAS.327..918T}, $\sim$100 stars
\citep[][]{2004AJ....127..840P} and $\sim$600 stars (B06). There have
also been measurements of Fe and Mg features in 2483 individual stars in
Fornax \citep[][]{Walker09} that have not so far been calibrated to
metallicity.  These studies have shown that Fornax contains a
relatively metal-rich stellar population, and that the metallicity
distribution covers a range from $-2.8 < $[Fe/H]$< 0.$ with a peak at
[Fe/H]$\sim -0.9$.  B06 clearly showed that the ``metal rich'' stars
([Fe/H]$> -$1.3) have a centrally concentrated spatial distribution,
with colder kinematics than the more spatially extended ``metal poor''
stars ([Fe/H]$< -$1.3). B06 also showed tentative evidence that the
ancient stellar population in the centre of Fornax does not exhibit
equilibrium kinematics (apparent as non-gaussian, double peaked
velocity distribution). This could be a sign of a relatively recent
accretion of external material, such as the gas accretion of the merger
with another small stellar system (galaxy or star cluster).

Alpha, iron-peak, and heavy-elements, in individual RGB stars of
different ages, can provide detailed information about the evolving
conditions of star formation and chemical evolution in a galaxy
through-out its history.  The most accurate way to determine the
elemental abundances of an RGB star is to directly measure the strength
of as many individual absorption lines as possible in a spectrum. This
typically requires a resolving power of, $\rm R \sim 40000$ (which is
$\sim$ 0.125 \AA\ at $\lambda$=5000 \AA), with a large wavelength
coverage ( $\sim$ 2000 \AA) and has traditionally been carried out one
star at a time with slit-spectrographs.  The large wavelength
coverage is required to obtain a minimum number of individual absorption
lines for an accurate abundance analysis of chemical elements of
interest. The wavelength range chosen is always a trade off between
number of lines available (strong enough to be detected and isolated
enough to be of any use) and instrument sensitivity.

Detailed abundance analyses using UVES high resolution spectroscopy have
been carried out for 3 individual RGB stars in Fornax
\citep[][S03]{2003AJ....125..707T, 2003AJ....125..684S}.  There have
also been high resolution UVES studies of 9 individual stars in 3 of the
globular clusters associated with Fornax
\citep[][]{2006A&A...453..547L}.  One of the globular clusters was
observed to have a mean metallicity of [Fe/H]=$-2.5$ and all of them
\citep[see also,][]{2003AJ....125.1291S} have metallicities quite
distinctly lower than the majority of the field stellar population,
although Fornax field also host a low-level metal-poor tail extending
down to $-2.8$ (B06).

The FLAMES multi-fibre spectrograph at the VLT has dramatically
increased the efficiency with which spectroscopic observations can be
made, by allowing more than 100 spectra to be obtained in one
shot. However, the observations are of lower resolution (R~$\sim 20
000$) than standard and with much a smaller wavelength coverage
(typically a series of $\sim$300\AA\ wide bands).  This trade-off of 
large samples for lower resolution and less wavelength coverage has
required the development of new automated data analysis procedures
which will be described in this paper.

In this paper we present detailed chemical abundances for a {\em large}
sample of 81 RGB stars in Fornax (30 times larger than S03),
randomly selected on the upper RGB in the center of the galaxy. This
allows us to quantify the detailed chemical evolution of this galaxy.
Because we selected stars randomly, and in the central parts of the
galaxy, we will see that the dominating population in our sample belongs
to the peak of  its star formation activity.  In
Sec.~\ref{sect:sampleselect}, we describe the sample selection,
observations, reduction and basic measurements, in
Sec.~\ref{sect:abundances} we describe the abundance analysis method, in
Sec.~\ref{sect:results} we discuss the results, and in
Sec.~\ref{sect:discussion} we discuss the implications of our results.

\section{Observations and data reduction}
\label{sect:sampleselect}

\begin{figure}
\includegraphics[width=\hsize]{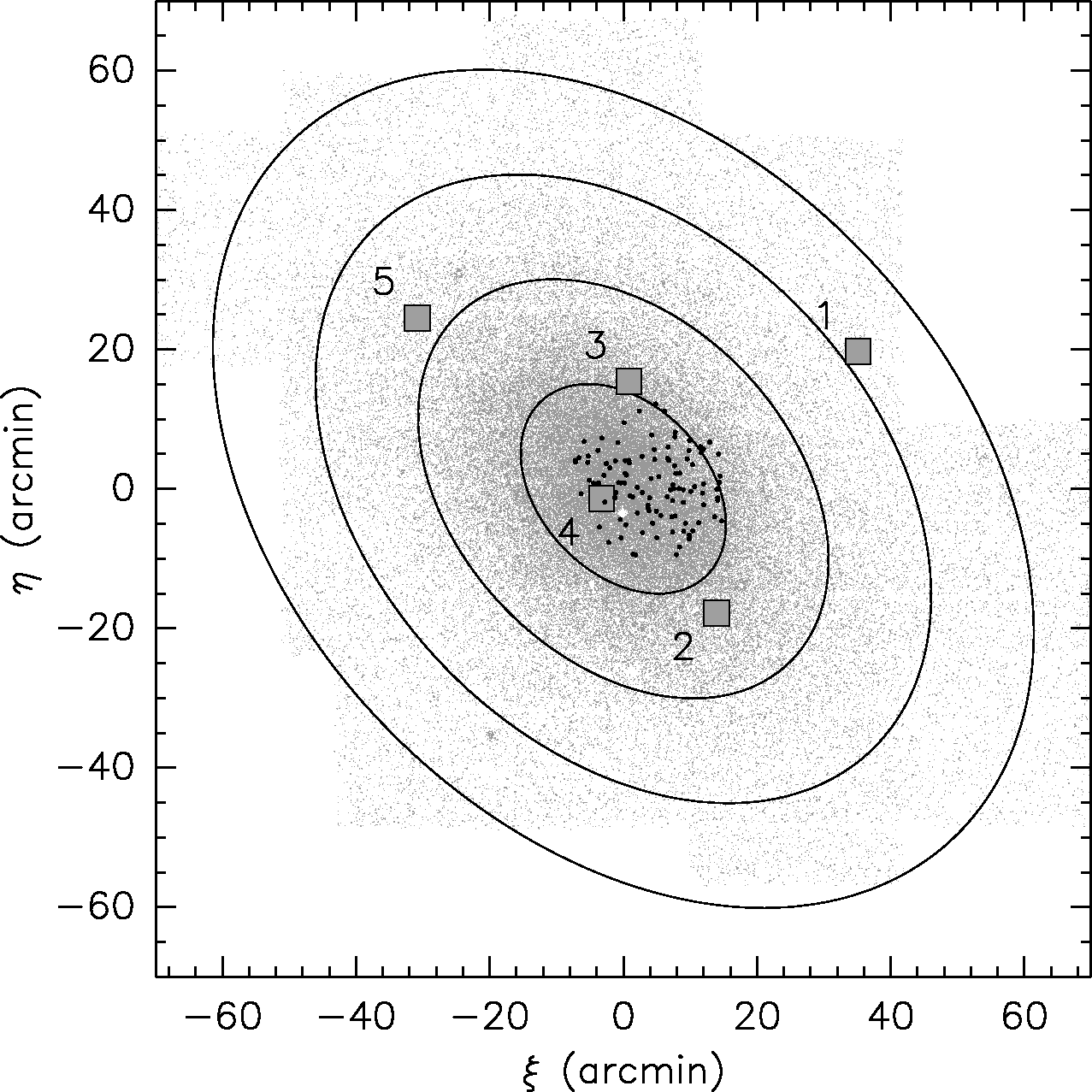}
\caption{The spatial distribution of Fornax dSph stars observed at high
resolution (centrally concentrated black dots), overlaid on our
photometric survey (small grey dots, from B06).  The five globular
clusters are also shown, as squares. The inner ellipse corresponds to
the core radius, and the outer ellipse to the tidal radius.
\label{fig:fnx-space}}
\end{figure}
\begin{figure*}[hbt]
\begin{center}
\includegraphics[width=\hsize]{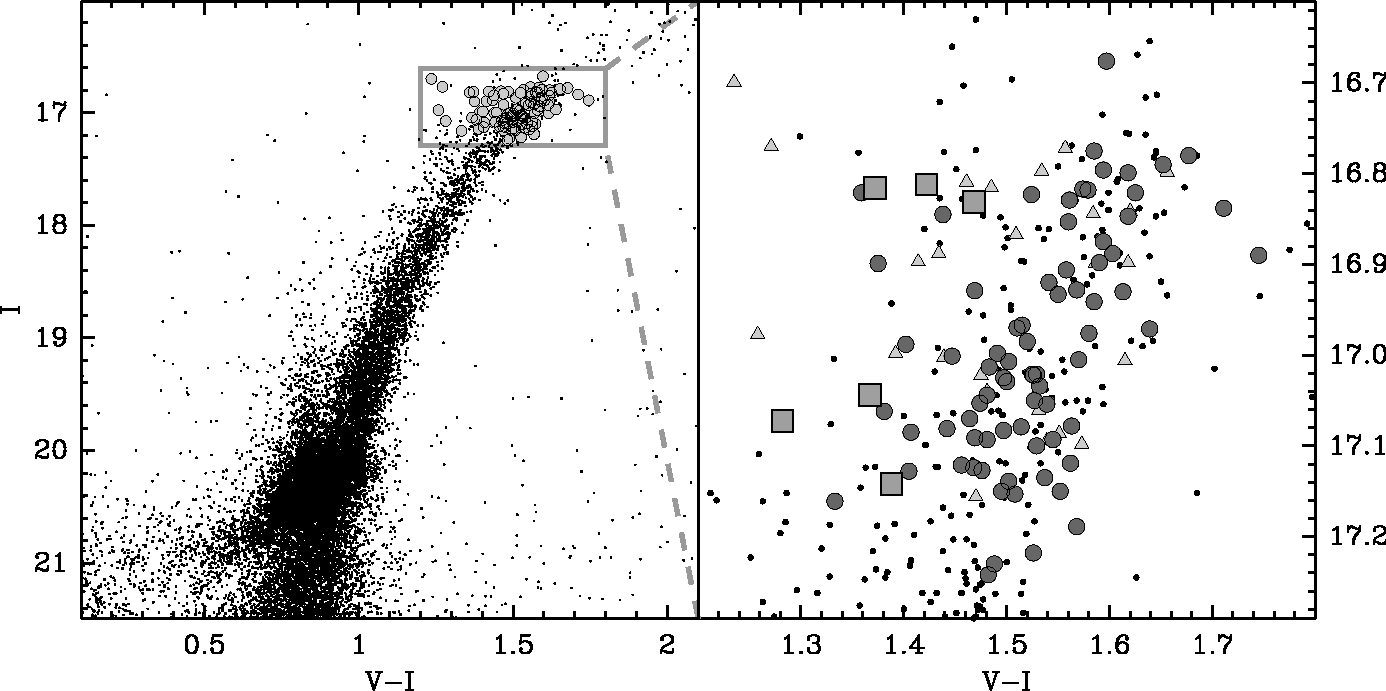}
\caption{The $I$ vs $V-I$  color-magnitude diagram of the Fornax centre
as derived from our ESO/MPG/2.2m WFI data (B06). It corresponds to the
field of view of FLAMES.  The spectroscopic targets analyzed in this
work are identified with larger grey circles in the box in 
the left panel. The right panel zooms-in to this region.
\label{fig:fnx-cmd-target}} 
\end{center}
\end{figure*}

Figure~\ref{fig:fnx-space} gives a spatial overview of the Fornax dSph
galaxy. We highlight the stars observed in this high resolution FLAMES
survey of individual stars in the central 25$'$ region.  The photometric
selection of the targets is presented in
Figure~\ref{fig:fnx-cmd-target}. We selected bright RGB stars from the
$V$,$I$ CMD (B06), to include a broad range of colour, and thus
the entire age and metallicity range present. We could only select the
brightest red giants to keep exposure times down to a reasonable level
(see Table~\ref{tab:fnxfi-FLAMES}). In order to minimize the number of
AGB stars in our sample, we cross-checked our list of potential targets
with known carbon stars (kindly provided by Serge Demers). The low
resolution spectroscopic survey of this field with FLAMES with the Ca~II
triplet setting (as part of B06) also helped to weed out as many as
possible obvious foreground stars from our high resolution sample.

\subsection{Observations}

FLAMES is a fibre-fed, multi-object instrument connected to GIRAFFE, a
spectrograph which has a low ($\rm R= 6000-9000$) and a high resolution
($\rm R = 17000-30\,000$) grating, covering a total wavelength range
$\lambda\lambda 3700-9000$\,\AA\, broken up into 21 high-resolution (or
8 low-resolution) setups \citep[][]{2002Msngr.110....1P}.  In its
high-resolution mode, each wavelength section (setup) is $\sim300$ \AA\
wide.  For our purposes we used FLAMES/GIRAFFE with the high resolution
setups HR10, HR13 and HR14.  Table~\ref{tab:fnxfi-FLAMES} summarizes the
different setups and gives the exposure times.  Note that the definition
(resolution and wavelength coverage) of HR14 changed after October
10$^{\mathrm{th}}$ 2003, when the HR14 grating was up-graded from its
engineering to operational version. HR14 was moved from order 9 to order
8 to match the new grating efficiency curve.

We used FLAMES/GIRAFFE in MEDUSA mode, which consists of up to 132
individual 1.2 arc-sec fibres, that can be placed independently on
targets in the 25$'$ diameter FLAMES field of view.  We were able to
place fibres on 107 stellar targets, and the rest were assigned to
blank areas, for an accurate determination of the sky light.

We encountered two technical problems. First, during our first run, at
the end of September 2003, one reference star turned out to have a
previously unnoticed proper motion.  As a consequence, a significant
fraction of the flux has been lost due to bad centering of the fiber
on the target. Second, GIRAFFE hosts a simultaneous calibration unit
that provides 5 fibers evenly spread across the spectrograph entrance
slit, that are illuminated by a ThAr lamp. They allow very high radial
velocity accuracies.  However, during the January~2004 run, these
simultaneous calibration fibers have been seriously over-exposed,
especially in setup HR14. As a consequence, the bright calibration
light leaked on to our stellar spectra.  Only a small fraction of our
sample was affected.  All contaminated lines were simply removed from
our abundance analysis.

\subsection{Reduction}

We used  the girBLDRS\footnote{{http://girbldrs.sourceforge.net/}}
(GIRAFFE Base-Line  Data Reduction Software)  pipeline  written at the
Observatory  of  Geneva by A.  Blecha and  G. Simond,  to  extract and
calibrate our spectra.  For the subtraction of  the sky emission lines
and continuum, we ran our own  specially developed software written by
M. Irwin (see \citet{2008MNRAS.383..183B} for details).

\begin{table}[!htp]
\begin{center}
\caption{Exposure Time Log
\label{tab:fnxfi-FLAMES}}
\begin{tabular}{lrrrr}
\hline \hline
       & HR10&   HR13& HR14$_{\mathrm{old}}$& HR14$_{\mathrm{new}}$ \\
\hline
$\lambda_{\mathrm{min}}$(\AA)& 5339&   6120&      6383&      6308\\
$\lambda_{\mathrm{max}}$(\AA)& 5619&   6406&      6626&      6701\\
$R$                          & 19800& 22500&     28800&     17740\\
\hline
date& \multicolumn{4}{c}{Exposure time (s)}\\
\hline
  2003-09-29 &      --&   14400&    6225&-- \\  
  2003-09-30 &    3600&      --&   14102&-- \\  
  2003-10-01 &   10800&      --&    6900&--\\  
  2004-01-14 &    3600&      --&      --&-- \\  
  2004-01-15 &      --&    3600&      --&-- \\  
  2004-01-19 &      --&    7200&      --&-- \\  
  2004-01-20 &      --&    3600&      --&-- \\  
  2004-01-21 &      --&      --&      --& 7200 \\  
  2004-01-22 &      --&      --&      --& 7503 \\  
  2004-01-23 &    3600&      --&      --& 3600 \\  
  2004-01-24 &    3600&      --&      --&-- \\  
  2004-01-26 &    3600&      --&      --&-- \\  
\hline 
  Total      &      8h&      8h&   7h34m& 5h05m \\  
\end{tabular}
\end{center}
\end{table}

\begin{figure}[!htp]
\begin{center}
\includegraphics[width=1.0\hsize]{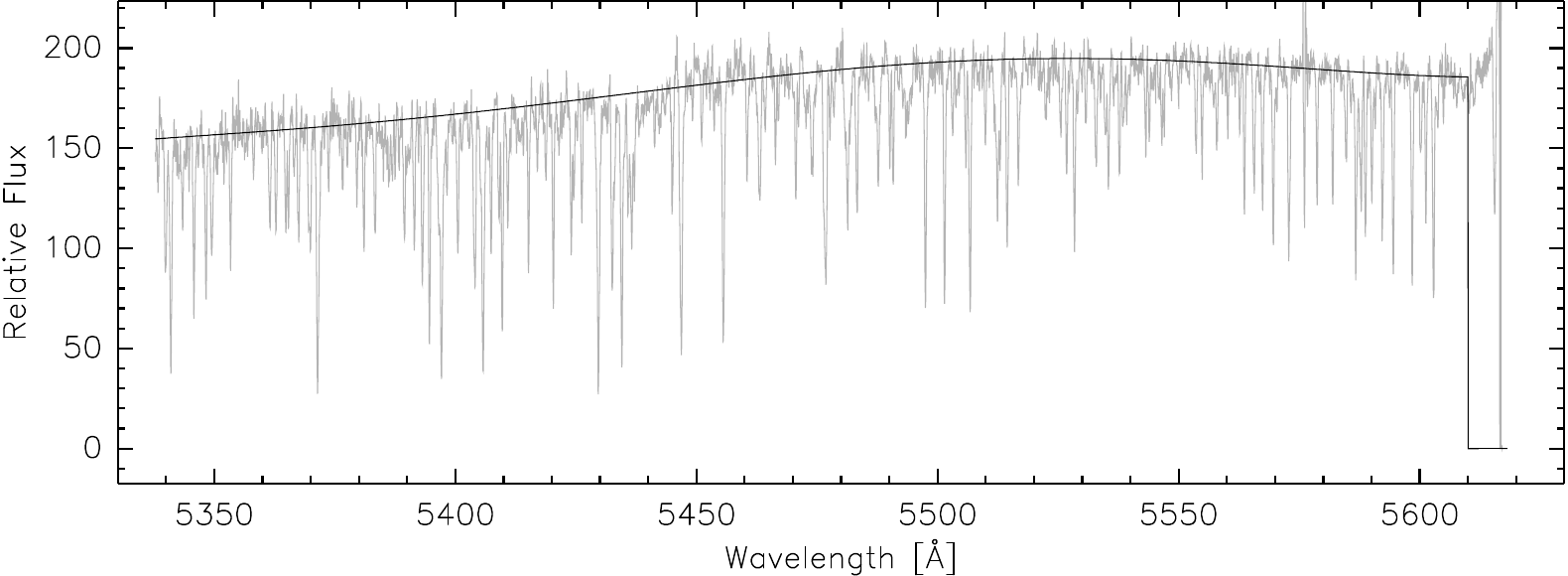}
\smallskip

\includegraphics[width=1.0\hsize]{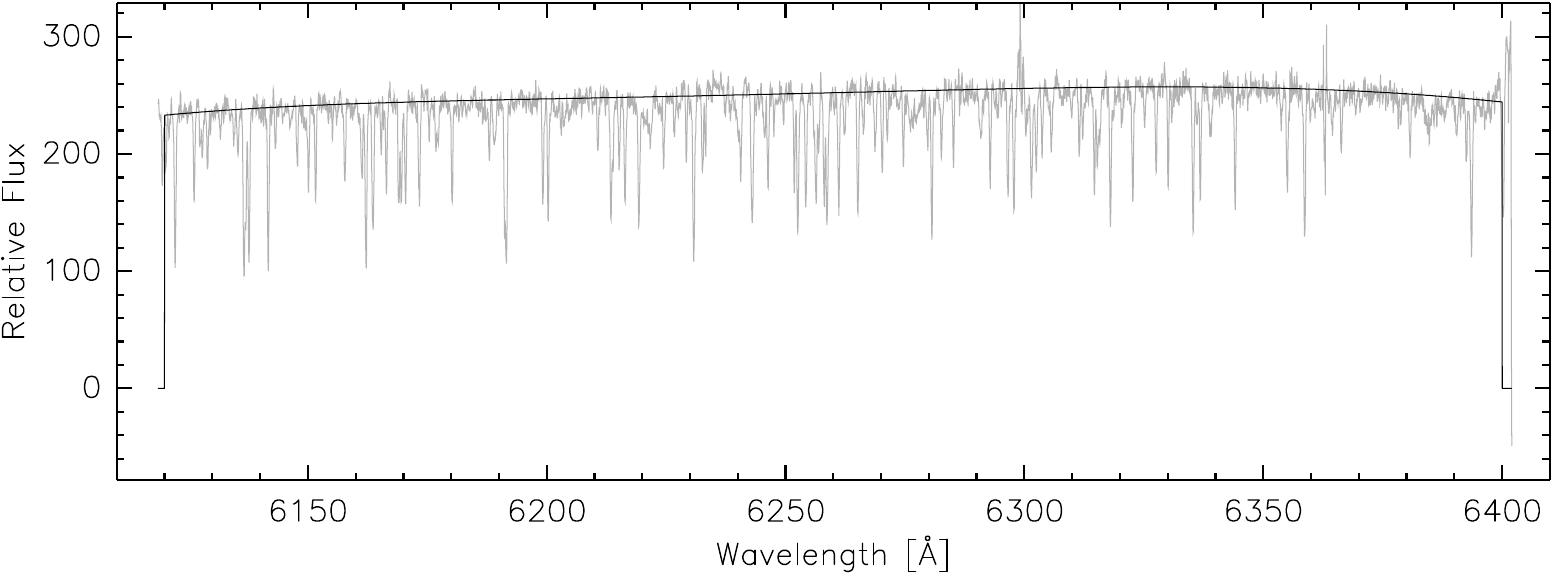}
\smallskip

\includegraphics[width=1.0\hsize]{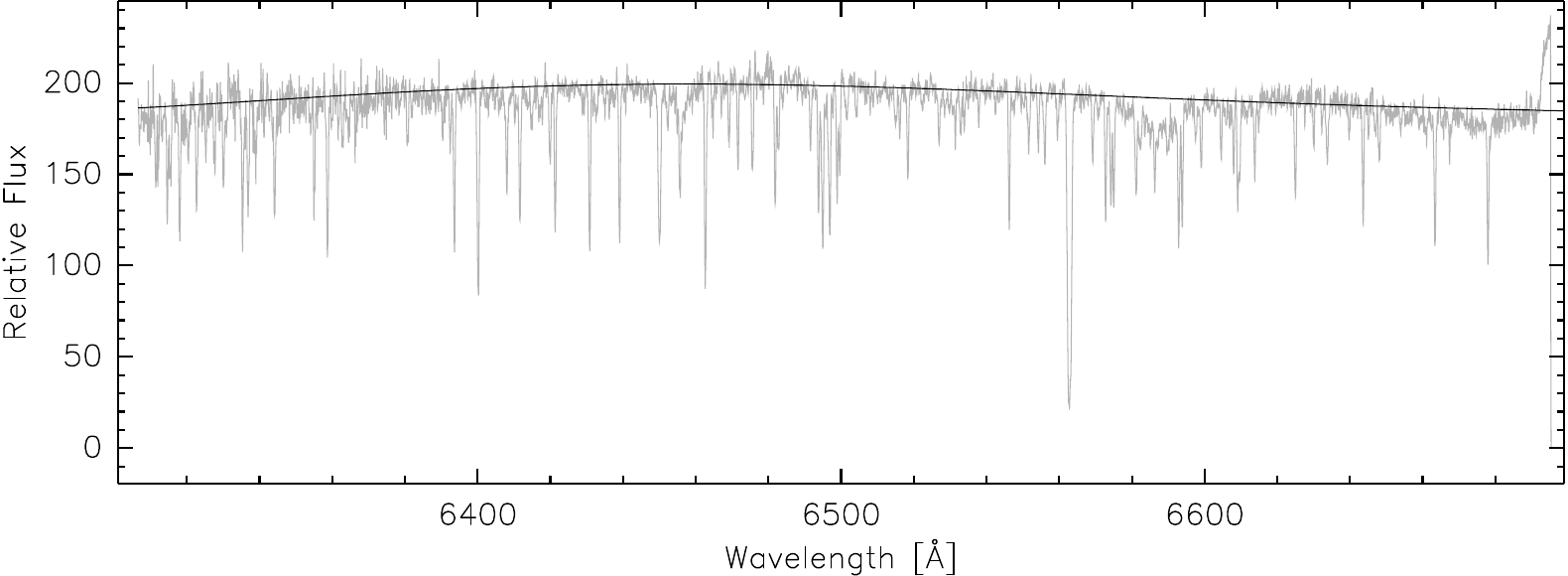}
\caption{Reduced spectra of star BL239, in each of the three GIRAFFE
setups, HR10, HR13 and HR14, from top to  bottom. The continuum fitted
by \texttt{DAOSPEC} is also shown.
\label{fig:H101314}}
\end{center}
\end{figure}

The heliocentric corrected multiple exposures of each setup were
combined with the \texttt{IRAF}\footnote{{http://iraf.noao.edu/}} task
\texttt{scombine}, using a flux weighted average with median sigma
clipping (for cosmic ray removal). This was done independently for
each period (September and January), due to the change of grating.
These two stacked spectra were finally combined in a flux weighted
average.  For the two HR14 set-ups with different resolution and
wavelength coverage, we only used their overlapping section ($\sim$
6400 -- 6600 \AA\ ). We convolved the higher resolution (old) HR14
spectra to the lower resolution HR14 one.  Figure~\ref{fig:H101314}
provides an example of the extracted spectra of BL239 in the three HR
setups. BL239, with V = 18.5 mag and S/N of 32 at 5500 \AA\, 37 at
6200\AA\ and 50 at 6500\AA\ is typical of our stellar sample .

\subsection{Radial velocity measurements (\vrad)}
\label{part:determiningVrad}

In order to verify   membership to Fornax, the radial velocity (\vrad)
of each star  has been measured independently  in  each setup (HR10,
HR13 and  HR14)  and for each   observing epoch  (September  and
January), producing six  independent measurements. These measurements
were made with  the girBLDRS routine \texttt{giCrossC.py},
cross-correlating the observed spectrum with  a mask. We  used a G2
spectral-type  mask that although hotter than our targets, was
found the most robust to yield the velocities of our metal-poor cool
giants.  Table~A.1 lists the final \vrad, calculated as the weighted
average of the six velocities, with their associated error,
corresponding to their standard deviation.  We found no systematic
difference between setups nor between epochs.  However, 8 stars were
found to have significantly different velocities between the September
and January runs ($>$ 3 km/s difference) and hence they were discarded
from any further analysis, as they might be binary stars.

\begin{figure}
\includegraphics[width=\hsize]{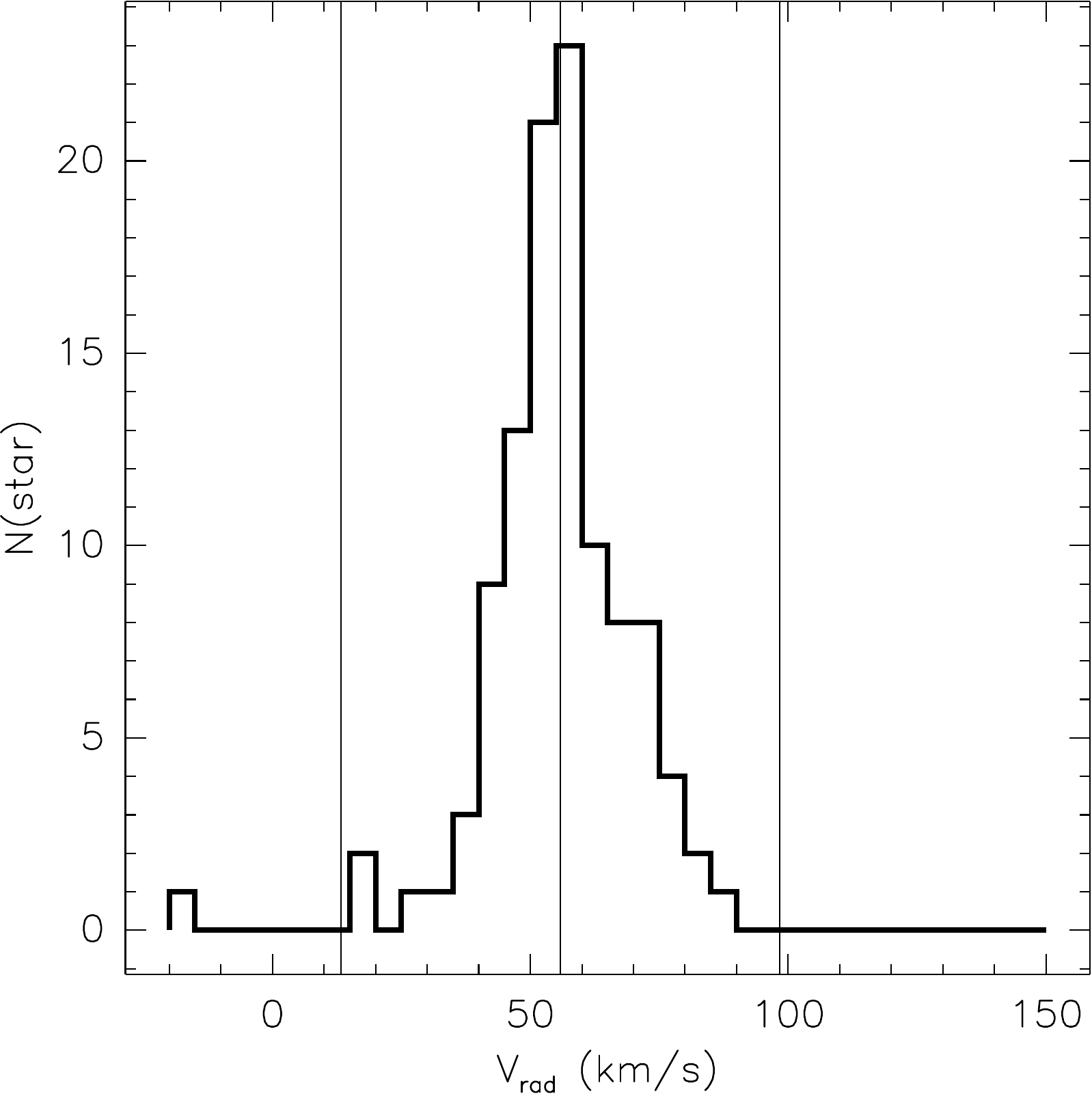}
\caption{Histogram of Fornax FLAMES/HR radial velocity measurements, \vrad.
The central value and the $\pm$ 3$\sigma$ cut-offs are shown with solid 
vertical lines. 
\label{fig:histovrad}}
\end{figure}

Figure~\ref{fig:histovrad} shows the final distribution of \vrad, and
the central velocity together with the 3$\sigma$ cut-offs defining
stellar membership of Fornax dSph.  Our criterion differs only slightly
from the 2.5$\sigma$ chosen in B06.  Only one star in our sample is an
obvious non member (BL109), with a negative \vrad\ and was excluded from
further analysis.  The final mean heliocentric velocity (\vrad) of the
stars identified as probable members is 55.9 km/s with a line of sight
velocity dispersion $\sigma$ = 14.2.  The typical (median) error on the
individual velocities measurement is $\simeq$ 0.55 km/s (see
Table~\ref{tab:fnx-vrad}).  Our results are very close to the values
derived in B06 from a Ca~II triplet measurements at lower resolution
(\vrad\ = 54.1 $\pm$ 0.5 km/s and $\sigma$ = 11.4 $\pm$ 0.4 km/s for
2$\sigma$ selection and $\sigma$ = 13.7 $\pm$ 0.4 km/s for a 3$\sigma$
selection).

Some of our stellar member candidates have been further discarded for
various reasons such as noisy spectra or problem with stellar
atmosphere model fitting (e.g., too large a dispersion in [Fe/H]).

\subsection{Measuring the Equivalent Widths}

The equivalent widths ($EW$) of stellar absorption lines are classically
measured individually by hand (e.g., using \texttt{SPLOT} in
\texttt{IRAF}). The large data sets that FLAMES produces effectively
requires the use of an automatic procedure.  This has been developed as
\texttt{DAOSPEC}\footnote{{http://cadcwww.dao.nrc.ca/stetson/daospec/}}
\citep[][]{Stetson08}, a software tool that has been optimized for
GIRAFFE HR spectra.  \texttt{DAOSPEC} iteratively detects all lines in
the stellar spectrum, fits gaussians of fixed FWHM (in $\rm
km.s^{-1}$) on all detected features, subtracts them from the observed
spectrum, fits a continuum over the entire residual spectrum and
corrects for it, until the residuals reach a flat minimum. These
iterations are meant to converge on the list of detected line, the FWHM
of the lines, and the global continuum. Once convergence is achieved,
\texttt{DAOSPEC} cross-correlates the list of detected lines with a
user provided reference line list to obtain a radial velocity and line
identification for its detections. For a thorough description of the
algorithm and its performances, see \citet[][]{Stetson08}. The use of
gaussian approximation of fixed FWHM ((in $\rm km.s^{-1}$) is fully
justified in R=20000 resolution spectra where the instrumental profile
dominates over the astrophysical broadening of spectral lines. The
placement of the continuum is a critical step, as it influences the
measurements of $EW$s.  Since \texttt{DAOSPEC} was run independently on
each of the three setups, the continuum levels in each setup were also
independently determined. As a consequence, any problem in the continuum
level determination would show up as a systematic difference in the
abundances deduced from lines of a given element in the different
setups.  Figure~\ref{fig:H101314} shows an example of the spectra for
one star in our sample (BL239) in the 3 setups and together with the
continuum placement.\\

\begin{figure}[!htp]
\begin{center}
\includegraphics[width=1.0\hsize,angle=0]{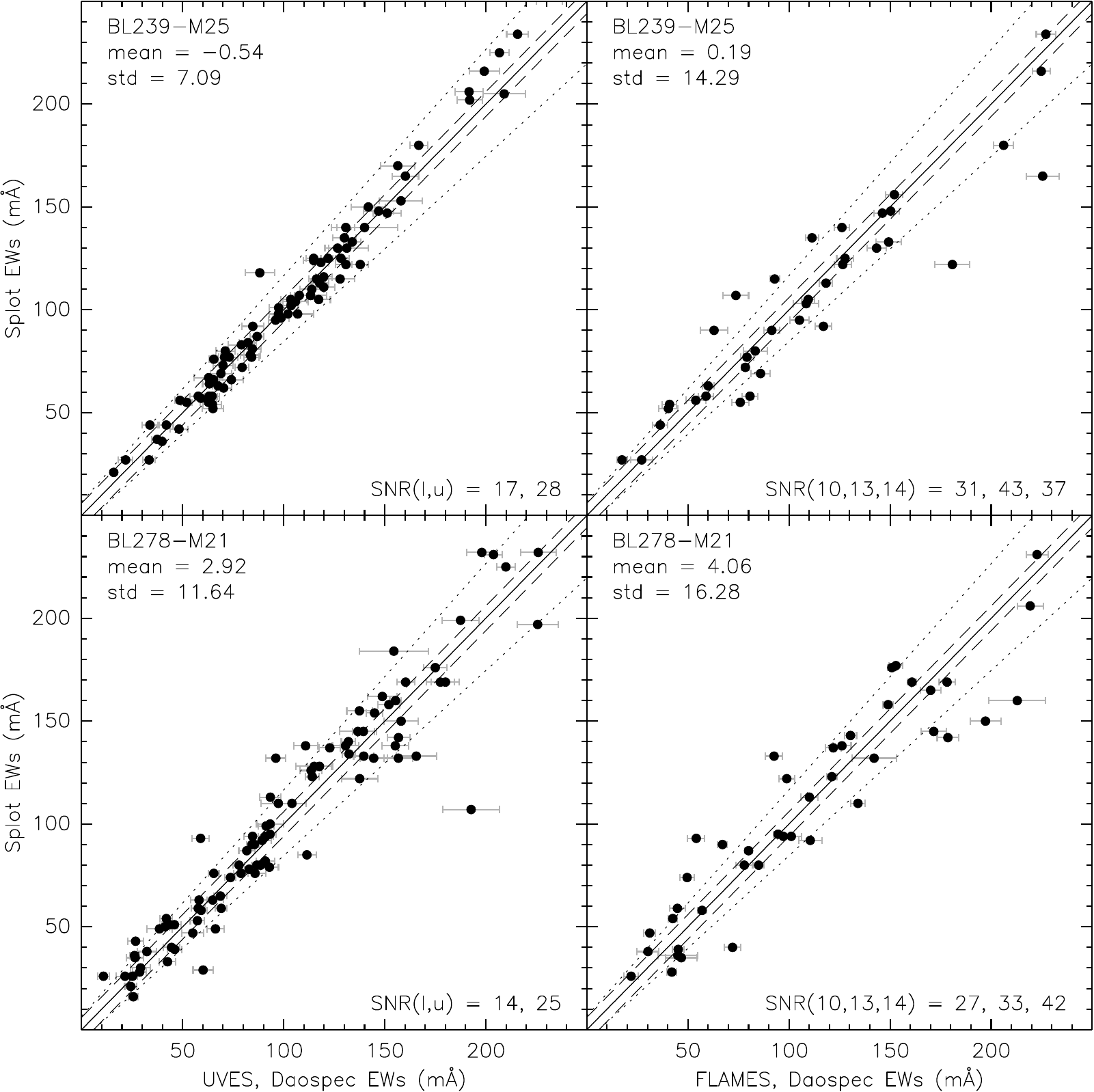}
\caption{A comparison of the SPLOT and \texttt{DAOSPEC} EW measurements for
GIRAFFE (right-hand side) and UVES (left-hand side) measurements of the
same two stars in Fornax dSph, BL239 (top panels) and BL278 (bottom
panels). The solid line shows perfect agreement between the two sets of
measurements. As in S03, the dashed lines delineate an $\pm$6~m\AA\
offset from this line, and the dotted lines represent a 10\% error
convolved with this 6~m\AA\ error.  These lines are representative of
the errors on \texttt{SPLOT}-based $EW$s.  We indicate in each panel in
the upper left-hand corner the mean difference between the UVES and
GIRAFFE measurements (mean) and the standard deviation (std).  The
signal-to-noise ratios (SNR) as given by \texttt{DAOSPEC} for the UVES
lower CCD (l, $\sim$4800--5800\AA), and the upper UVES CCD (u,
$\sim$5800--6800\AA) or the three different orders of the GIRAFFE
spectra are given in the bottom right-hand corner of each panel.
\label{fig:daospl}}
\end{center}
\end{figure}

To test for the reliability of \texttt{DAOSPEC} measured $EW$s,
Figure~\ref{fig:daospl}  compares, for a given star, the equivalent
widths   measured    on  the UVES      spectrum  with \texttt{SPLOT} by
S03 to  our \texttt{DAOSPEC} measurements on the  same UVES spectrum
and, finally, to the $EW$s measured on the GIRAFFE spectrum.  This
exercise has been performed for two different stars, that we have in
common with S03 (resolution $R \sim$ 43000 and $\Delta \lambda$ = 4800
-- 6800 \AA).  The mean difference between the measurements performed
with \texttt{SPLOT} and \texttt{DAOSPEC} on the  UVES spectra is minimal
and is not increased by decreasing  signal-to-noise  (SNR),  as can
be judged from the left column of the figure.   Conversely, as
expected,  the dispersion of  the   measurements increase  with
decreasing  spectral quality, i.e.   with decreasing  SNR  at the  lower
resolution  of the GIRAFFE spectra.  Nevertheless,  we find that
\texttt{DAOSPEC} can  be used with confidence for $EW \lesssim 200$
m\AA\. We note that in the comparison of EWs measured from UVES and
GIRAFFE spectra, the error is dominated by the GIRAFFE measurement, as
expected from the lower SNR {\it} per \AA\ of these spectra. For the two
stars exemplified in the right column of Figure~\ref{fig:daospl}, the
corresponding SNR per \AA\ are respectively:
(UVES,GIRAFFE)=(231,118),(238,163),(198,119) for BL239-M25 in H10,H13
and H14, and (UVES,GIRAFFE)=(175,103),(161,125),(186,135) for BL278-M21
in H10,H13 and H14.

We did not observe a fast rotating hot star, hence we were unable to
directly correct  our  spectra  for  telluric  absorption. Instead  we
removed from our abundance analysis the  atomic lines that fell within
one   resolution  element of a telluric  line.\\

\section{Abundance measurements}\label{sect:abundances}

\subsection{Stellar Models}
\label{sect:StellarModelsUsed}

The release of the spherical MARCS
models\footnote{{http://marcs.astro.uu.se/}} represents a major
improvement in the modeling of stellar atmospheres
\citep{2003ASPC..288..331G,2008A&A...486..951G}. Prior to this, most of
the  abundance analyses of RGB  stars  were based on  the models of
\citet{1975A&A....42..407G}, in plane-parallel  approximation, meaning
that  radiative  transfer were solved   in only  one  depth  variable,
neglecting  the curvature of  the atmosphere.  The new MARCS spherical
models  have  been further expanded  by  B. Plez to  cover the extreme
range of parameters of our RGB sample, i.e., \teff  $< 4000$ K, [Fe/H] $
\sim -3.0$\,dex,  and low gravities at  all metallicities (B.  Plez,
private communication).\\

Due to their relative novelty, spherical models have not yet been widely
adopted in the literature (however see \citet{1992A&A...256..551P} and
\cite{1992A&AS...94..527P}).  A thorough comparison of the models with
spherical and plane-parallel geometries has been conducted by
\citet{2006A&A...452.1039H}.  One of their main conclusions is that the
structure of the spherical models is very different to the
plane-parallel ones, and this leads to significantly different stellar
abundances.  However geometry has much less impact on line formation
(spectral synthesis).  The discrepancies between the fully spherical
(s\_s) and fully plane-parallel (p\_p) models increase with temperature
and decreasing gravity, reaching a maximum around 5000K after which they
decrease slightly.  More specifically, at \teff\ = 5000K,  the
difference in FeI  abundances can reach 0.15\,dex for \logg\ = 1.0, and
0.25 dex for \logg\ = 0.5, for equivalent widths of the order
$\sim$120m\AA\ as encountered in Fornax. By comparison, at \teff\ =
4000K and \logg\ = 1.0, the difference between (s\_s) and (p\_p) models
does not exceed $\sim$0.02\, dex.  Following
\citet{2006A&A...452.1039H}, we computed that a spherical model
atmosphere combined with a plane-parallel calculation for the line
formation (the so-called s\_p case) leads to maximum abundance deviation
of the order of 0.02--0.03\,dex, for temperatures \teff\ around 4000K
and \logg\ $\sim$ 0.5 (typical of our sample). These differences are
negligible in view of   our measurement errors, therefore we  have
carried out our abundance analyses in the s\_p framework, using the code
CALRAI, first  developed  by \citet{1967AnAp...30..211S} and
continuously updated over the years.

\subsection{The line list}
\label{sect:linelist}

The  choice of  a  line  list  is a   critical  step of  the abundance
analyses. To be selected, the lines must  have reliable \gf-values and
be sufficiently isolated from other lines to be accurately measured at
the resolution of the observations.  Dealing with cool stars (the mean
temperature  of our  sample is  $\sim$4000K)  impose to  check for the
presence of molecular  lines which  could contaminate the atomic ones,
hence biasing the measurements. The main molecular contaminant in
the wavelength domain and stellar parameters covered here is in fact CN,
while TiO has been checked to be negligible: the strongest TiO features
in our domain are around 6300\AA, but although TiO does form in
these cool atmospheres, it strongly decrease with both the gravity
and metallicity, making it a negligible contaminant to neighboring
atomic lines.

We started with the reference line list of S03, which was optimized for
UVES and metal poor stars ($-3.0 \leq$ [Fe/H] $\leq -1.5$), and adapted
it to the higher average metallicity of our Fornax field stars and the
smaller wavelength coverage of FLAMES.  To do this we selected new lines
from \citet{2003A&A...404..187G}, \citet{2000A&A...364L..19H}, and
\citet{2004A&A...423..507Z} with \gf-values on the same scale as S03. We
checked for possible systematic differences in \gf-values using a high
resolution ($R \approx  120\,000$), high signal-to-noise, UVES spectrum
of Arcturus\footnote{{http://archive.eso.org/wdb/wdb/eso/uves/form}},
from which we derived the abundances from all our lines.  Finally, we
removed lines with possible blends (molecular or atomic) at the
resolution of GIRAFFE (R=20000), using synthetic spectra computed for
our range of stellar parameters.  The final line list is shown in
Table~\ref{tab:fnxfi-ewlist1}

\subsection{Stellar parameters}

We  started with   photometric  estimates for  \logg\ and  \teff,  and
set [\fei/H]=$-$1.0\,dex  and \vmic\  =  2.1\,km/s.  We subsequently
modified these  initial   values until  a good  fit   to a model was
obtained, as summarized in the following sections.

\subsubsection{Photometric effective temperature (\teff)}
\label{sec:phototeff} 

Optical photometry ($V$, $I$) was available for the full sample (B06)
and was complemented by infrared photometry ($J$, $H$,  $K$) from
\citealt{2007A&A...467.1025G}, for 60\% of our sample.  We considered
four different colours, $V-I$, $V-J$, $V-H$ and $V-K$ to estimate
\teff, following the calibration of \citet{2005ApJ...626..465R},
with  a reddening  law of $(A(V)/E(B-V)   =   3.24)$  and  an
extinction $E(B-V)$  =   0.03 \citep{2000ApJ...543L..23B}.

The three Fornax stars in common with S03 were used to check the
photometric temperature scale against excitation (as constrained
from their UVES spectra). Namely, we determined, for these three stars,
the spectroscopic effective temperature for which the iron abundance
deduced for lines of various excitation potentials (\kiex) was constant
and showed no variation with \kiex.  The \teff($V-I$) were in perfect
agreement with the spectroscopic \teff,  while   the   temperatures
derived   from  the  IR     colours, \teff($V-\{J,H,K\}$), were slightly
offset. This could be due to various effects, including a possible
zero point on our WFI V and I photometry. Because of the agreement of
the \teff($V-I$) with the excitation temperature scale, we decided to
shift the \teff($V-\{J,H,K\}$) onto the \teff($V-I$) scale by applying
constant shifts of 87\,K, 109\,K   and 99\,K  respectively for
\teff($V-\{J,H,K\}$). This procedure  is  equivalent to
zero-pointing our photometry to make it agree with the excitation
temperature scale.  For a given  star, the different \teff\ then agree
within  $\pm$ 50\,K. \\

The colours predicted by the MARCS 2005 atmosphere models at different
temperatures, gravities and metallicities were   also compared to  the
temperature calibration of \citet{2005ApJ...626..465R}, and found to
agree well. This is comforting, since it   means that the  model
atmospheres  we used  to deduce abundances also produce the correct
colours for  our stars.  However we did notice that  the \teff\ derived
on  the $V-\{I,  J\}$ colours are much more sensitive to the gravity of
the model than $V-\{H, K\}$, an   effect   that    is  neglected     in
the  calibrations     of \citet{2005ApJ...626..465R} and may account for
some of the systematic differences  between  temperatures deduced  from
$V-I$  and  infrared colours.

 Excitation temperatures were also examined for the whole sample,
but found a relatively high uncertainty associated with the slope of
measured iron abundance with the line \kiex when using GIRAFFE spectra.
The median error associated with the slope measurement for the total
sample (0.034\,dex.eV${-1}$) corresponds to a \teff\ uncertainty of
$+130/-250$\,K. We therefore decided to rely the photometric
temperatures for the analysis. The relatively high \teff\ uncertainty of
200\,K quoted in Table~\ref{tab:fnxfi-errors} reflects a conservative
errorbar associated with the color temperature scale and photometric
zero-point uncertainties. 

The final \teff\ we used is the average of the four \teff, coming from
the four photometric colours and they are presented in
Table~\ref{tab:fnx-stelparmod}.

\subsubsection{Surface gravity (\logg)}
\label{sec:photologg}

We used the photometric colours to estimate the surface gravity of the
sample stars using the standard relation:\\

\noindent
$\logg_{\star} = \logg_{\odot} +
\log{\dfrac{\mathcal{M}_{\star}}{\mathcal{M}_{\odot}}} + 
4 \times \log \dfrac{T_{\mathrm{eff}\star}}{T_{\mathrm{eff}\odot}} + \\
~~~~~~~~~~~~0.4 \times (M_{\mathrm{Bol}\star} - M_{\mathrm{Bol}\odot})$
\\

We adopted the distance modulus $(M-m)_V = 20.65$ of
\citep{2000ApJ...543L..23B} and the mass of the RGB stars was assumed to
be 1.2 \msol, in agreement with the mean young age for are sample,
around $\sim 2-5$\,Gyrs (Fig.\ref{fig:agez}). Because the sample also
contain a small fraction of older stars of lower masses, we iterated the
analysis for those stars by modifying the gravity using the
corresponding isochrone mass, once the metallicity and age of the star
had been determined.

The bolometric  corrections were  computed  for  each star  using  the
calibration of \citet{1999A&AS..139..335A}.   The gravity had  a minor
effect on our abundances deduced from neutral ions: lowering \logg\ by
0.5\,dex decreases  \fei\ by $\approx$ 0.1\,dex and \feii\ by $\approx$
0.3\,dex.  We determined that it was preferable to use the same
photometric \logg\ scale for all stars rather than using the ionization
balance of Fe, because our \feii\ abundances were deduced from too few
and too weak lines to warrant a precise ionisation balance
determination. This is illustrated in the bottom panels of figure
\ref{fig:slopes_sigma} where the distribution of the errors on the mean
\ion{Fe}{i} and \ion{Fe}{ii} are reported.  \ion{Fe}{ii} is clearly
quite uncertain, with a median error of 0.13\,dex, and extending up to
0.2dex or more. The corresponding uncertainty on \logg\ determined from
ionisation balance would be of 0.3\,dex or more. The same holds true
for Ti, for which the \ion{Ti}{II} abundance was too uncertain (derived
from 2-3 lines).

\subsubsection{Microturbulent velocity (\vmic)}
\begin{figure}
\includegraphics[width=\hsize]{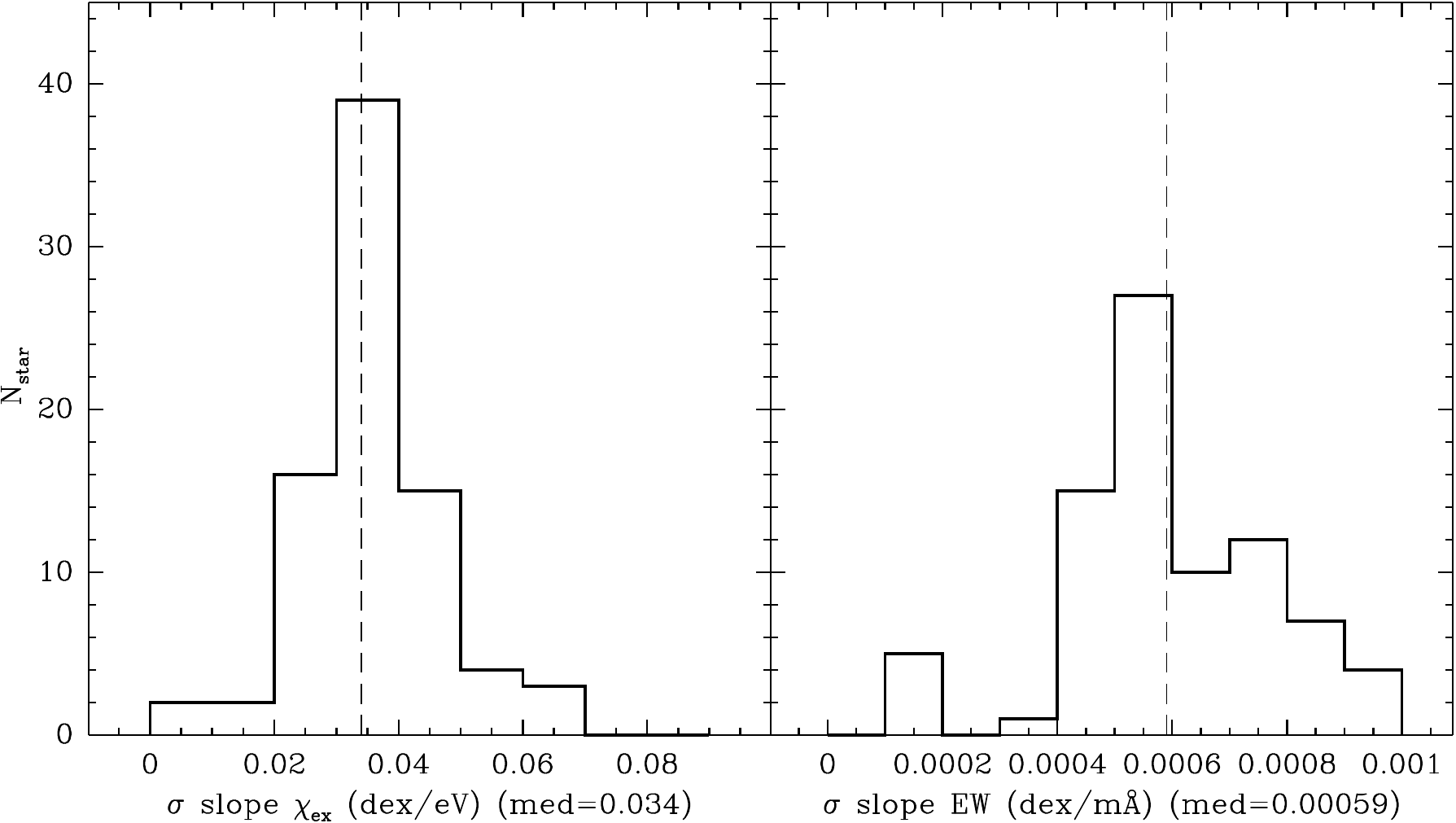}
\includegraphics[width=\hsize]{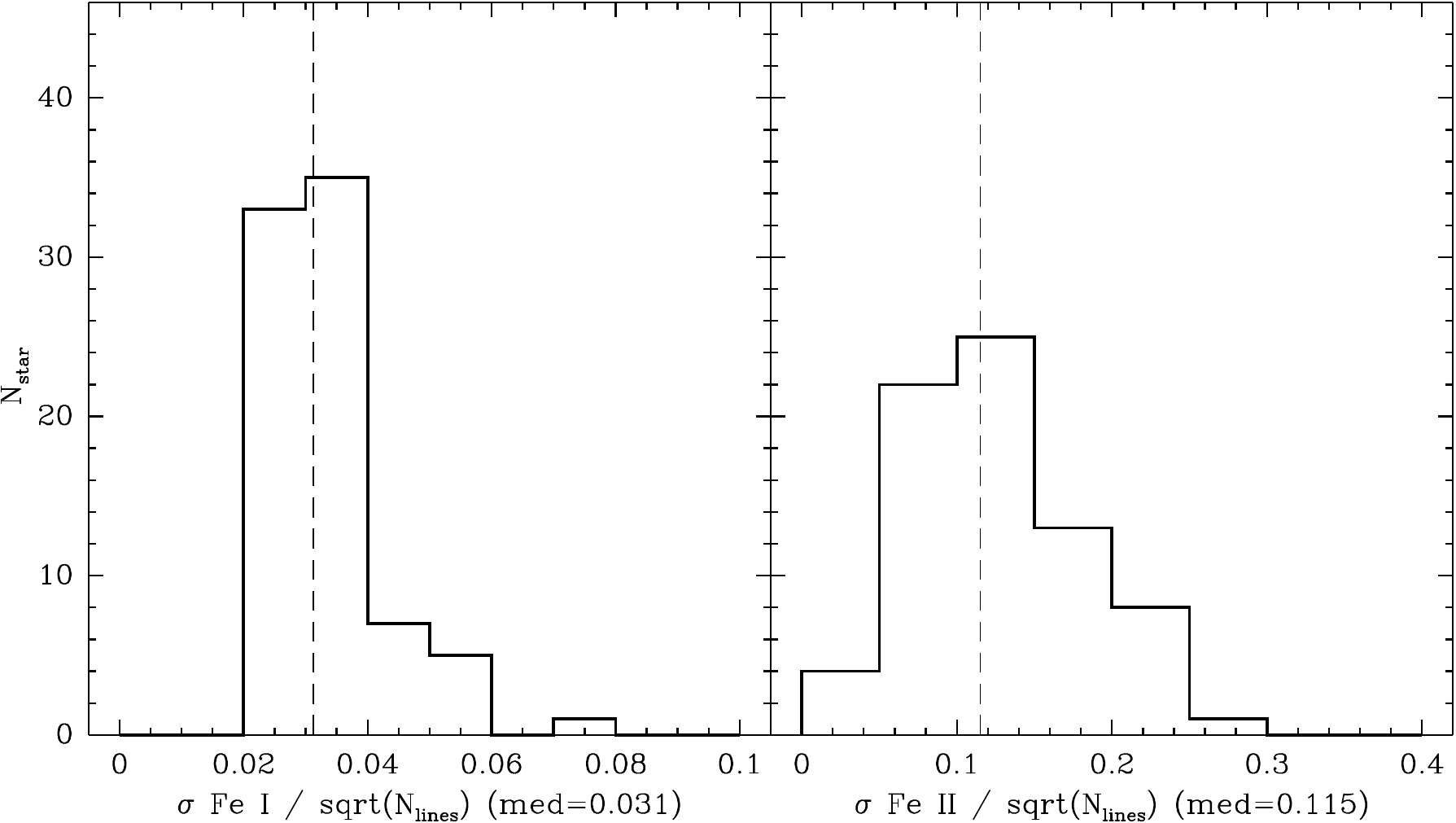}
\caption{ Distribution of errors of the stellar parameter diagnostics 
for the total sample. 
{\em upper panels:} propagated error on the measurement of the slope of 
\ion{Fe}{i} abundances versus the excitation potential of the line 
({\em \kxbs\ left}), and versus the line strength ({\em \ewbs\ left}).
{\em lower panels:} error on the mean \ion{Fe}{i} and \ion{Fe}{ii} 
abundances as measured by the dispersion around the mean divided by 
the square root of the number of lines of each species.
\label{fig:slopes_sigma}}
\end{figure}

The \vmic\ was corrected until the  slope measured between [\fei/H] and
$EW$ (\ewb) was  zero (within its $1\sigma$ error).  We  took
advantage of the linearity (and symmetry) of \vmic\ on \ewb\ in order to
converge rapidly, applying the linear relation between the \ewb\ and the
\vmic\, empirically measured to be  $\delta\vmic\ = \delta~\ewb /
0.0055$.  The median uncertainty on the determination of \ewb\ of
$0.00059$\,dex.m \AA$^{-1}$ (see Fig.~\ref{fig:slopes_sigma}) corresponds
to $0.11\, \mathrm{km.s}^{-1}$, and the complete sample has formal
uncertainties below $0.001\, \mathrm{dex.m \AA}^{-1}$, corresponding to
$0.2\, \mathrm{km.s}^{-1}$.

\subsection{Precision and error estimates}
\label{sect:precision}

\texttt{DAOSPEC} provides an error estimate, $\delta EW$, for each
measured $EW$ \citep[see description of the $\delta EW$ measurements in
][]{Stetson08}. We propagated this $EW$ uncertainty throughout the
abundance determination process, thus providing for each line the
abundance uncertainty corresponding to $EW \pm \delta EW$.  The
abundance uncertainty need not be symmetric (and in general it is not),
so the largest of the two (upper and lower) uncertainties was adopted as
our \texttt{DAOSPEC} abundance error, $\delta_{\mathrm{DAO}}$.  We also
considered  the abundance dispersion around the mean ($\dfrac{\sigma
(\mathrm{X})}{\sqrt{(N_{\mathrm{X}})}}$) for the element X for which
N$_X$  lines could be measured (when N$_X$ $>3$), which provides an
estimate of error including both the EW mesurement and \gf-values
uncertainties.  Finally, to avoid biased estimates of the abundance
dispersion due to small number statistics, we also considered the
dispersion in \fei\  abundances as a lower limit for the expected
abundance dispersion around the mean of any element, leading to an error
estimate of $\dfrac{\sigma (\mathrm{Fe\,I})}{\sqrt{(N_{\mathrm{X}})}}$.
For the final error on each [X/H], we adopted the maximum of these three
values:

\begin{equation}
\delta(\mathrm{[X/H]}) = \mathtt{MAX} \left( \delta_{\mathrm{DAO}},
\dfrac{\sigma (\mathrm{Fe\,I})}{\sqrt{(N_{\mathrm{X}})}},
\dfrac{\sigma (\mathrm{X})   }{\sqrt{(N_{\mathrm{X}})}}\right) 
\end{equation}

\noindent
The error  on  the abundance  ratios,   [X/Fe] was calculated  as  the
quadratic sum  of the errors on  [X/H] and  [Fe/H].  This conservative
estimate includes  all sources of error  due  to the measurements, and
will   be    used   throughout  the    figures   and    discussions  of
Sect.~\ref{sect:results} and \ref{sect:discussion}.

The uncertainties caused by our choice of stellar parameters were
estimated by varying the stellar parameters of each star by their
uncertainties, and comparing the abundances computed with these modified
parameters to the nominal abundance. Those errors, averaged over our
sample and combined quadratically, are presented in
Table~\ref{tab:fnxfi-errors}. Note that combining those errors
quadratically explicitely ignores covariances between parameters, and is
therefore an upper limit for the total error \citep[see e.g.
][]{1995AJ....109.2757M,2002ApJS..139..219J}.

\begin{table}[!htb]
\begin{center}
\caption{Errors due to uncertainties in stellar parameters.
\label{tab:fnxfi-errors}}
\texttt{
\begin{tabular}{lrrrr}
\hline
\hline
Element & $\Delta$ \teff\,=&  $\Delta$ \logg\,=& $\Delta$ \vmic\,=& Combined\\
 & +200\,K&  -0.5\,dex & +0.2\,km/s & \\
\hline
\,[Na\,{\sc i}/H] & -0.18&  0.01&  0.01&  0.18\\
\,[Mg\,{\sc i}/H] & -0.06&  0.00&  0.06&  0.08\\
\,[Si\,{\sc i}/H] &  0.14&  0.12&  0.02&  0.19\\
\,[Ca\,{\sc i}/H] & -0.23&  0.04&  0.07&  0.24\\
\,[Ti\,{\sc i}/H] & -0.33&  0.08&  0.04&  0.34\\
\,[Ti\,{\sc ii}/H]&  0.10&  0.21&  0.04&  0.24\\
\,[Cr\,{\sc i}/H] & -0.32&  0.11&  0.09&  0.35\\
\,[Fe\,{\sc i}/H] & -0.04&  0.12&  0.08&  0.15\\
\,[Fe\,{\sc ii}/H]&  0.32&  0.28&  0.04&  0.43\\
\,[Ni\,{\sc i}/H] & -0.03&  0.14&  0.05&  0.15\\
 \,[Y\,{\sc ii}/H]&  0.06&  0.19&  0.02&  0.20\\
\,[Ba\,{\sc ii}/H]& -0.03&  0.15&  0.26&  0.30\\
\,[La\,{\sc ii}/H]& -0.05&  0.23&  0.04&  0.23\\
\,[Nd\,{\sc ii}/H]&  0.00&  0.20&  0.02&  0.20\\
\,[Eu\,{\sc ii}/H]&  0.03&  0.22&  0.04&  0.22\\
\hline
\end{tabular}
}
\end{center}
\end{table}

\subsection{Hyperfine splitting correction}

\begin{figure}
\includegraphics[width=\hsize]{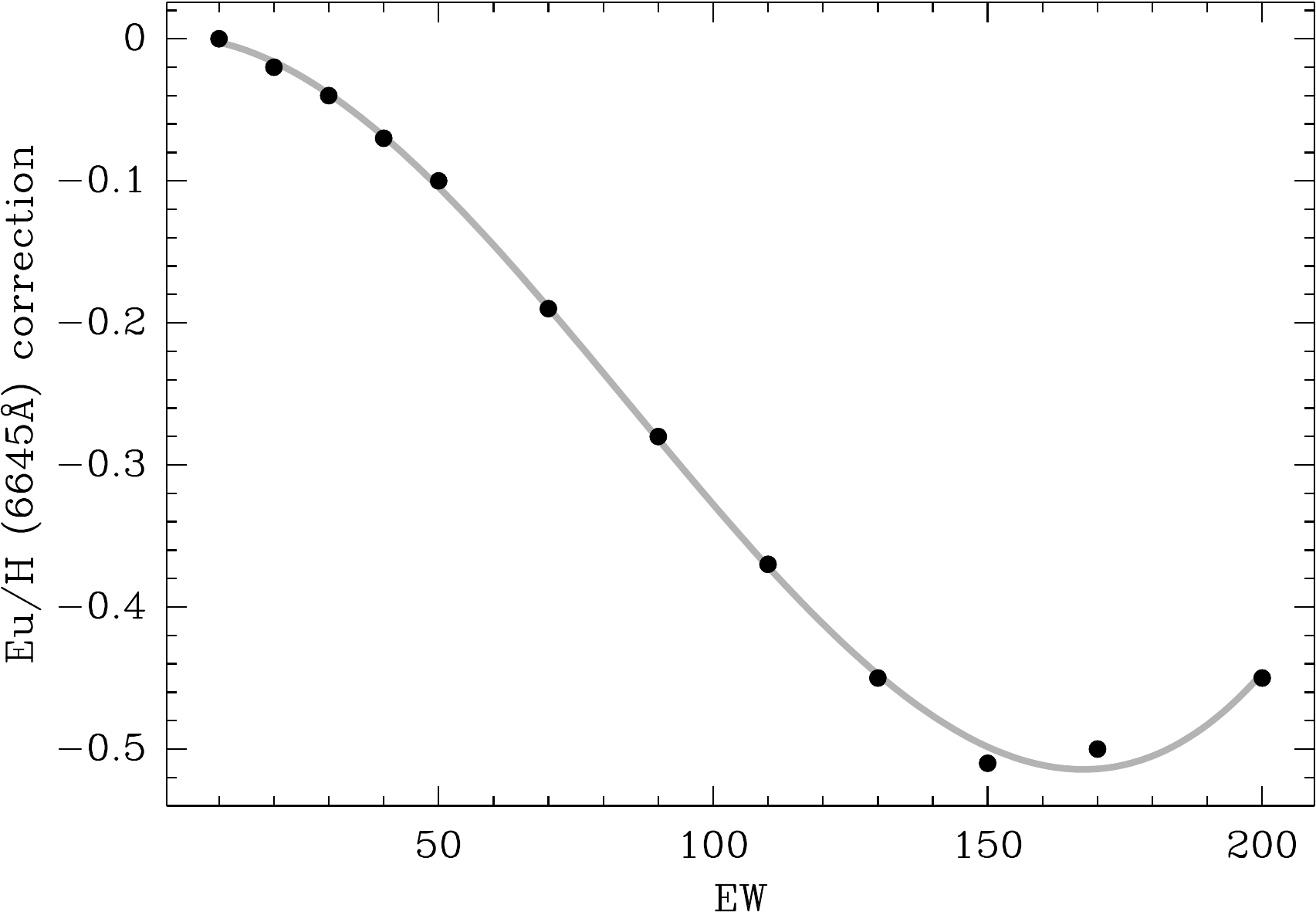}
\caption{HFS correction (dex) for the Eu line at 6645.1\AA, tested on a
plane-parallel model of \teff\ = 4300, \logg\ = 0.6, [Fe/H] = -1.5 and
\vmic\ = 1.7.
\label{fig:eucor_hfs}}
\end{figure}

Ignoring the hyperfine structure (HFS) of a line can lead to
overestimate the abundance of the element responsible for this line,
 as HFS acts to desaturate lines that are above the linear part of
the curve of growth. Given our large sample size, instead of computing
all lines one by one including HFS, we defined a generic {\em HFS
correction} per line to be applied to the abundance deduced without
HFS from the EW of each line.  We found that, for the range of stellar
parameters and abundances of our sample, this correction depends
predominantly on the $EW$s, and very little on any other parameter. As
an example, Figure~\ref{fig:eucor_hfs} displays the HFS correction for
Eu ($\lambda = 6645.1$ \AA ).  Similar corrections were derived for the
La ($\lambda = 6320.4$ \AA) line and applied to the sample.

\subsection{Systematics}
\label{sect:systematics}

Since the present work uses new automatic procedures,  new stellar
atmosphere  models  and a new line list  adapted  to the resolution and
wavelength range of GIRAFFE, it is  important to estimate possible
systematic errors that these changes would trigger.

Our sample includes the 3 stars observed by S03 with UVES that were
analyzed in a classical high resolution scheme. We ran checks in 5
steps each tackling a specific ingredient of the analysis presented in
this paper. At each step, we re-determined the stellar parameters
providing the best fit model.  The results are listed in
Table~\ref{tab:3shet}.\\


\begin{table}[!htp]
\begin{center}
\caption{The results of our investigation into possible systematic errors
in our procedures. The names of the three stars common to
the present work and to S03 are given as are their coordinates.
For each check we give the atmospheric parameters of the best
fit models, \teff, \logg, [Fe/H] and \vmic.
\label{tab:3shet}}
\ttfamily
\begin{tabular}{llrr}
\hline \hline

ID&     Shetrone&  RA(J2000)&    DEC(J2000) \\
\hline
BL239 &     Fnx-M25&      02 39 47.09&  -34 31 49.8 \\
BL266 &     Fnx-M12&      02 40 10.00&  -34 29 58.8 \\
BL278 &     Fnx-M21&      02 40 04.38&  -34 27 11.3 \\
\hline
\hline
\end{tabular}
\begin{tabular}{lrrrr}
\multicolumn{5}{c}{}\\
\multicolumn{5}{c}{Check 1}\\
\hline
Star & \teff& \logg& [Fe/H]& \vmic\\  
\hline
BL239& 4100&  0.00& -1.30&  2.2\\
BL266& 4300&  0.00& -1.60&  2.0\\
BL278& 4000&  0.60& -0.60&  1.7\\
\hline
\multicolumn{5}{c}{Check 2}\\
\hline
Star & \teff& \logg& [Fe/H]& \vmic\\  
\hline
BL239& 4100&  0.00& -1.30&  2.1\\
BL266& 4200&  0.00& -1.50&  2.1\\
BL278& 4000&  0.60& -0.60&  1.7\\
\hline
\multicolumn{5}{c}{Check 3}\\
\hline
Star & \teff& \logg& [Fe/H]& \vmic\\  
\hline
BL239& 4100&  0.60& -1.00&  2.0\\
BL266& 4200&  0.70& -1.50&  1.9\\
BL278& 4000&  0.60& -0.70&  2.0\\
\hline
\multicolumn{5}{c}{Check 4}\\
\hline
Star & \teff& \logg& [Fe/H]& \vmic\\  
\hline
BL239& 4100&  0.60& -1.00&  2.0\\
BL266& 4200&  0.70& -1.50&  1.9\\
BL278& 4000&  0.60& -0.70&  2.0\\
\hline
\multicolumn{5}{c}{Check 5}\\
\hline
Star & \teff& \logg& [Fe/H]& \vmic\\  
\hline
BL239& 4123&  0.68& -0.91&  2.1\\ 
BL266& 4212&  0.83& -1.44&  2.0\\ 
BL278& 4072&  0.64& -0.72&  2.3\\ 

\hline
\end{tabular}
\end{center}
\end{table}

\begin{figure*}[!Hb]
\centering
\includegraphics[width=0.49\textwidth]{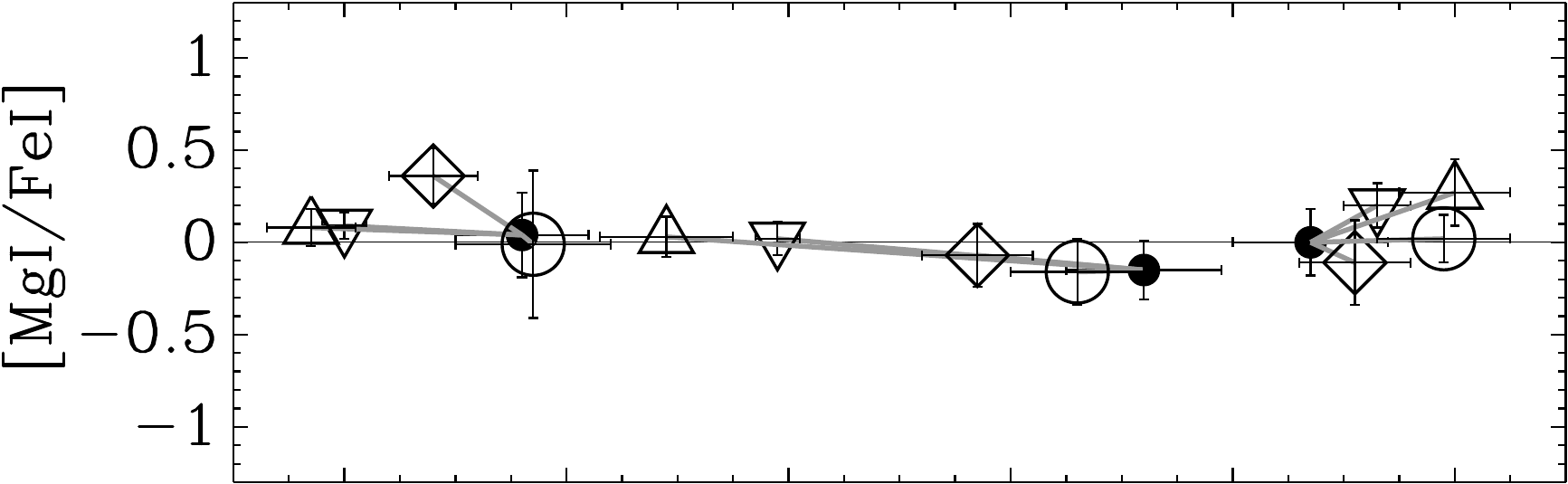}
\includegraphics[width=0.49\textwidth]{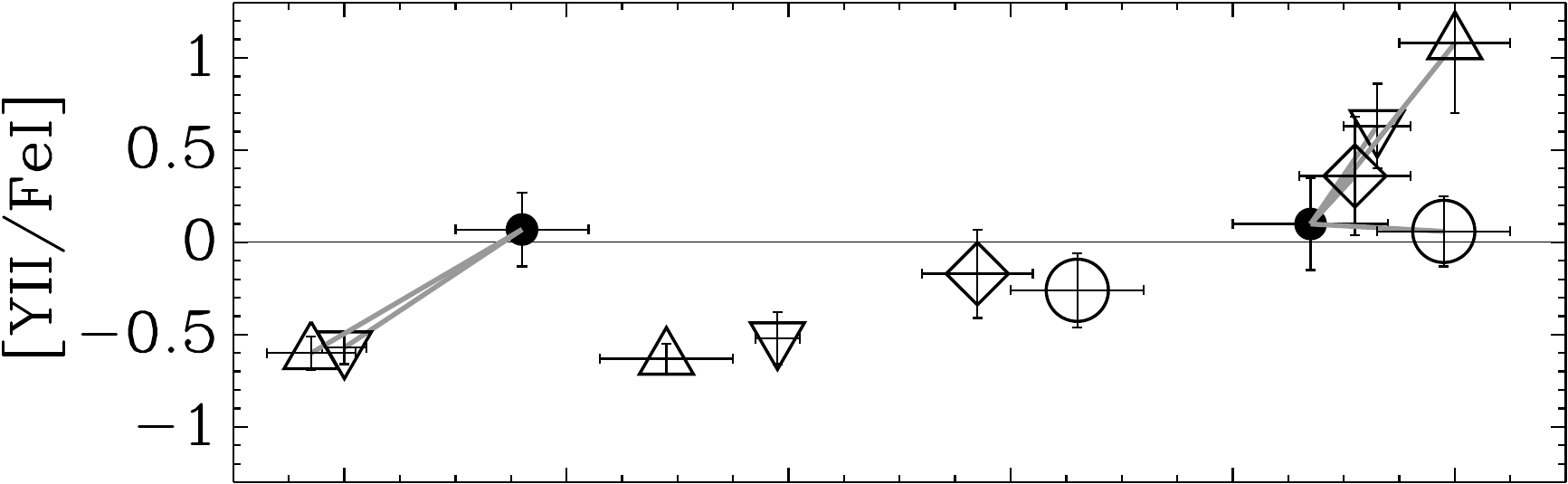}
\includegraphics[width=0.49\textwidth]{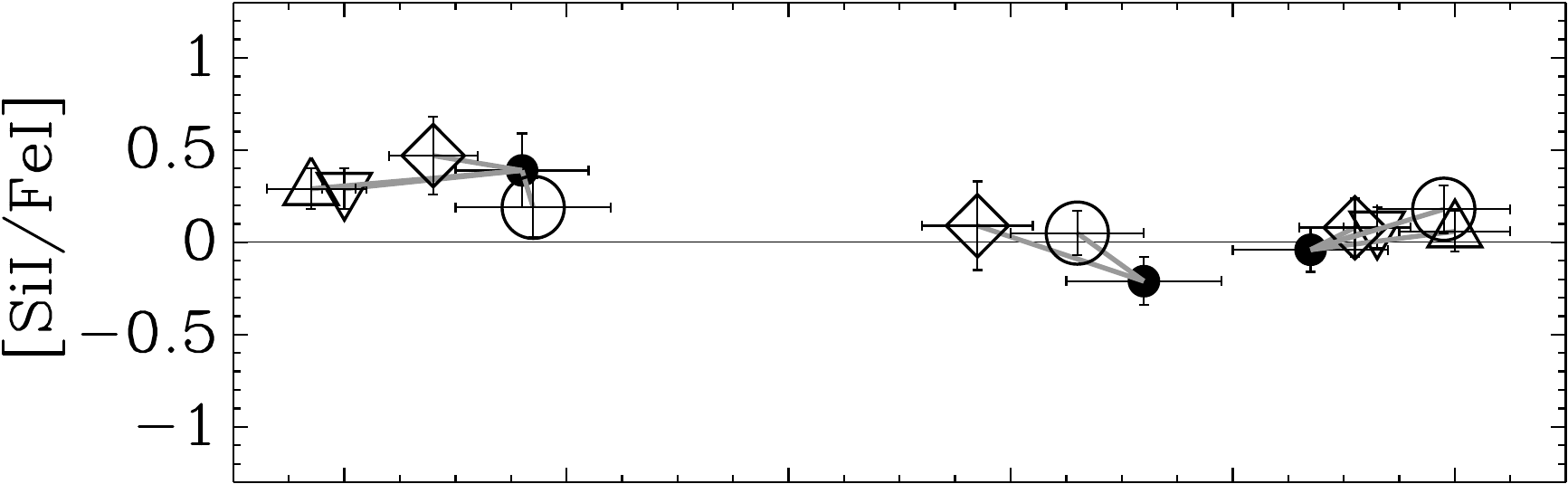}
\includegraphics[width=0.49\textwidth]{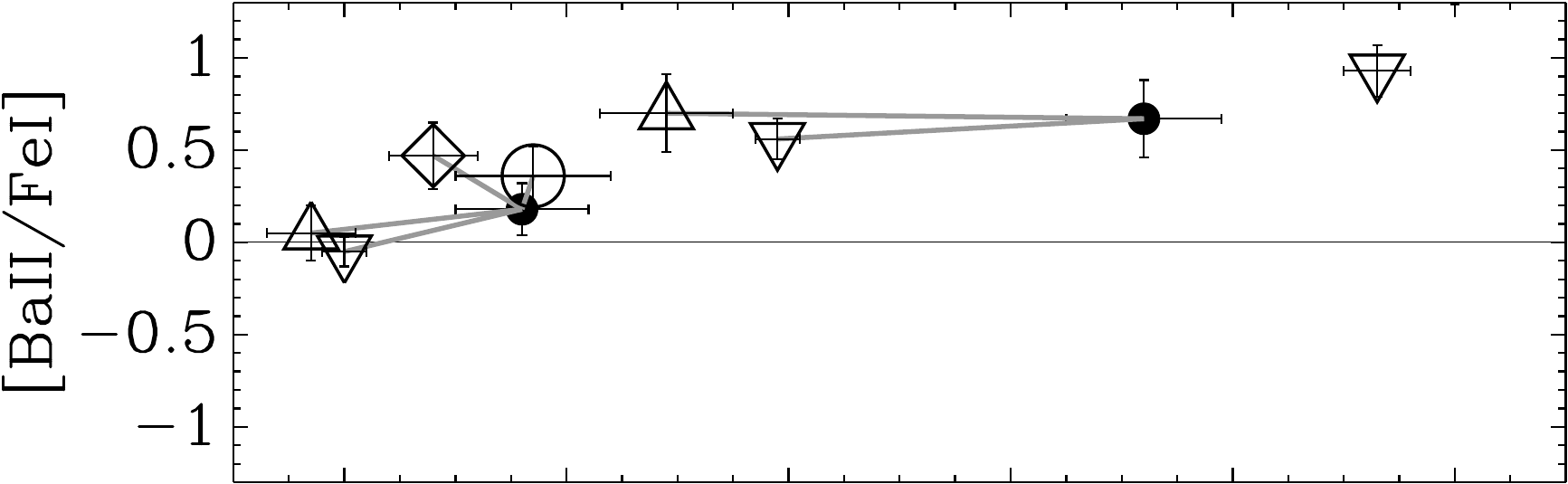}
\includegraphics[width=0.49\textwidth]{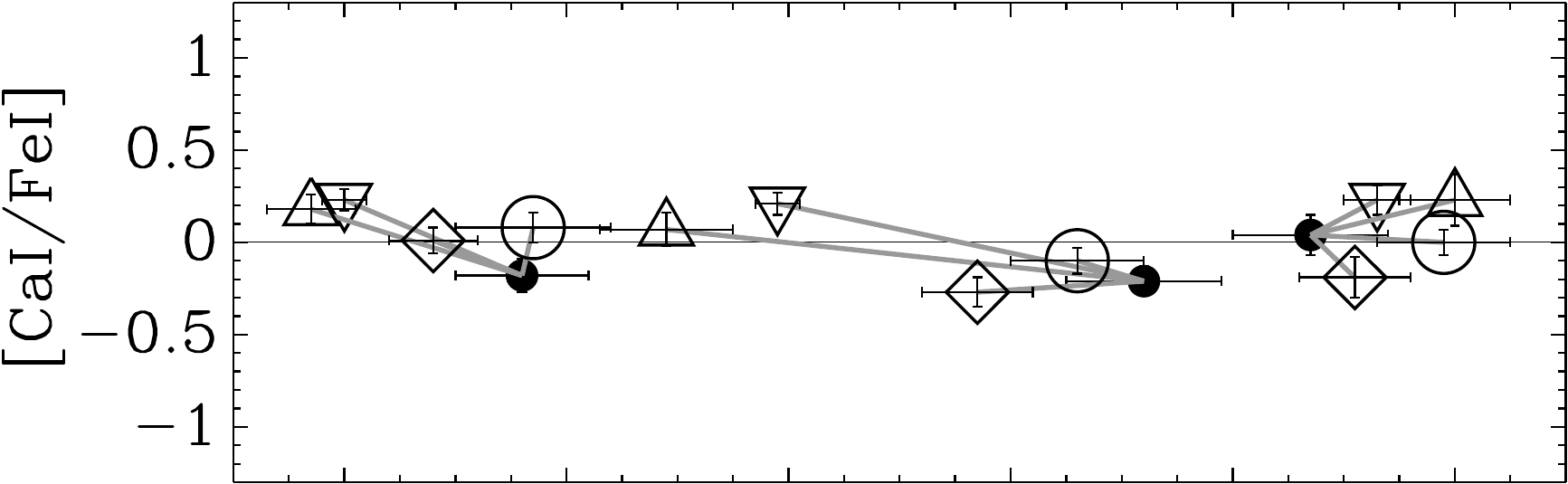}
\includegraphics[width=0.49\textwidth]{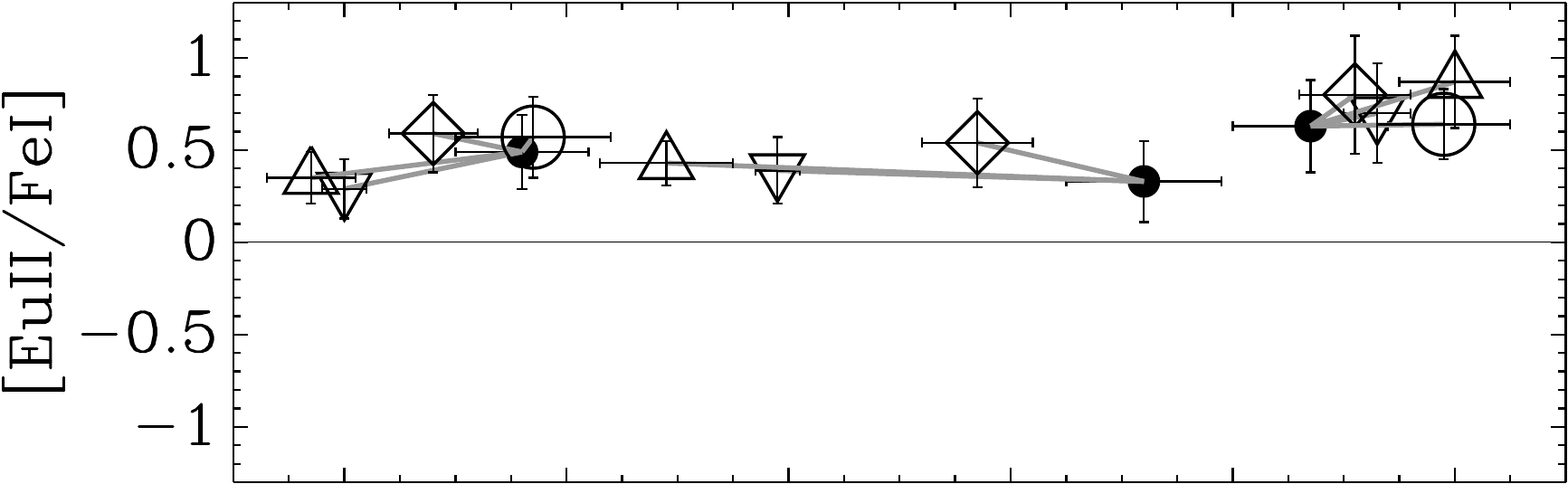}
\includegraphics[width=0.49\textwidth]{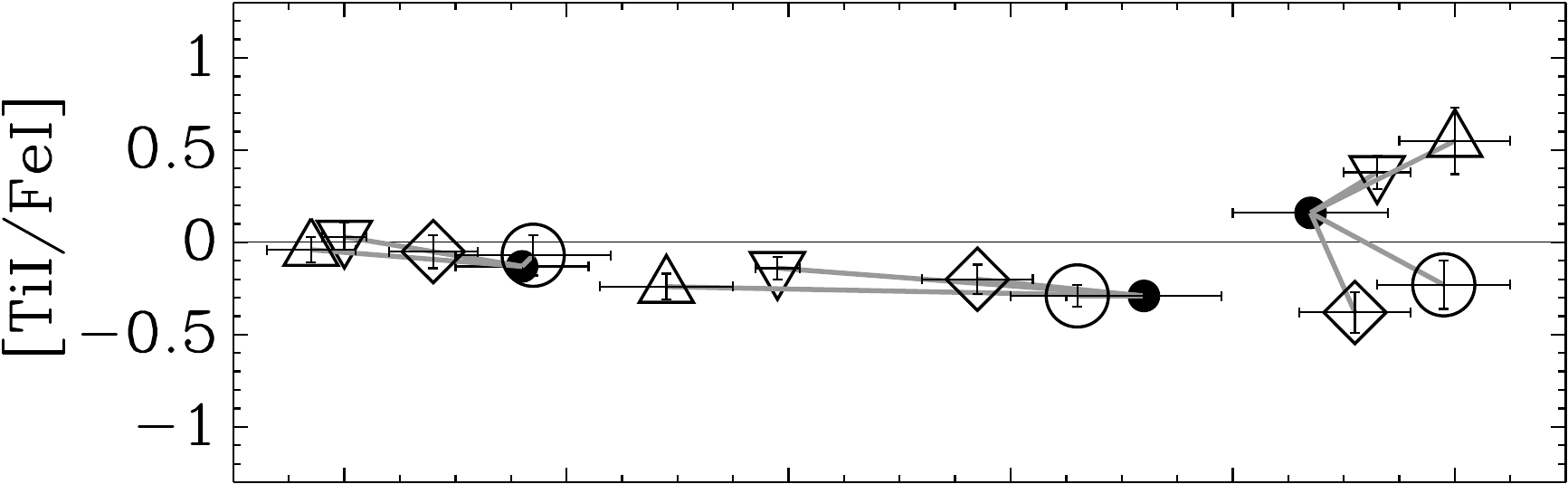}
\includegraphics[width=0.49\textwidth]{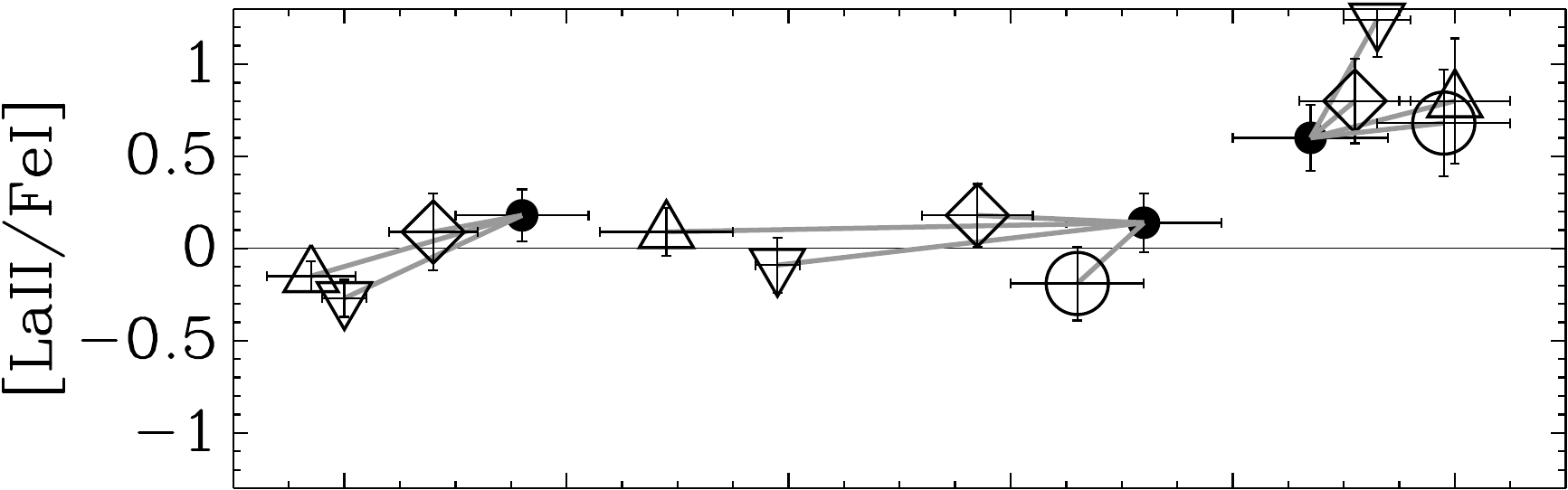}
\includegraphics[width=0.49\textwidth]{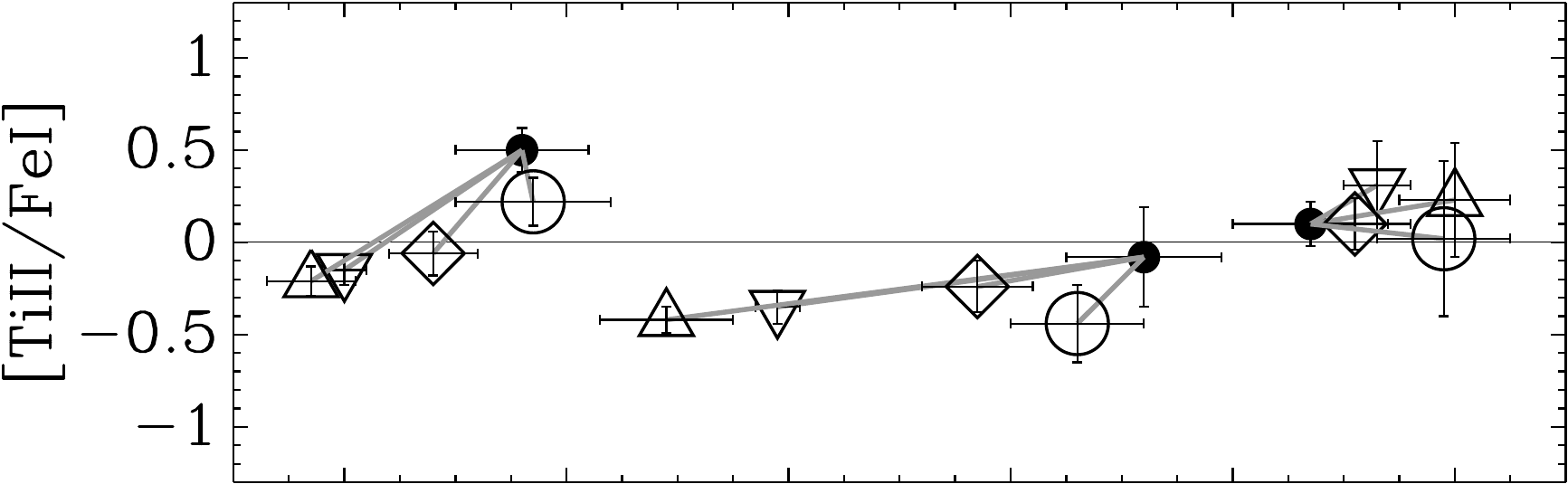}
\includegraphics[width=0.49\textwidth]{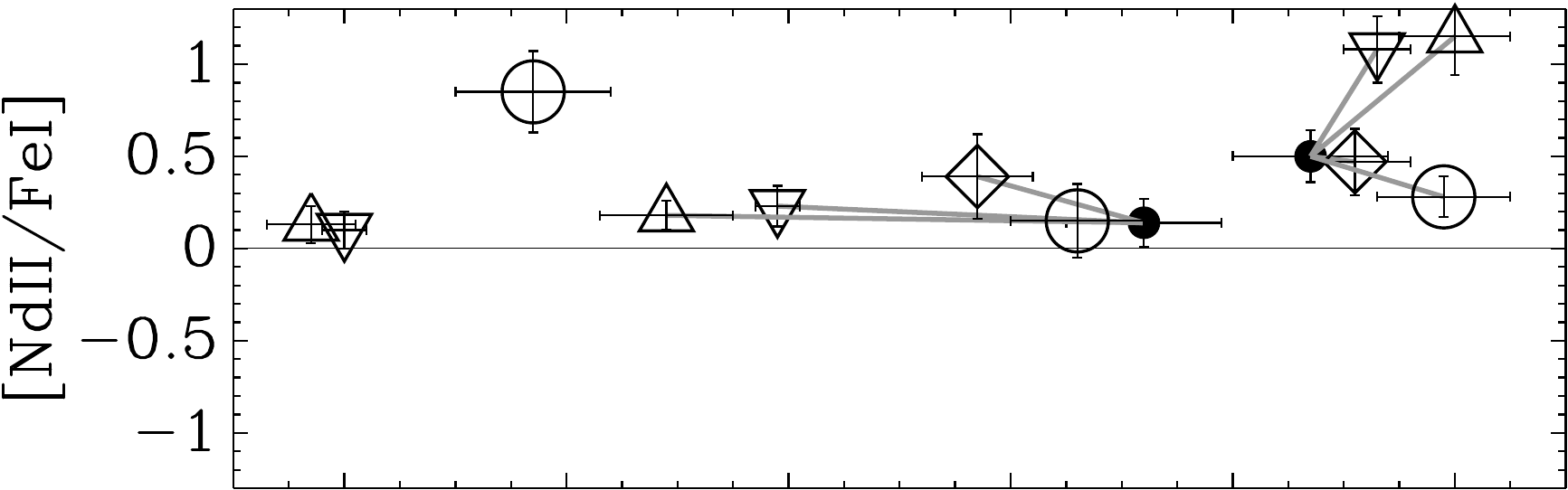}
\includegraphics[width=0.49\textwidth]{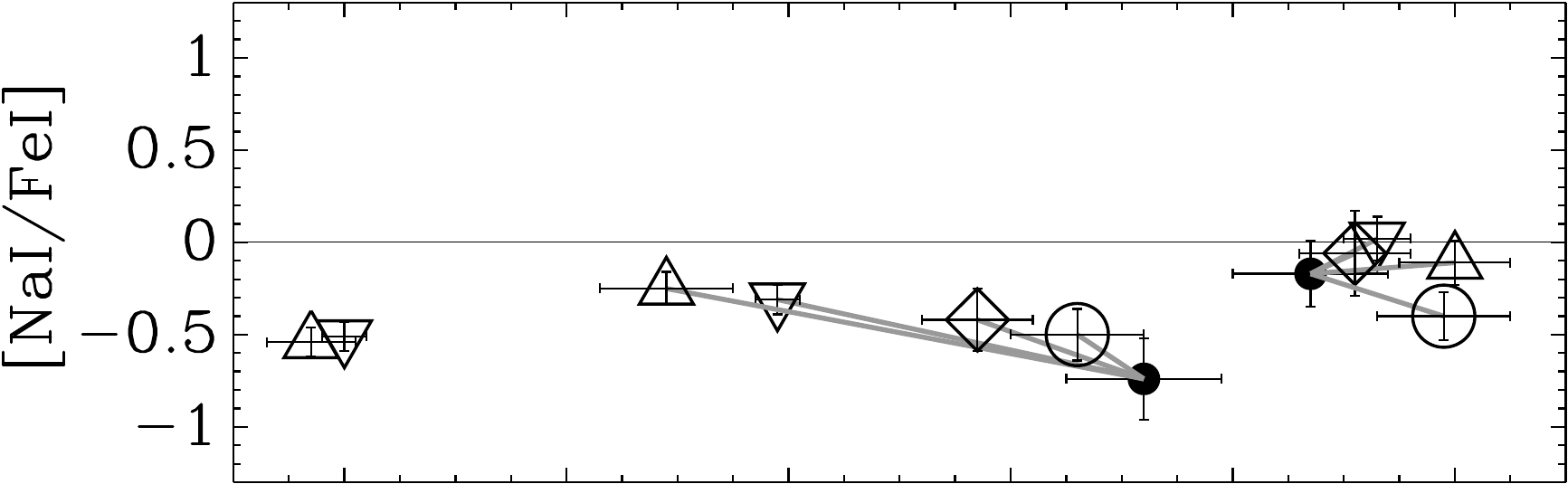}
\includegraphics[width=0.49\textwidth]{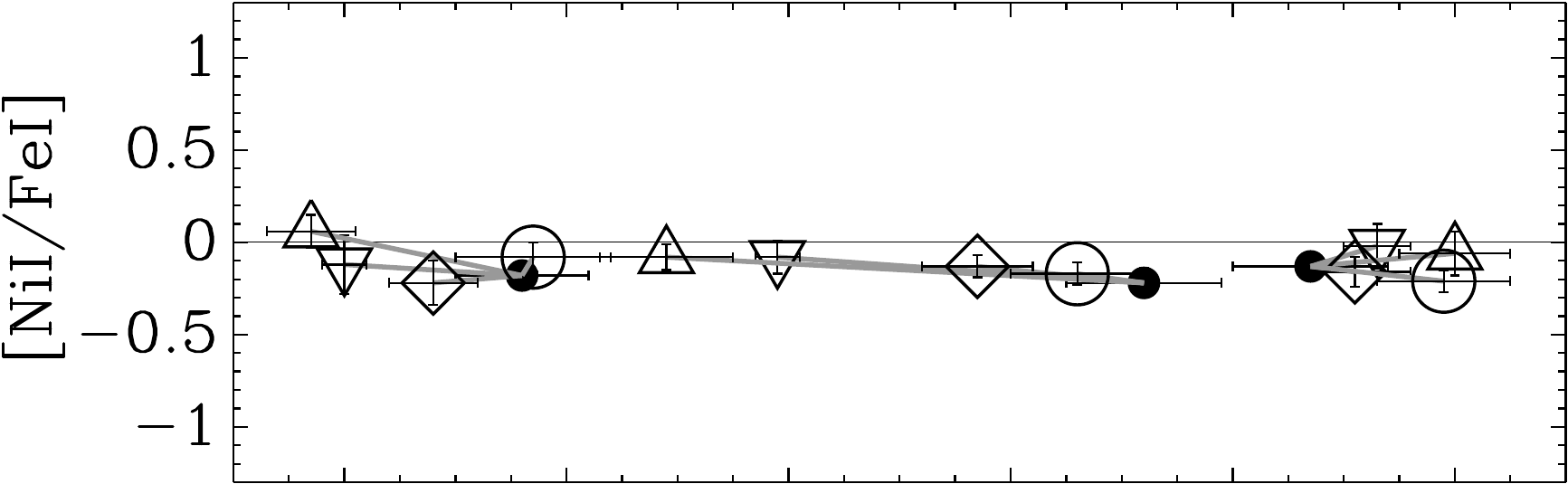}
\includegraphics[width=0.49\textwidth]{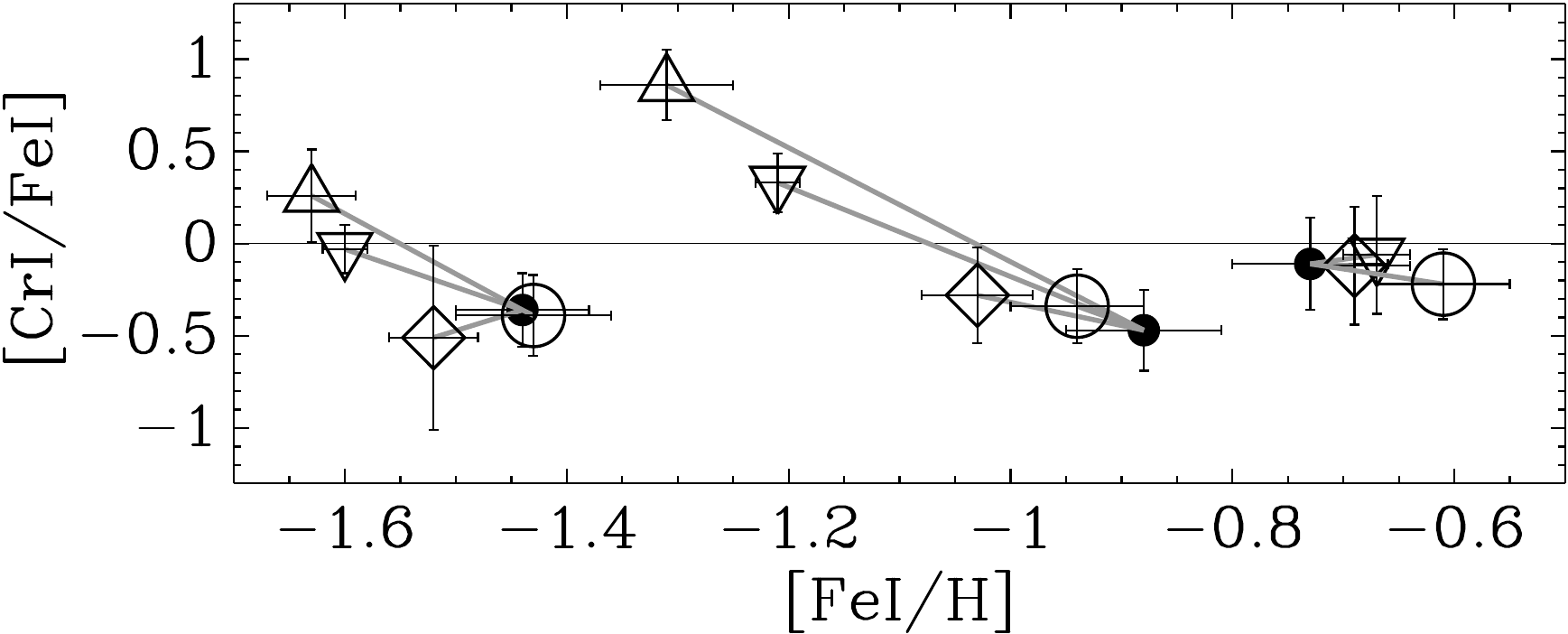}
\includegraphics[width=0.49\textwidth]{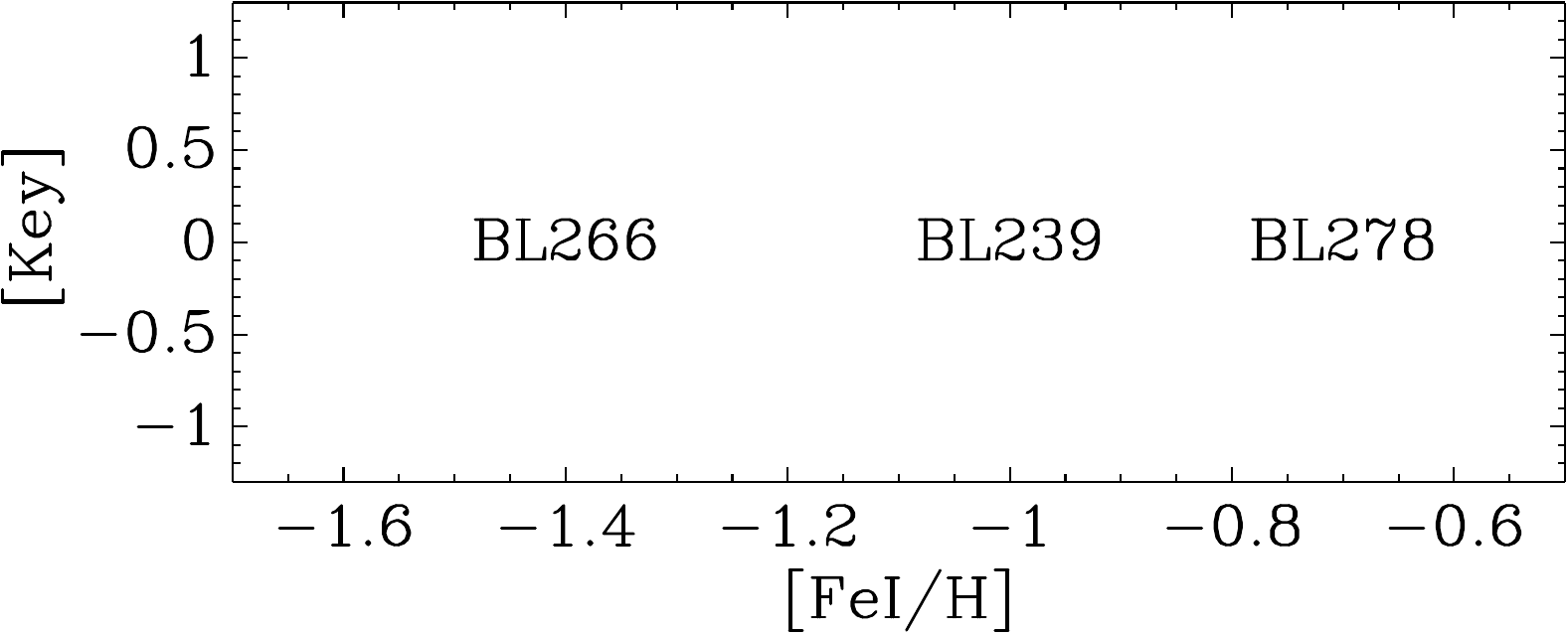}
\caption{To check the effect of systmeatics in our analysis the
abundance ratios for the 3 stars observed with both UVES (S03) and
GIRAFFE (this work) are shown here for the 5 different checks discussed
in section 3.7.  To see which star is which, see the bottom right-hand
panel.  The GIRAFFE data points are shown as filled circles with lines
joining each different check for the same star. The UVES abundances are
shown  as open symbols.  Check 1 results are shown as inverted
triangles; Check 2 as triangles; Check 3 as diamonds; Check 4 as circles
and Check 5 as small filled dots.
\label{fig:diffabo3fnx}}
\end{figure*}

\noindent
We carried out 5 different analyses, as follows:

\begin{itemize}
\item Check 1, basic sanity check: we reproduce the
 original  abundance analysis of S03, starting from the
 UVES spectra, using the S03 line list, measuring the $EW$s 
 with \texttt{splot}  and working with the  plane-parallel models of 
 \citet{1975A&A....42..407G}.
\item Check 2, the influence of the geometry: the 
same set-up as in Check 1, except the plane-parallel models are  
replaced by the spherical MARCS models.
\item Check 3, the influence of the line list and wavelength coverage: 
the same set-up as in Check 2, except {\it (a)} the original
line list was replaced by the one used in this work; {\it (b)} we
restricted the analysis to the wavelength coverage of the FLAMES HR
10, 13 and 14 setups.
\item Check 4, the influence of the automatic measurement of $EW$s: 
the same set-up as in Check 3, except the $EW$s are measured 
automatically with \texttt{DAOSPEC} instead of manually 
with \texttt{splot}.
\item Check 5, the influence of the spectral resolution: the 
same set-up as in Check 4, except the $EW$s are now taken
from the new GIRAFFE spectra instead of the S03 UVES spectra, and so 
the spectral resolution is reduced by a factor of two, 
from $R \simeq$ 40\,000 to $R \simeq$ 20\,000.
\end{itemize}

\noindent
Figure~\ref{fig:diffabo3fnx} displays the results of these 5 checks and
compares the abundances of 13 different elements and/or ionisation
states independently determined at each stage.  One conspicuous result
of our checks is the different [Fe/H] obtained for BL239 in our new
analysis compared to S03 (a difference of +0.4\,dex).  This effect is
due to the fact that historically the S03 line list was optimized for
metal-poor stars and hence lacks weak lines.  This is appropriate for
BL266.  In the case of BL278, since this star was obviously more
metal-rich, S03 were naturally required to add weak lines to their line
list.  Conversely, BL239 falls between these two regimes and no weak
lines were added. This had the effect of driving the micro-turbulence
velocities to artificially high values, and thus biasing metallicities
towards lower values.

There  are  several differences between the \citet{1975A&A....42..407G}
and  \citet{2008A&A...486..951G} models. The two most  important are the
different geometry used  (plane-parallel versus spherical) and the
physics involved in  the opacities.  Abundances calculated  with these
two methods are similar, most of  the time within  the error bars.  As
discussed  in Sec.\ref{sect:StellarModelsUsed}, the effect of geometry
alone  is expected  to account for  up to  $\sim$0.05\,dex and can  be
partially  responsible for  the difference  in  abundance between  the
triangles and inverted triangles symbols of
Figure~\ref{fig:diffabo3fnx}.

The effects of using different line lists can be seen in
Figure~\ref{fig:diffabo3fnx} comparing the triangles and the diamonds.
This is usually the largest source of differences, showing the
importance of having a common line list for comparing abundance results.

The method used to measure the $EW$s also affects the abundance results
but only by a small amount, and in most cases well within the errors.
This can be seen by comparing circles and diamonds in
Figure~\ref{fig:diffabo3fnx}. This check reinforces our confidence in
using the automatic \texttt{DAOSPEC} $EW$ measurement code, allowing us
to go from hand-measurement to a much faster automated processing.

Finally, comparing UVES and GIRAFFE results shows that even with a loss
of a factor two in resolution and a factor  three in wavelength
coverage, it is possible to determine accurate abundances. By comparing
the empty circles to the solid dots in Figure~\ref{fig:diffabo3fnx}, it
is clear that not only [Fe/H] but also most of the abundance ratios are
identical within the errors.

\section{Results}\label{sect:results}

We were able to determine detailed abundances for 81 stars from our
original sample of 107 FLAMES spectra.  Their atmospheric parameters are
presented in Table~\ref{tab:fnxfi-photo}.  The $EW$s are listed in
Table~\ref{tab:fnxfi-ewlist1}.  Finally, all abundance ratios are given
in Table~\ref{tab:fnxfi-tababo1}.  In the following, the solar
abundances of \citet{1989GeCoA..53..197A} are adopted, with the
exception of Ti, Fe and La \citep{1998SSRv...85..161G}.

\subsection{Iron abundance}
\label{sect:iron}

\begin{figure}
\begin{center}
\includegraphics[width=\hsize]{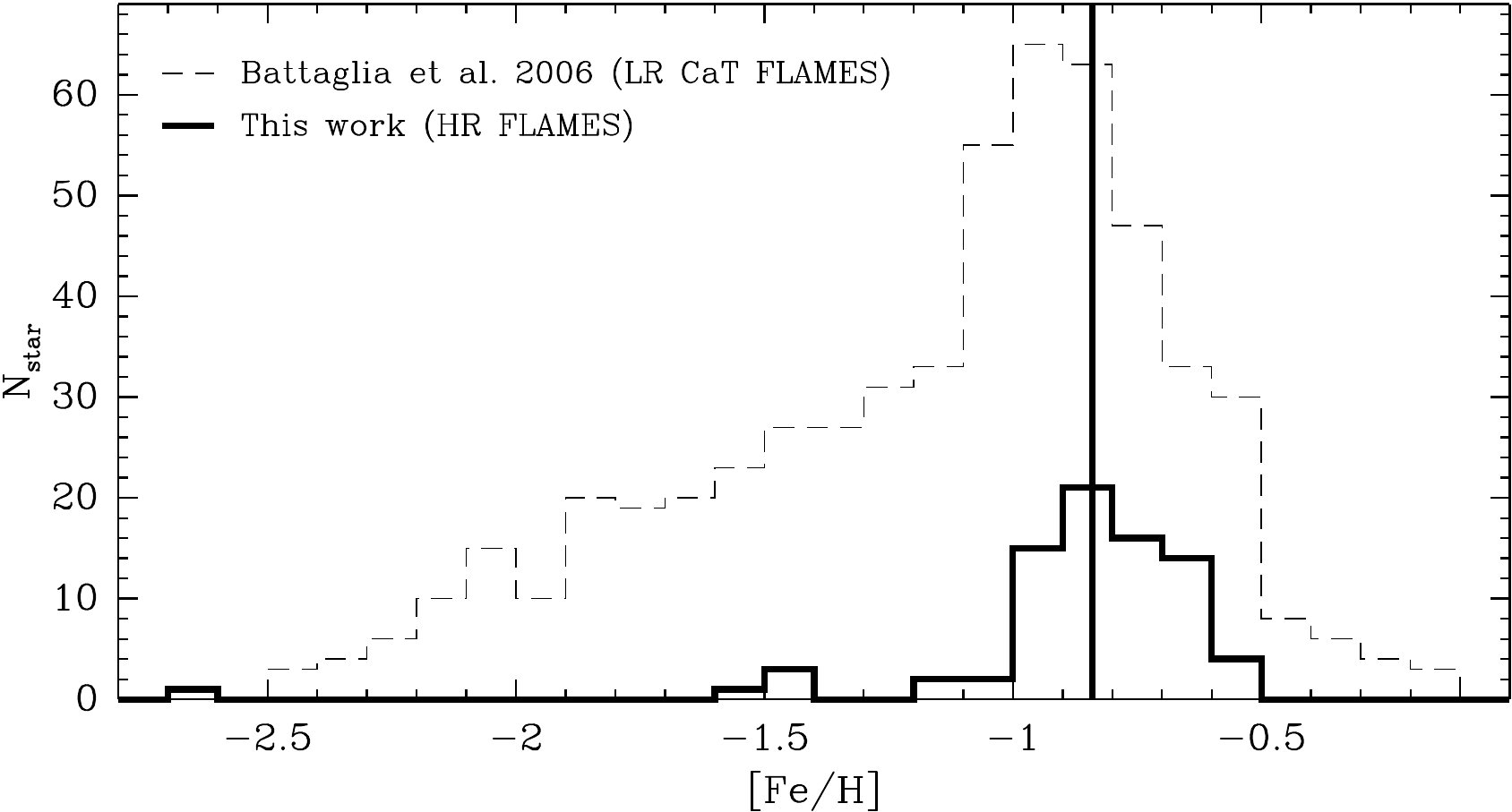}
\caption{The [Fe/H] distribution of our FLAMES high resolution
abundance analysis (solid line), peaking at [Fe/H]
$\approx$ -0.8. The metallicity distribution obtained from the 
larger set of Ca~II triplet measurements (B06) is shown in
dashed line for comparison.
\label{fig:fnx-histofeh}}
\end{center}
\end{figure}

Figure~\ref{fig:fnx-histofeh} compares our FLAMES HR [Fe/H] distribution
to the FLAMES LR [Fe/H] distribution derived by B06, using low
resolution ($R$ = 6500) \ion{Ca}{II} triplet measurements . The overlap
between the two samples have allowed \citep[][]{2008ApJ...681L..13B} to
verify  the \ion{Ca}{ii} infrared triplet calibration to [Fe/H].
Although B06 have larger uncertainties on their [Fe/H] estimates, they
benefited from a much larger sample of 600 stars spread over a much
larger area of Fornax than the present work.  Our distribution peaks at
[Fe/H] $\simeq -0.8$ and is clearly skewed towards more metal rich stars
than B06.  There are a few outlying stars on the metal poor side, but
the centre of Fornax is definitely dominated by stars with [Fe/H] $>
-1.2$.  Albeit unevenly, we still sample two orders of magnitude in
[Fe/H] ($-$2.5 $\lesssim$ [Fe/H] $\lesssim$ $-$0.5).\\

Although our field contains one of the Fornax globular clusters, Cluster
4 \citep{1961AJ.....66...83H}, we don't seem to have selected any of its
stars. Cluster 4 is moderately metal poor, [Fe/H]$\sim -1.5$
\citep[e.g.,][]{2003AJ....125.1291S}.  \citet{2006A&A...453..547L}
suggested that the metallicity scale of \citet{2003AJ....125.1291S}
tends to over estimate [Fe/H] by 0.3 to 0.5 dex, hence any [Fe/H] $\sim
-1.8$ to $-2$ dex star in our sample, spatially close to Cluster 4 is a
potential cluster member.  The closest stars to Cluster 4 all have
[Fe/H] $\approx$ $-0.8$ dex, and are therefore significantly too metal
rich to be members.

The single truly metal-poor star in our sample, BL085 at [Fe/H] =
$-2.58$, is not located near Cluster 4 and should thus be representative
of the oldest and most metal-poor field population in Fornax. However,
the \vrad\ for this star places it at the very edge of our membership
boundary, and in the slightly more strict membership criteria of B06 it
is considered a non-member.  We did not consider this a reason to
exclude it from our analysis, but care should be taken in attributing it
too definitively to Fornax.

\subsection{Alpha Elements}
\label{sect:alpha}

The $\alpha$ elements are  predominantly produced by  high mass, ($> 8
\msol$) short lifetime Type II supernovae explosions (SNe~II).
[$\alpha$/Fe] is thus a way of tracing the relative contribution of
SN~II and SN~Ia products that were available when the stars formed
\citep[e.g.,][]{Gilmore91}.  Stars forming when the interstellar medium
(ISM) has only been enriched by SNe~IIs have high [$\alpha$/Fe], while
those forming after the SNe~Ia have significantly enriched the ISM in
iron have lower [$\alpha$/Fe].

\begin{figure}[htp]
\centering
\includegraphics[width=\hsize]{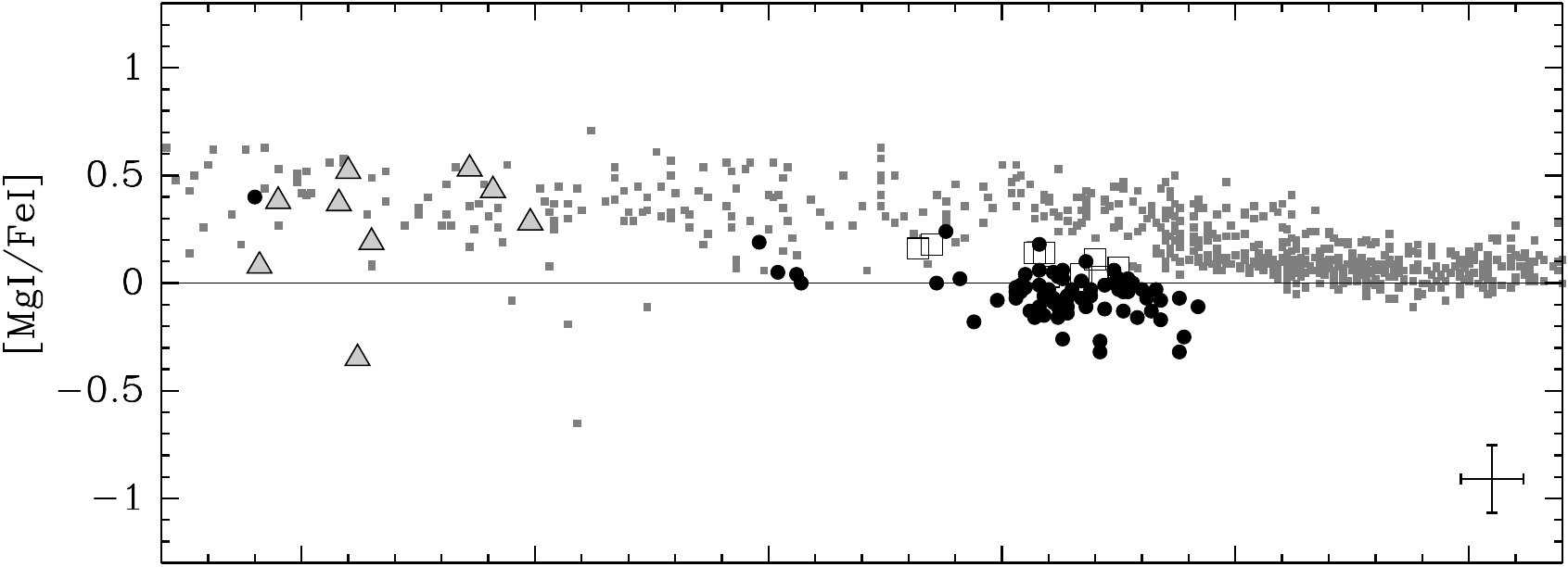}
\includegraphics[width=\hsize]{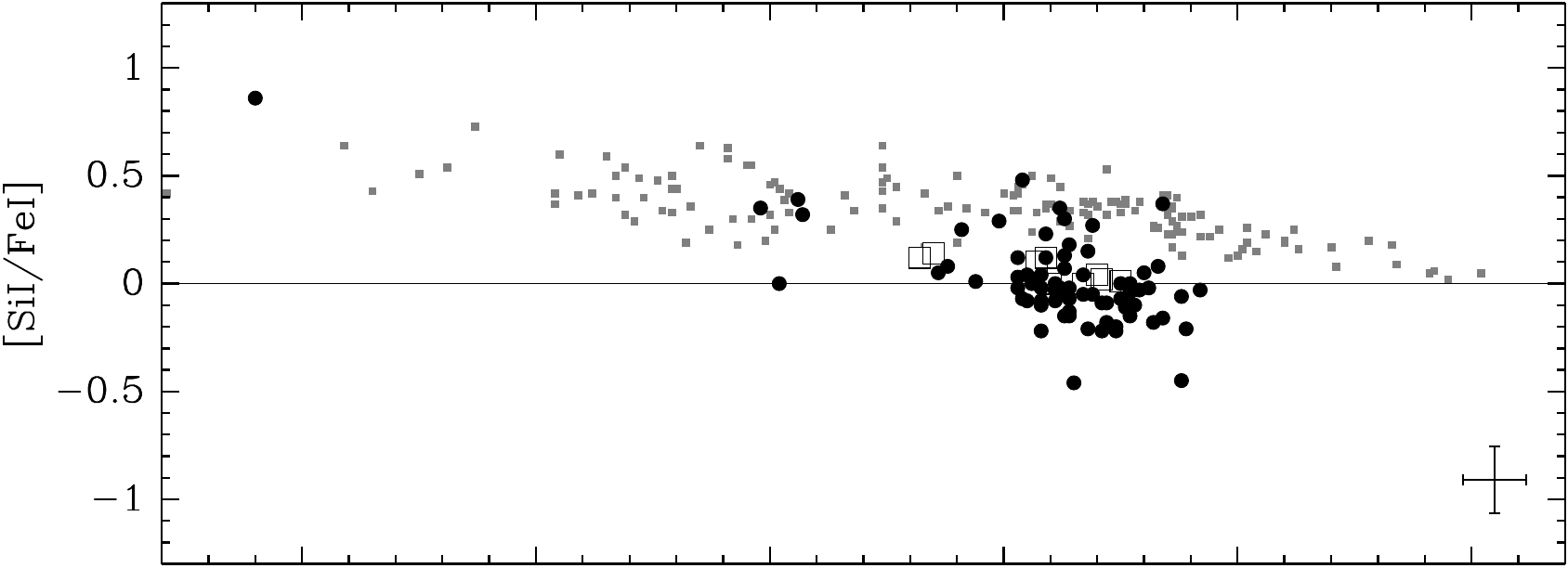}
\includegraphics[width=\hsize]{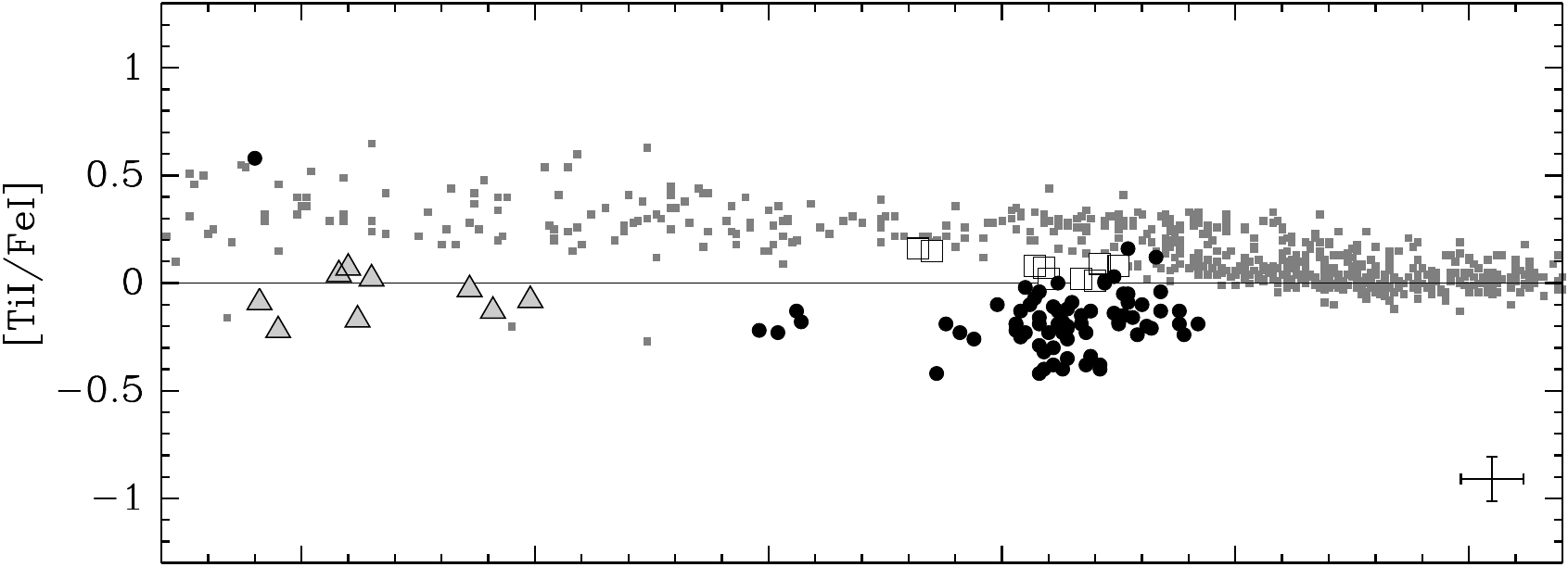}
\includegraphics[width=\hsize]{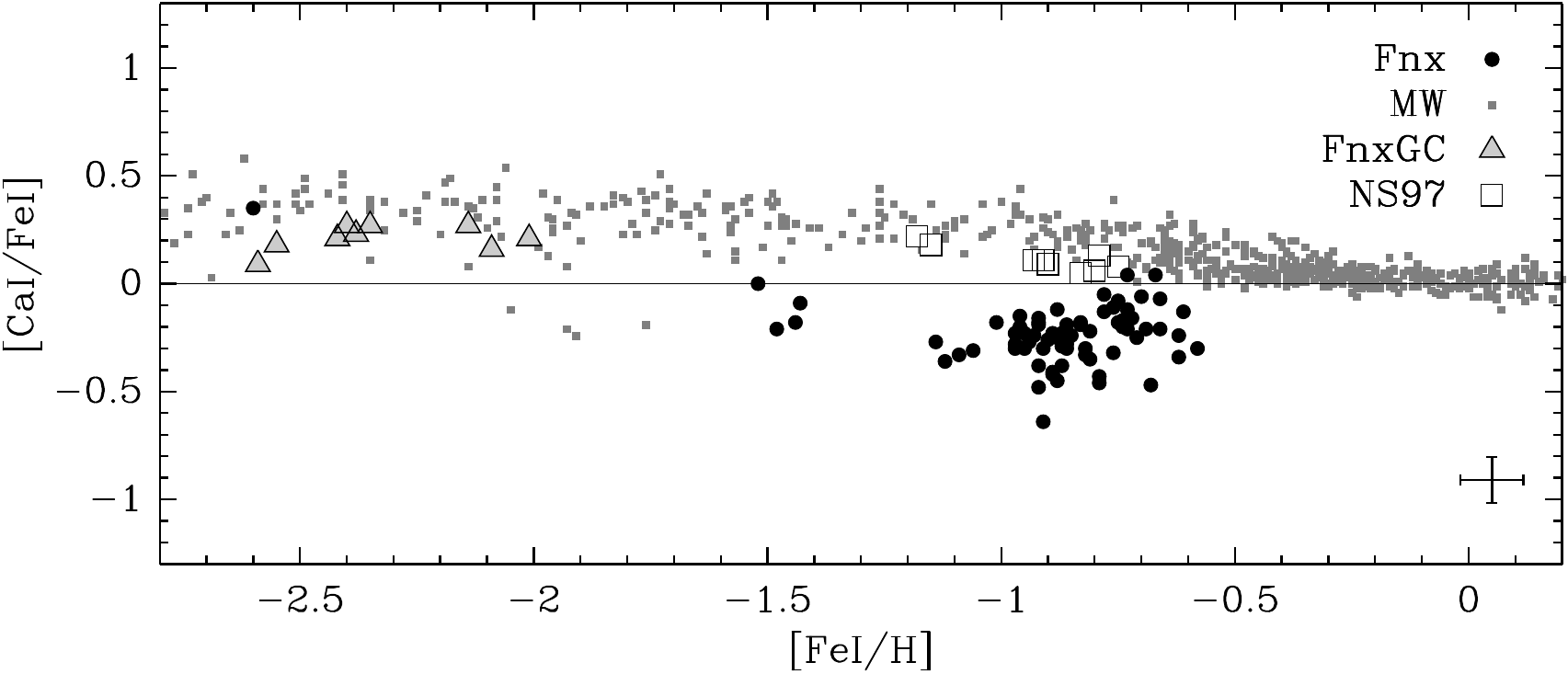}
\caption{
The abundance ratios [Mg/Fe], [Si/Fe] and [Ca/Fe] versus [Fe/H].  Our
observations of Fornax field stars are shown with solid circles. Also
plotted for comparison are Galactic stars \citep[original references
in][]{2004AJ....128.1177V} with small grey squares, the Fornax globular
clusters \citep{2006A&A...453..547L} with triangles and eight peculiar
halo stars  \citep{1997A&A...326..751N} with  empty squares.  There is
a  representative (average) error  bar  for the Fornax field star
abundances in the bottom right corner of each panel.  This is the
quadratic  sum of [X/H]    + [Fe/H], (measurement  errors) taken  from
Table~\ref{tab:fnxfi-tababo1}.
\label{fig:mgsica}}
\end{figure}

\begin{figure}[htp]
\centering
\includegraphics[width=\hsize]{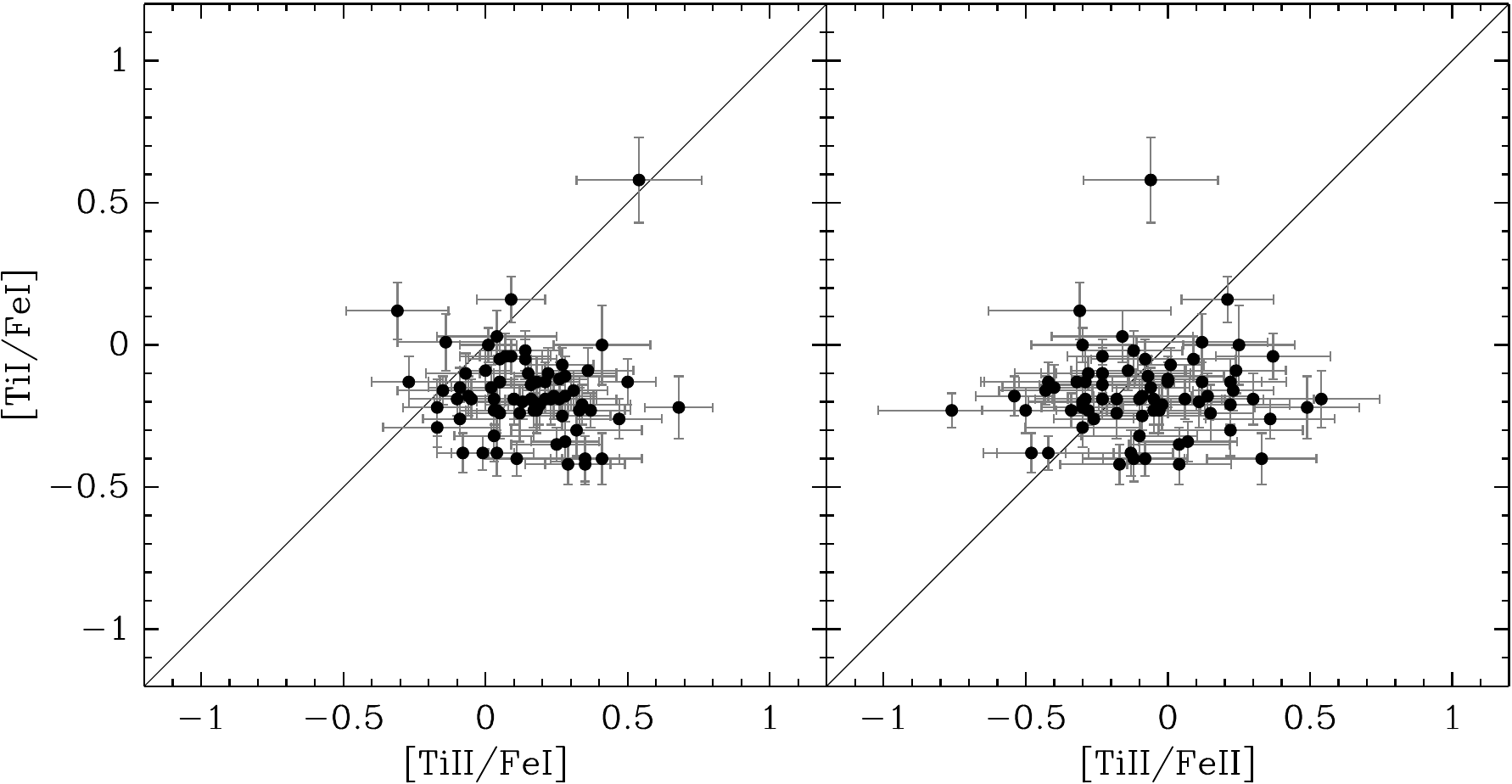}
\caption{The two ionisation states for Titanium, \tii\ and \tiii\
are compared to our HR sample of Fornax field stars. On the 
{\em left panel,} there is a clear offset of [\tii/\ion{Fe}{i}] 
versus [\tiii/\ion{Fe}{i}], while on the {\em right panel,} 
[\tii/\ion{Fe}{i}] and [\tiii/\ion{Fe}{ii}] are on a more common scale, 
showing that ionisation balance is probably not very well achieved in these stars. 
\label{fig:ti1ti2}}
\end{figure}

The individual $\alpha$-element ratios for our sample of Fornax field
stars [Mg/Fe], [Si/Fe], [TiI/Fe] and [Ca/Fe] are shown in
Figure~\ref{fig:mgsica}. For comparison, we also present high resolution
abundances of 9 stars in 3 Fornax globular clusters
\citep{2006A&A...453..547L}, and a compilation of Milky Way halo, thick
and thin disk stars \citep{2004AJ....128.1177V} , as well as eight
peculiar halo stars \citep[][hereafter NS97]{1997A&A...326..751N}.
These eight stars from NS97 have unusual kinematics and orbital
parameters, including a large maximum distance from the Galactic centre
($R_{\mathrm{max}}$) and a large distance from the Galactic plane
($z_{\mathrm{max}}$).  They were also found to display low [$\alpha$/Fe]
and [Ni/Fe] (along with other chemical peculiarities) compared to the
majority of MW halo stars.  These differences have led to the suggestion
that these stars might have been accreted by the MW from a dwarf galaxy,
and this is why we include them in our figures.

\begin{figure}
\centering
\includegraphics[width=\hsize]{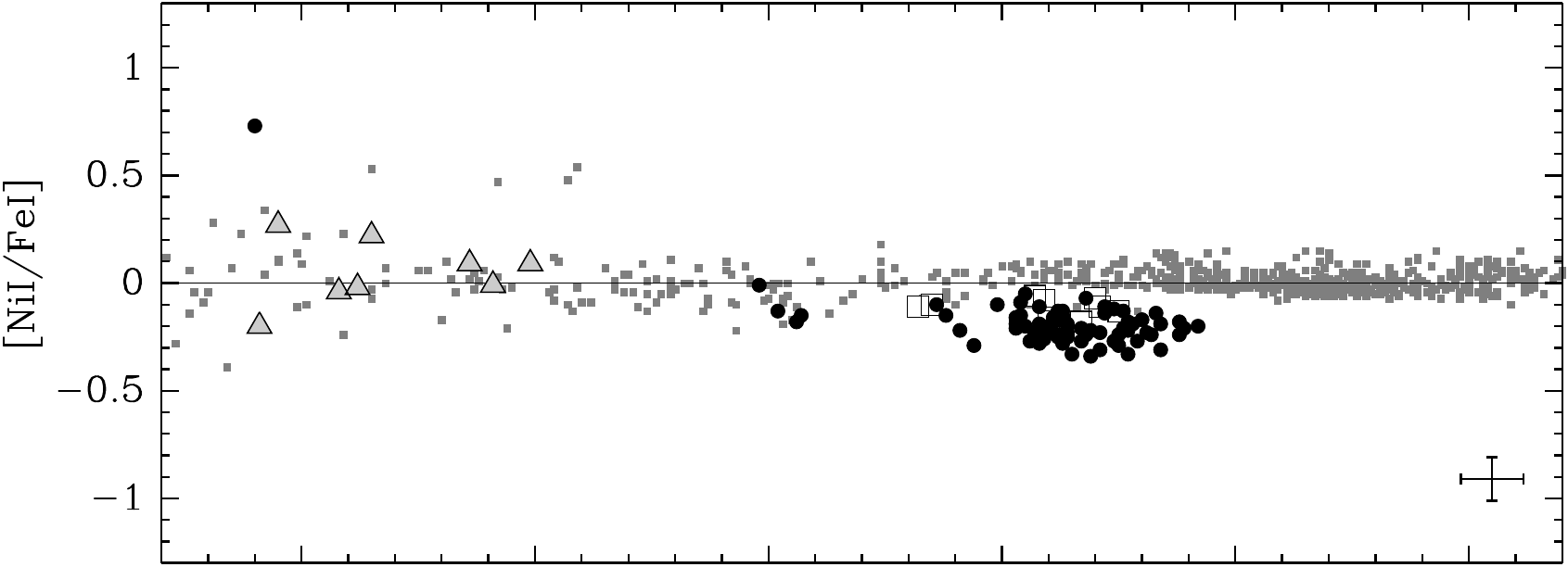}
\includegraphics[width=\hsize]{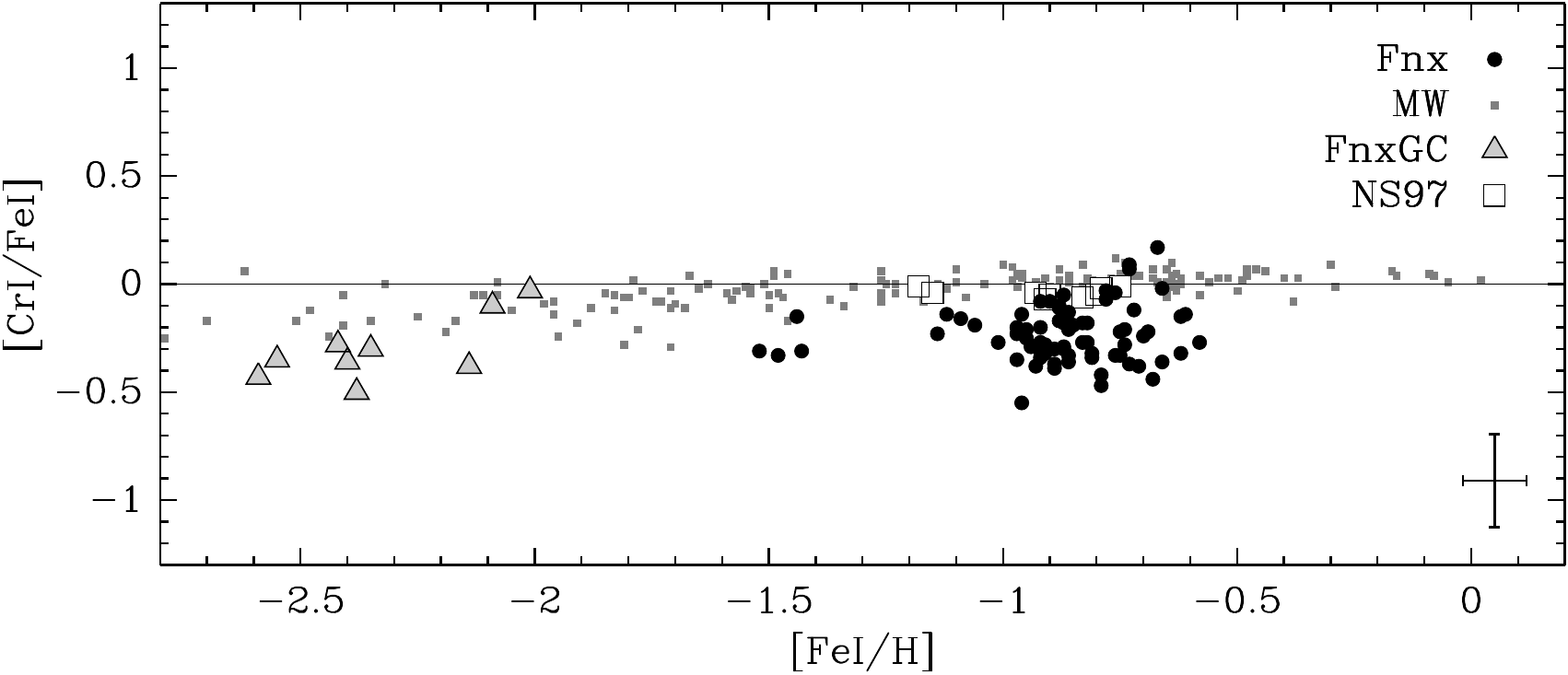}
\caption{The Iron peak elements [Ni/Fe] and [Cr/Fe] versus
[Fe/H] for the Fornax field stars, the Fornax globular clusters, and MW 
stars.  The references and symbols are the same as in Figure~\ref{fig:mgsica}.
\label{fig:nicr}}
\end{figure}
The single metal poor star in our new sample of Fornax field stars,
BL085 ([Fe/H] = $-2.58$), has $\alpha$-element abundances very similar
to the Galactic halo stars and also (Fornax) globular cluster stars at
similar [Fe/H]. In contrast at higher [Fe/H], the rest of the Fornax
field stars are significantly under-abundant in [$\alpha$/Fe] with
respect to the MW. [Mg/Fe] is typically lower than the level of the
Galactic stars by between 0.2 and 0.5\,dex and the highest values just
overlap with NS97 stars. The [Si/Fe] ratios are the highest for the
\alfe\ in Fornax with a mean value close to solar and this corresponds
to the same level as the eight peculiar halo stars of NS97. [Ca/Fe] and
[Ti/Fe] are typically lower than [Si/Fe] and [Mg/Fe] especially for the
4 stars at [Fe/H] $\sim -1.5$. In addition, both [TiI/Fe] and [CaI/Fe]
are constant in value, contrary to [MgI/Fe] and [SiI/Fe]. This trend was
first noticed in dSphs by \citet{2004oee..symp..218S}.
\citet{2004AJ....128.1177V} hypothesized that it was due to different
nucleosynthetic origins (Mg and Si form through hydrostatic C and O
burning in the cores of massive stars, whereas some isotopes of Ca and
Ti form in the $\alpha$-process).  \citet{2007ApJ...661.1152F} have
also seen this effect in the galactic bulge, where the evidence for fast
chemical enrichment clearly favours massive stars for the main
contributors to these elements.  Another possibility in Fornax where
SNe~Ia have clearly have time to contribute strongly the chemical
enrichment, is provided by the sensitivity of the SNe Ia ejecta to the
metallicity of their progenitors
\citep{2003ApJ...590L..83T,2006A&A...453..203R}. As we will see in
Section \ref{sect:ironpeak} it  applies to Ni, but it could also affect
the intermediate mass elements.   SNe Ia are thought to produce little
Mg, while they  are  able to produce significant amounts of Ca and Ti : for example, the ratio of Ca/Fe is only a factor four larger in a 
SNe II of 50$\rm M_{odot}$ than in a SNe Ia 
\citep[]{Tsujimoto95}, and the total mass of Ca produced is actually larger in the latter.
\citet{2006A&A...453..203R} show that for a SNe Ia progenitor metallicity
[Z/Z$_{\odot}$] lowered from 0.  to $-0.5$, the  production of some Ca
and Ti isotopes can diminish by  a factor $\sim$ 3.  Hence, if a large
fraction of Ca and Ti are produced in SNe  Ia   in Fornax, the  low  [Ca/Fe]
and [Ti/Fe]  could then be a  consequence of  the  low metallicity  of their
progenitors compared to the Milky Way.

These are however still conjectures which require a few words of
caution. Figure~\ref{fig:ti1ti2} illustrates that [\tii/Fe] differs from
[\tiii/Fe] and is lower by $\sim$0.4 dex.  This {\em could be a sign of
significant non-LTE effects, affecting predominantly \tii\ though both a
departure from ionisation equilibrium and NLTE acting most on the \tii\
lines that have significantly lower \kiex than \tiii\. However, NLTE may
not be the only culprit to blame, as any temperature biais in the \teff\
scale also both shifts ionisation equilibriae and affects more low
excitation lines such as those of \tii.}

We would prefer to determine the Ti abundance using \tiii\ lines,
however the number of available lines ($\sim2-3$) is small, and smaller
than the number of \tii\ lines ($\sim8-9$),  as reflected in the larger
error bars.  Therefore, as [Ti~I/Fe] is statistically more reliable, the
abundance of titanium is calculated as the average of results from the
\tii\ lines. This leaves some, quite large, uncertainty as to the true
value of [Ti/Fe], although we also point out in
Figure~\ref{fig:ti1ti2} that [\tiii/\ion{Fe}{ii}] agrees reasonably well
with [\tii/\ion{Fe}{i}] albeit with a large dispersion, lending strength
to a low [Ti/Fe] ratio.  CaI could be similarly affected, although we
have no way to check.

Of the two O lines available in our wavelength range the most reliable
one at 6300\,\AA\ is not suitable, as the typical \vrad\ of the Fornax
stars means that this matches a telluric absorption line rendering it
unusable for most of our stars.  The second line is weak at the limit of
detection with \texttt{DAOSPEC} at this resolution.  The continuum in
its immediate vicinity is also not well defined, because it is blended
with CN lines and Ca\,{\sc i} auto ionisation features.  Thus the $EW$s
seem to be frequently over estimated by \texttt{DAOSPEC}.  For all these
reasons, we have decided not to include the O abundances.\\

\subsection{Iron peak elements}
\label{sect:ironpeak}

According to nucleosynthetic predictions, iron peak elements like Iron
(Fe), Chromium (Cr), and Nickel (Ni) are believed to be formed
predominantly from explosive nucleosynthesis in SN~Ia
\citep{1999ApJS..125..439I, 2005A&A...443.1007T}. In
Figure~\ref{fig:nicr}, we present the [Ni/Fe] and [Cr/Fe] abundance
ratios for the Fornax stars plotted against [Fe/H].  Both ratios appear
to behave in a similar way.  [Ni/Fe] has much smaller error bars than
[Cr/Fe], due to the larger number of lines available ($\sim$15 versus
1).  Hence the additional scatter in [Cr/Fe] compared to [Ni/Fe] is only
due to larger error bars.\\

To test the validity of our [Cr/Fe] and [Ni/Fe] abundance ratios, we
carried out tests on a high S/N spectrum of Arcturus, for which we
confirmed [Cr/Fe] $\simeq$ [Ni/Fe] $\simeq$ 0.0. Restricting the [Cr/Fe]
analysis to the same single line available to our Fornax sample, we
obtained [Cr/Fe] $= -0.2$. We performed the same analysis for Ni, using
only the Ni lines available for Fornax, and recovered the expected
[Ni/Fe] = 0.0.  These two results suggest that the Cr line we have
access to ($\lambda = 6330.09$\AA) has an erroneous $\log$~\gf, leading
to a lower abundance.  We therefore updated the $\log$~\gf for this
Cr transition to reproduce [Cr/Fe] $=0.0$ in Arcturus, thereby insuring
a proper comparison to the Milky-Way samples in Figure~\ref{fig:nicr}.

It is interesting to note that our most metal poor Fornax star,
BL085([Fe/H]=$-2.58$), has a significantly higher [Ni/Fe] value than the
other Fornax field stars, and is comparable to MW halo stars and Fornax
GC stars.  However, at this low metallicity, we only detect 4 Ni lines
(instead of 15, as for the more metal rich stars), leading to a much
larger error bar than the average shown in the bottom right hand corner
of the plot ( see Table~\ref{tab:fnxfi-tababo1}). \\

The underabundance in [Ni/Fe] which we see in Figure~\ref{fig:nicr} for
Fornax stars with [Fe/H] $> -1.2$ has been previously observed in some
of the Milky Way halo stars (NS97, \citealt{2009IAUS..254..103N}) and in
the Large Magellanic Cloud (LMC) disk stars \citep{2008A&A...480..379P}.
This   effect is   even stronger in   the Sagittarius  dSph where  the
iron-peak elements   all  exhibit sub-solar ratios for stars with
[Fe/H]$>-1.5$ \citep{2007A&A...465..815S}.  \citet{2009IAUS..254..103N}
noticed that their low [Ni/Fe] stars also have low [Mg/Fe], similar to
what is seen in Fornax.

These low values of [Ni/Fe] and [Cr/Fe] cannot be easily explained with
our current understanding of nucleosynthesis.  The [Ni/Fe] ratios should
be zero and constant for all values of [Fe/H] since both Ni and Fe are
believed to be predominantly created by the same production mechanism in
SN~Ia \citep{2005A&A...443.1007T}.  The different behaviour seen in
Fornax and the NS97 stars from the Galactic trend is an indication that
the production factors for each iron-peak element is not the same and
depends upon the evolutionary history of the parent population.  For
example, it is possible that SNe~Ia Ni yields are linearly dependent on
the original metallicity of the white dwarf progenitor
\citep{2003ApJ...590L..83T,2005A&A...443.1007T}.

This under-abundance of Ni (and Cr) was \emph{not} observed by
\citealt{2003AJ....125..684S}, for the three stars we have in common.
We attribute this difference to a combination of systematic effects,
where the most important one is the use of a different line list.  See
section~\ref{sect:systematics} and Figure~\ref{fig:diffabo3fnx} for more
details on systematic effects in abundance determinations.

\subsection{The Na-Ni relationship}
\begin{figure}
\centering
\includegraphics[width=\hsize]{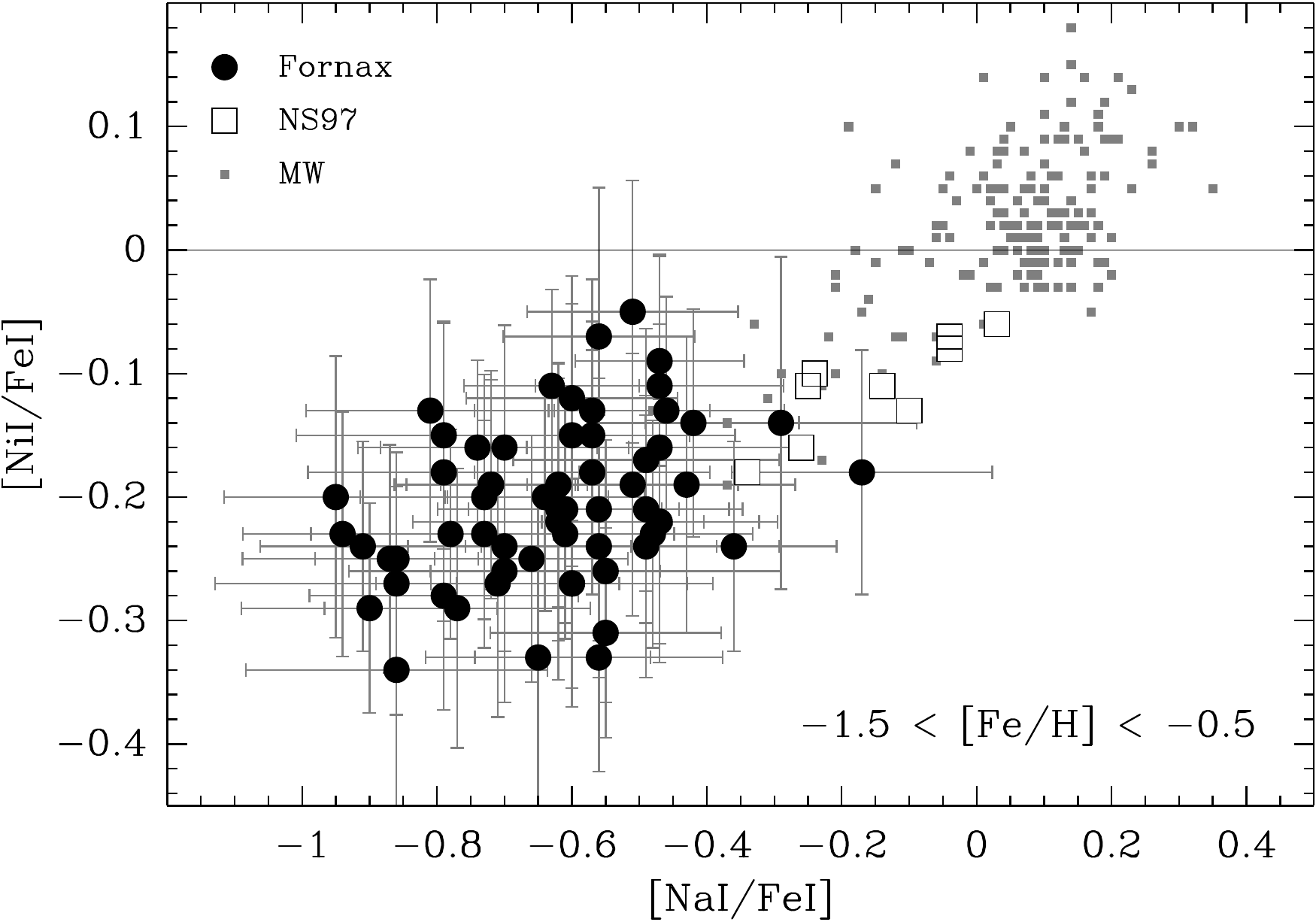}
\caption{
The relation between Na and Ni. The references and symbols are given in  
Figure~\ref{fig:mgsica}.
\label{fig:nina}}
\end{figure}

As  mentioned  in  section~\ref{sect:ironpeak},    Ni  is  assumed to
originate predominantly from SNe~Ia.   However, the production  of  Ni
might also   be    linked to    the  production   of  Na   in   SNe~II
(\citealt{1990ApJ...349..222T,1995ApJS...98..617T}).   The amount of Na
produced is controlled by the  neutron excess, where $^{23}$Na is  the
only stable  neutron-rich isotope  produced  in significant   quantity
during  the C and O burning  stage.  During the SNe~II explosions, the
elements are   photo-dissociated into   protons and  neutrons, further
recombining    to form $^{56}$Ni, which   in  turn $\beta$ decays into
$^{56}$Fe, the dominant isotope   of iron. $^{54}$Fe  and/or $^{58}$Ni
can also be produced at this stage, depending on  the abundance of 
neutron-rich  elements (e.g. $^{23}$Na). The  amount of $^{54}$Fe made
is  small compared to  the total  yield of  iron  (which is dominated by 
$^{56}$Fe production),  but this is  the main source of $^{58}$Ni, the
stable isotope of nickel \citep{1983psen.book.....C}.

In summary, the production of Ni  depends on the neutron excess 
and the   neutron excess  will   depend  primarily  on the amount   of
$^{23}$Na previously produced.  Hence,  a  Na-Ni correlation is  expected
when the chemical enrichment is dominated by SNe~II.  The advent
of the SNe~Ia explosions can break (or flatten) this relationship, as Ni
is produced without Na in the standard model of SNe~Ia
\citep[e.g.,][]{1999ApJS..125..439I}.

We have restricted the metallicity range of Figure~\ref{fig:nina} to
$-1.5 <$ [Fe/H] $< -0.5$, corresponding to the region in which the
Na-Ni relation can be seen in Fornax.  The region covered by our
sample of Fornax stars extend the relation known for the MW to lower
[Na/Fe] and [Ni/Fe] values with a similar slope. In SNe~II, the
dominant source of Ni is independent of the dominant source of
Fe. Consequently, low [Ni/Fe] is possible at low metallicities, before
the SNe~Ia contribute to the ISM chemical patterns. If the first
SNe~II in Fornax were not neutron rich, then there would be a lack of
Na (and therefore also Ni) production. The next generations of stars
could carry this signature, even at higher metallicity.  The fact that
we still see the SNII signature of the correlation between Na and Ni
in our sample of Fornax stars, means that the slope of the relation is
not diminished due to the subsequent addition of large amounts
of Ni. This is consistent with our previous discussion suggesting
that low SNe~Ia Ni yields are due to lower metallicity progenitors in
dSph compared to the Milky Way.

\subsection{Heavy elements}

\begin{figure}
\centering
\includegraphics[width=\hsize]{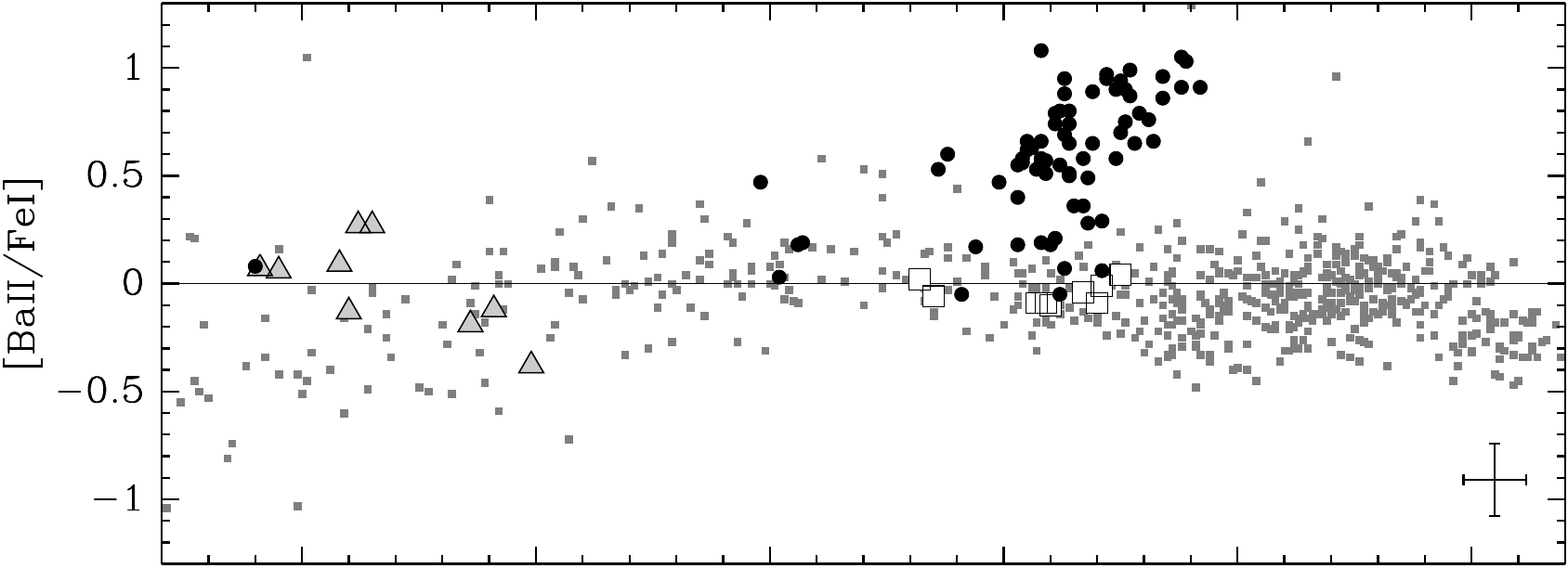}
\includegraphics[width=\hsize]{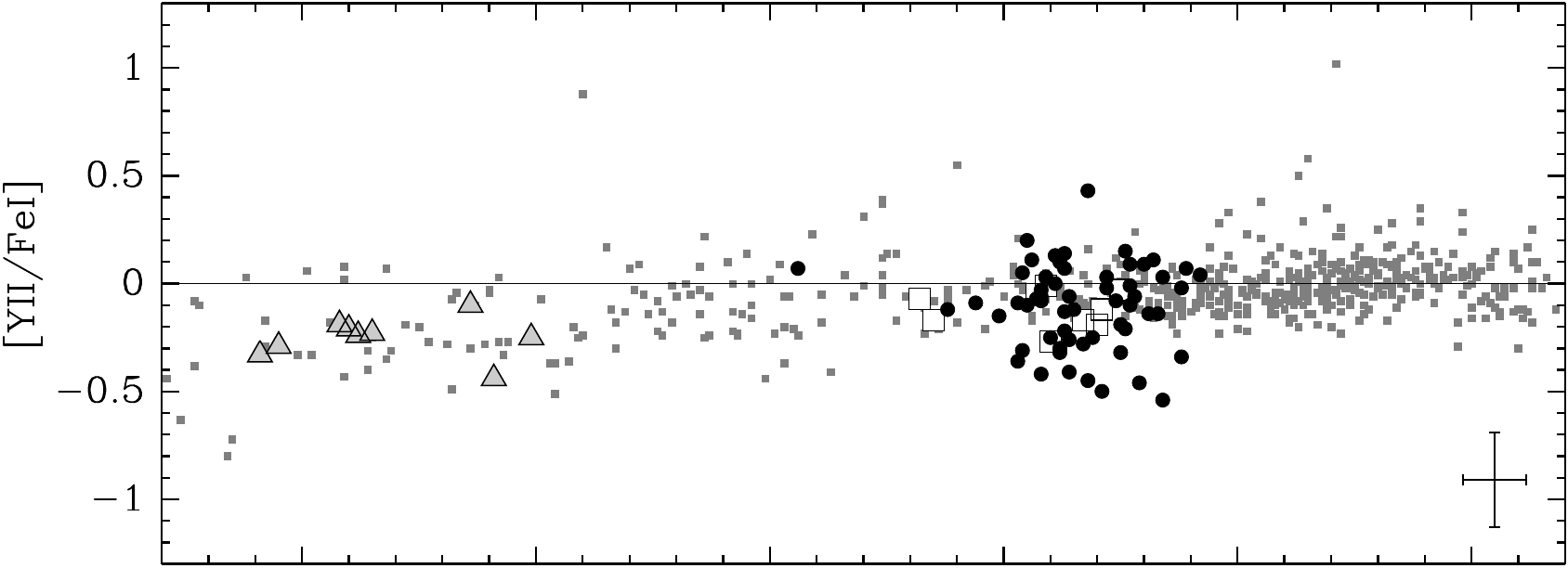}
\includegraphics[width=\hsize]{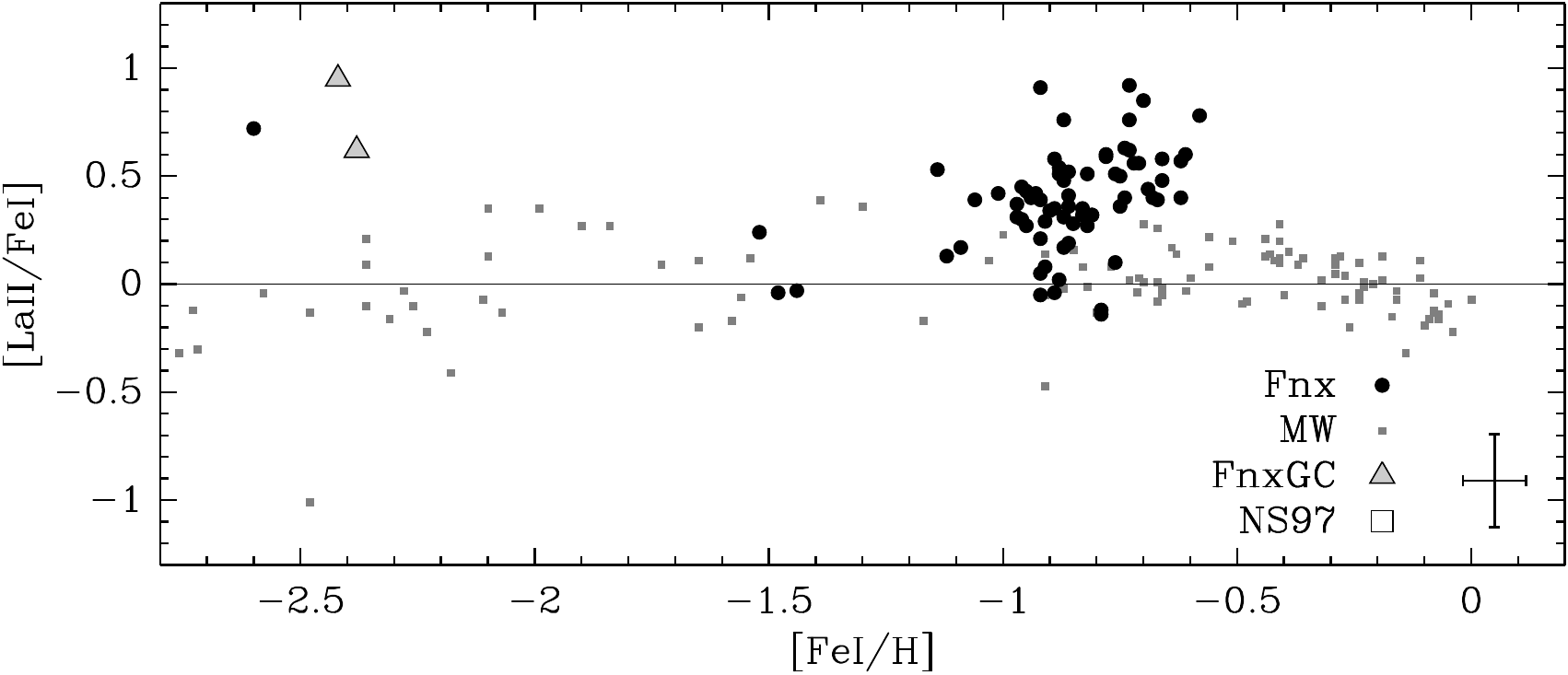}
\caption{The \spr\ elements [Ba/Fe], [Y/Fe] and [La/Fe]
plotted against [Fe/H] for the Fornax field stars, the Fornax globular
clusters, the MW compilation and the stars from NS97.  The references
and symbols are the same as given in Fig 9, and in addition we use
data from \citet{2002ApJS..139..219J} and \citet{2004ApJ...617.1091S}
for [La/Fe].
\label{fig:heavy1}}
\end{figure}

\begin{figure}
\centering
\includegraphics[width=\hsize]{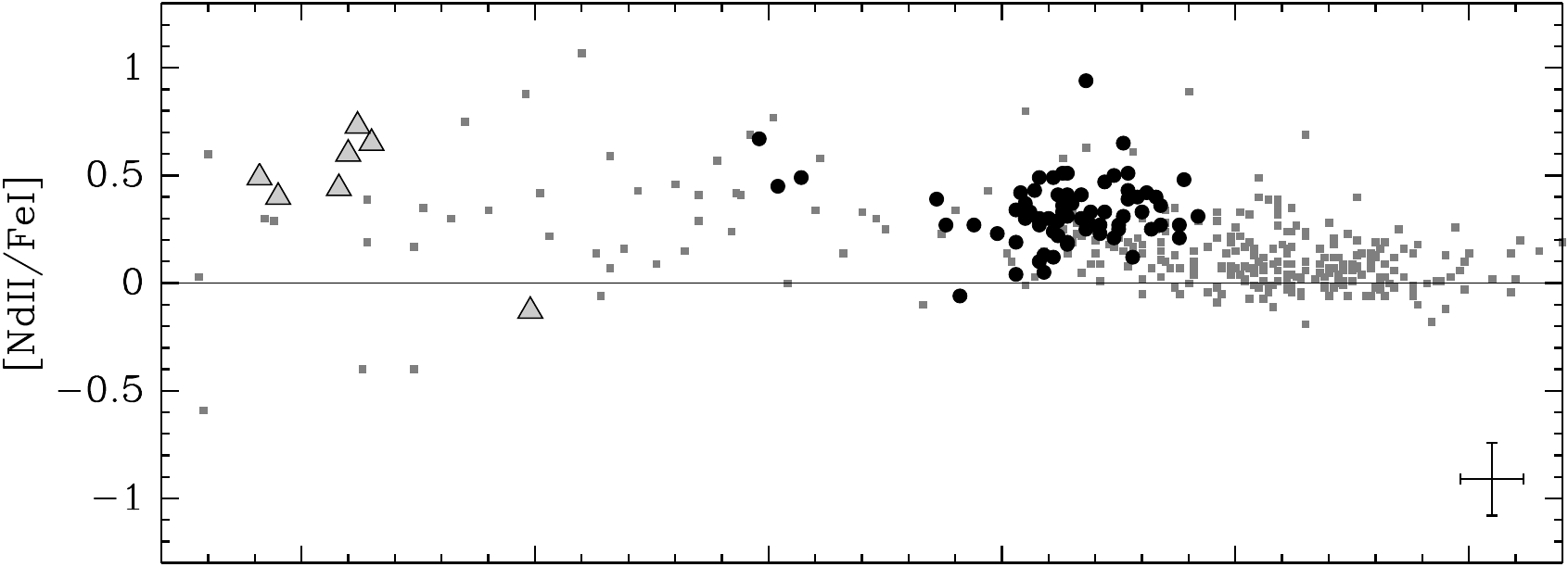}
\includegraphics[width=\hsize]{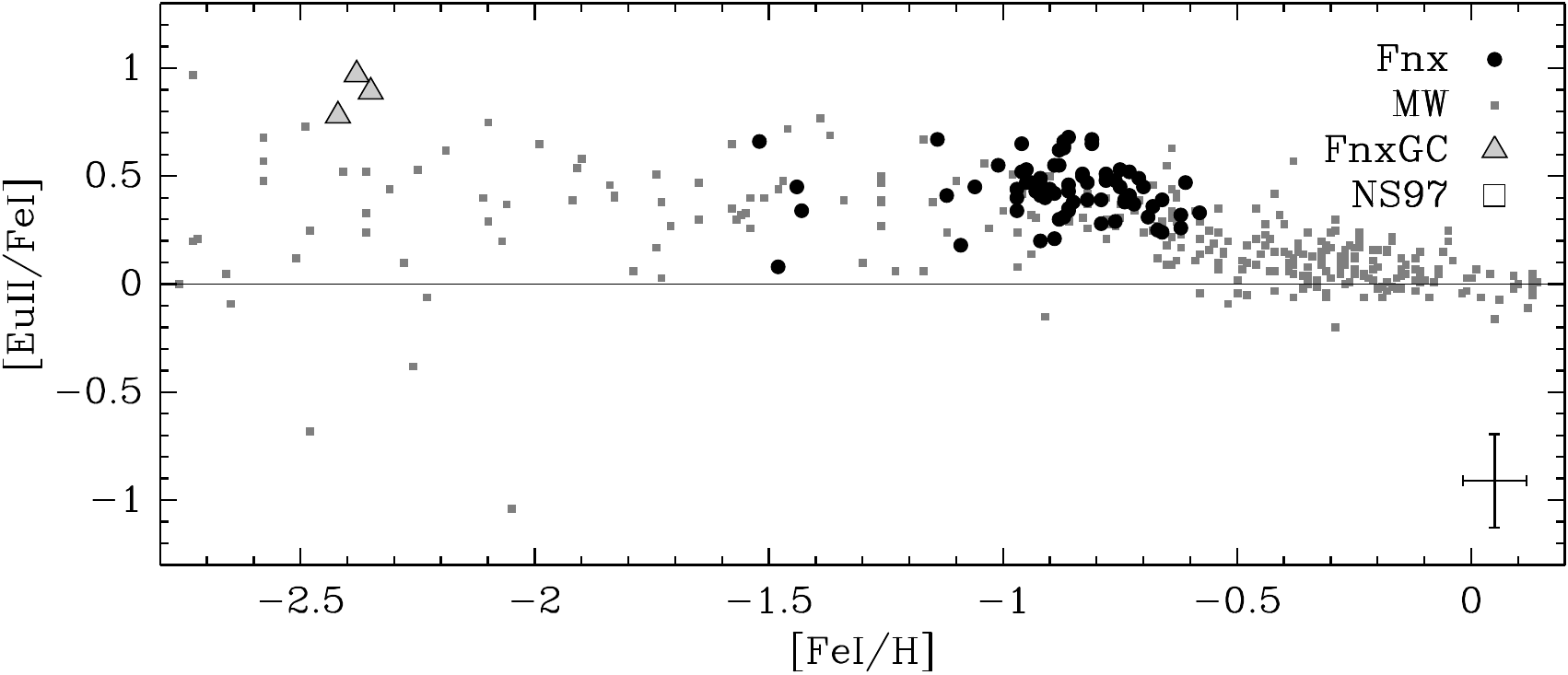}
\caption{The \rpr\ elements [Nd/Fe] and [Eu/Fe] plotted
against [Fe/H] for the Fornax field stars, the Fornax globular
clusters, and the MW.  The references and symbols are those of
Fig. 9. In addition the [Nd/Fe] values for the Milky Way stars are
taken from \citet{2000ApJ...544..302B} and
\citet{2003MNRAS.340..304R,2006MNRAS.367.1329R}.
\label{fig:heavy2}}
\end{figure}

Heavy elements (Z $>$ 30) are neutron capture elements.  One
distinguishes the \spr\ (or slow process) and the \rpr\ (rapid process),
with respect to the neutron-capture time-scales.  The \spr\ is believed
to predominantly take place relatively low mass stars ($2-4 \msol$). The
\rpr\ by contrast  requires more extreme conditions, for example the
very high temperatures and neutron densities found in SNe~II explosions.

\subsubsection{Tracing the \spr\ and \rpr\ contribution}

\begin{figure}
\centering
\includegraphics[width=\hsize]{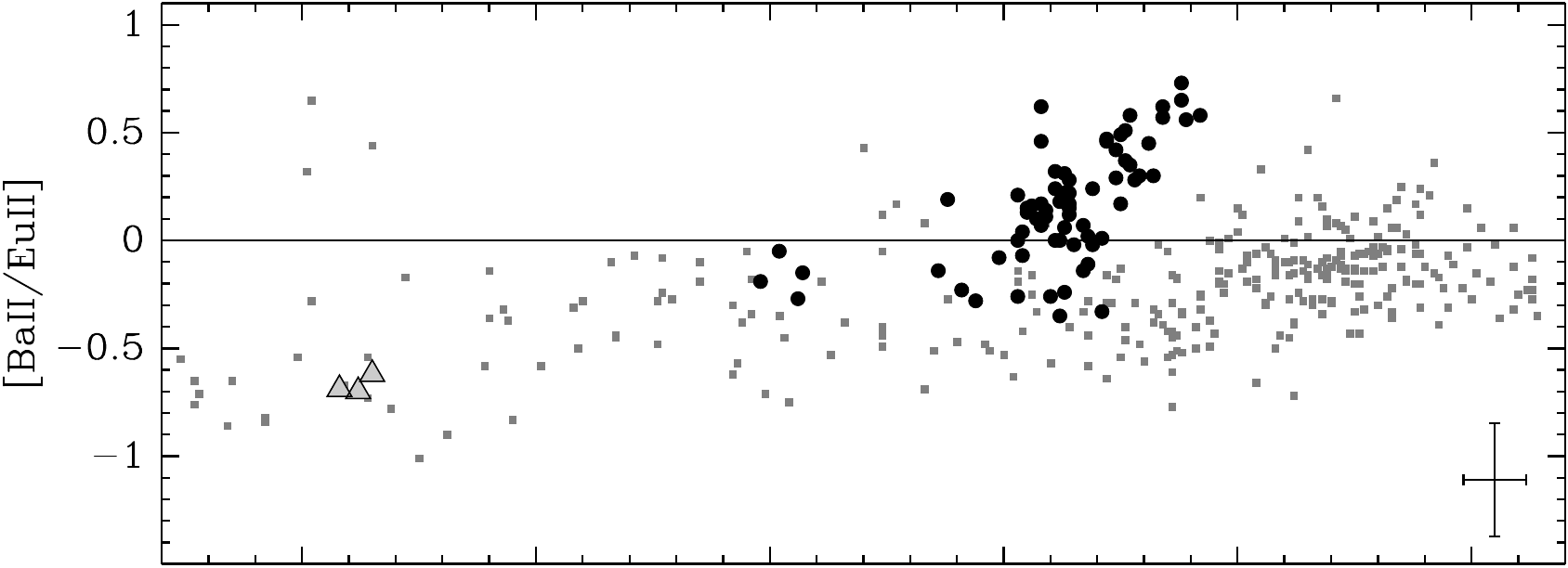}
\includegraphics[width=\hsize]{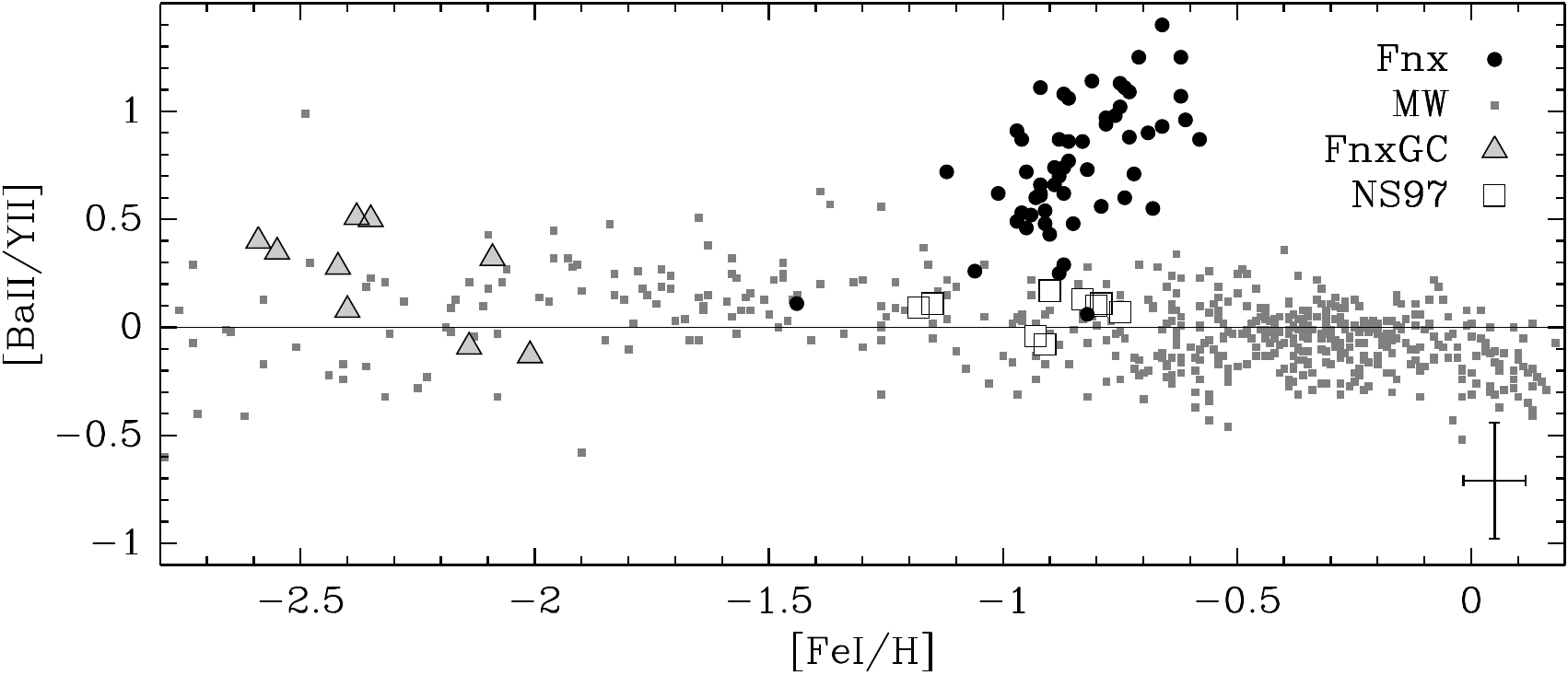}
\caption{[Ba/Eu] and [Ba/Y] as a function of [Fe/H] 
for the Fornax field stars
as well as Fornax globulars and MW compilation.
The references are given in Figure~\ref{fig:mgsica}.
\label{fig:fnx_ba2yy2}}
\end{figure}

Figures~\ref{fig:heavy1} and \ref{fig:heavy2} display our results for
[Ba/Fe], [Y/Fe], [La/Fe], [Nd/Fe] and [Eu/Fe], in decreasing order of
\spr\ contribution in the Sun. [Ba/Fe] and [La/Fe] display significant
excesses at   [Fe/H] $>  -1$.   Both [Ba/Fe] and [La/Fe] rise steeply
with increasing [Fe/H], however [Ba/Fe] reaches much higher levels than
in the Milky Way.   Conversely, [YII/Fe] is stable, with a  large
dispersion, ranging from the  solar values, like in MW stars,  down to
$\sim -0.7$.  [Eu/Fe]  and [Nd/Fe] are constant in our sample at the
same levels observed in  the MW for  the same [Fe/H].  In  summary, the
\rpr\ elements are overabundant   by  $\sim$0.5 dex  compared  to  iron
and stay constant.
The \spr\ elements are also overabundant but this increases  with
increasing [Fe/H].  The \spr\  origin  of Ba (and  hence La),  over  the
full   [Fe/H] range   of  our sample  is shown in
Figure~\ref{fig:fnx_ba2yy2}, where Ba   is   compared to Eu, a    97\%
\rpr\ dominated element in the sun \citep{1989RPPh...52..945K,
1999ApJ...521..691T}.

The exception  to the rule  of the  dichotomy between  \rpr\ and \spr\
elements in Fornax is [Y/Fe], which has a flat distribution, and falls
on the locus of the MW observations, at same [Fe/H].  In  order to  shed
light on this matter,  Figure~\ref{fig:fnx_ba2yy2} also presents [Ba/Y]
as a function of  [Fe/H].  Ba is clearly more abundant than Y and well
above the level of the MW at same [Fe/H].

There are three s-peaks of very small neutron capture cross sections
resulting in large \spr\ abundances.  The first one corresponds to Sr,
Y, Zr, the second one corresponds to Ba, La, Ce, Pr, Nd and the third
one terminates the \spr\ involving Pb
\citep{1961AnPhy..12..331C,1965ApJS...11..121S}.
\citealt{2004ApJ...601..864T} have shown that the \spr\  production in
AGB  stars  is  strongly  dependent  on [Fe/H]. At  solar metallicity,
AGB  stars produce large amounts of  \spr\    elements from  the
first  peak. At lower [Fe/H], more neutrons   per  Fe seed  are
available, thus bypassing  the first peak  and  progressively feeding
elements in the second neutron magic number peak.

As a consequence, the high [Ba/Y]  could be explained  if Fornax has a
larger  contribution from  metal poor  AGB stars  compared to  the MW,
favouring the creation of heavier  \spr\ elements.

\section{Discussion}\label{sect:discussion}

\subsection{Ages}

Having measured the abundances for our sample of 81 Fornax RGB stars
we can now use isochrones of the appropriate metallicities and
$\alpha$-element content to determine their ages.  Following B06, we
base our analysis on our ESO/MPG/2.2m WFI $V$ and $I$ photometry and
on the isochrones of \citet{2001ApJS..136..417Y} and
\citet{2002ApJS..143..499K}.

\begin{figure}
\centering
\includegraphics[width=\hsize]{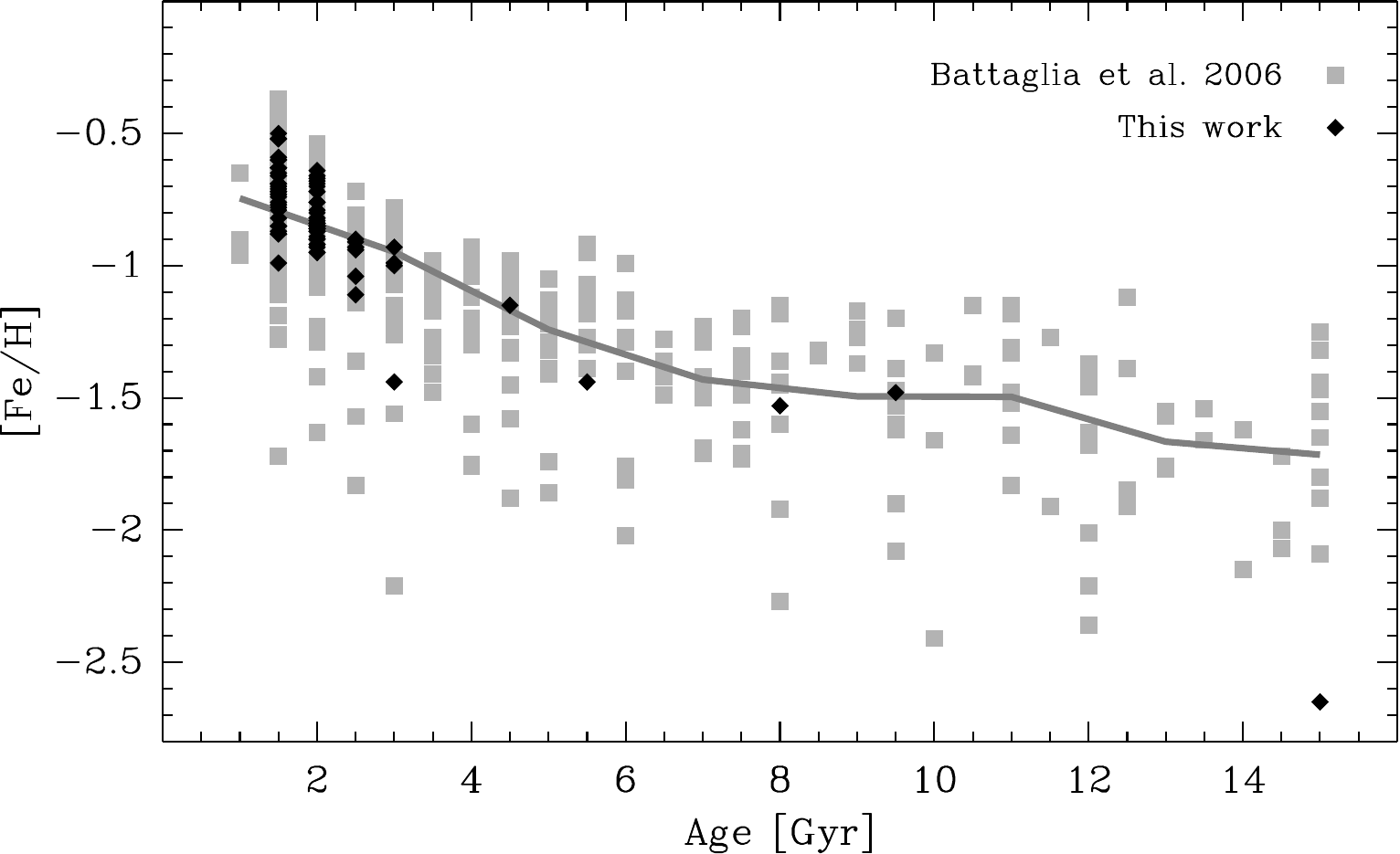}
\caption{The age-metallicity relation for the Fornax dSph, as derived from
the 81 stars from the HR sample (black diamonds) and the LR sample of
B06 (grey shaded squares). The central line follows the average [Fe/H] value of 
 each age bin.
\label{fig:agez}}
\end{figure}

In Figure~\ref{fig:agez}, we combine the results of this work with Ca~II
triplet measurements (B06) of a larger data set to derive a global
age-metallicity relation for the Fornax dSph.  There is a very nice
match between B06 and the present work.
It also stands out clearly that the majority of the stars we observed
here at high spectroscopic resolution in the center of Fornax have ages
between 1.5 and 2 ($\pm$1) Gyr old.  We note however, all these
ages have been derived from solar-scaled chemical composition
isochrones.
The inclusion of alpha-enhancement effects has been shown to reduce
stellar ages by $\sim$ 1 - 1.5 Gyr if one takes [$\alpha$/Fe]=0.3
instead of [$\alpha$/Fe]=0.0 \citep{2002ApJS..143..499K}.  If this
behaviour extrapolates to low $[\alpha$/Fe], the true ages of our sample
stars might be slightly higher, between 1 and 4 Gyr old. In any case,
it can be appreciated that this central sample of stars mostly probes
the population born in the most recent and most intense star formation
episode in the history of Fornax, although it also bears the imprints of
the full chemical evolution of the galaxy.

\cite{2008ApJ...685..933C} present deep photometry over the full surface
of Fornax and provide age-metallicty relations at different
galactocentric distances.  The results of this study agree qualitatively
with our results: the youngest stellar population is the most centrally
concentrated and the most metal-rich.  However, going into finer
details, our spread in metallicity is nearly twice as large as in
\cite{2008ApJ...685..933C}, which also seems to suffer from a systematic
shift in [Fe/H] of $\sim$0.5 dex. These discrepancies can be attributed
to the different techniques of analyses. In particular,
\cite{2008ApJ...685..933C} use the solar-scaled isochrones of
\cite{2002A&A...391..195G}. It is not clear how under solar
$\alpha$-element abundances, as revealed in the present  study, would
affect the colors of the  isochrones.  If we  extrapolate the effects of
passing  from $\alpha$ over-abundance to  solar-scaled models,  for
which the colors get redder due to higher mean opacity,
\cite{2008ApJ...685..933C} might have compensated for the effect of
$\alpha$-element under-abundance by increasing their overall
metallicity.

\subsection{Chemical Evolution}

\begin{figure}[t]
\centering
\includegraphics[width=\hsize]{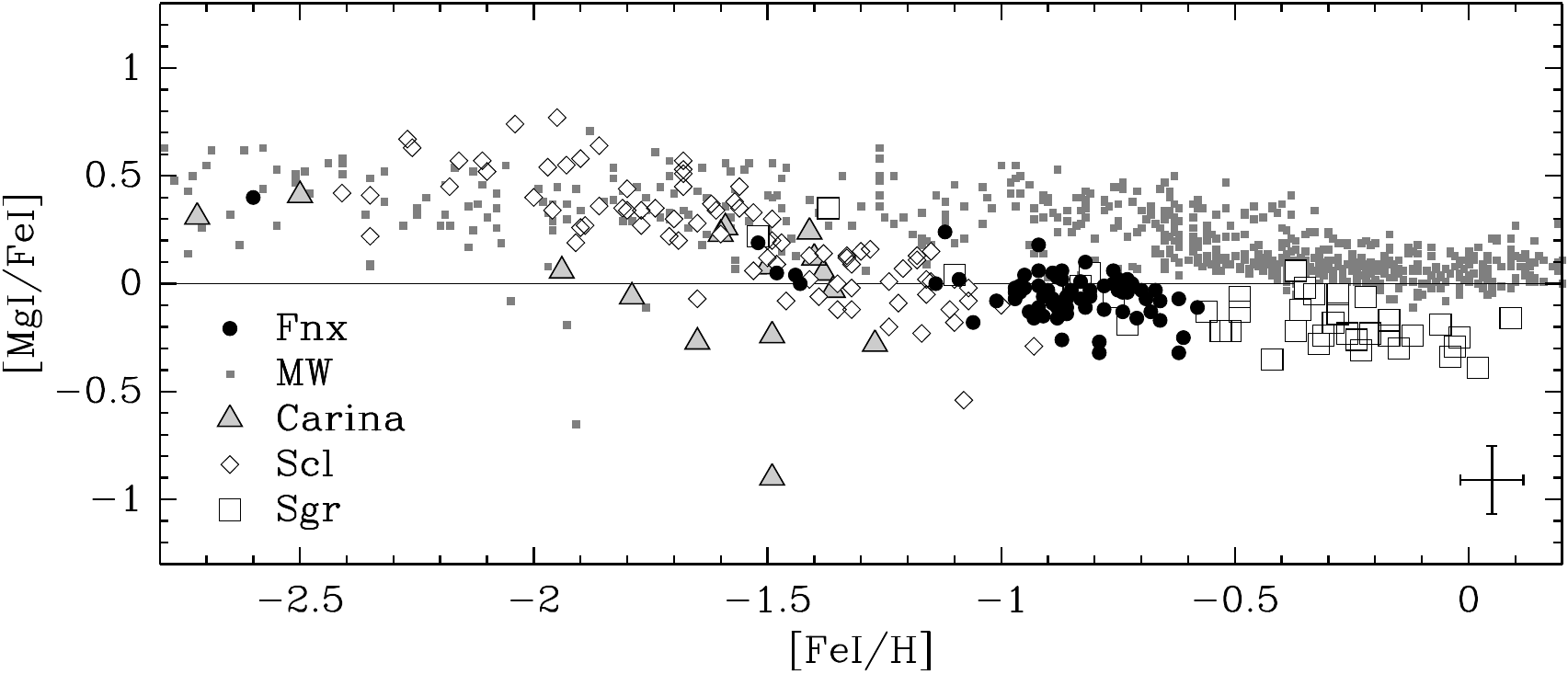}
\caption{The relation between [Mg/Fe] and [Fe/H]  for our sample of RGB
stars observed in Fornax, for the Milky Way halo and disk population
\citep[original references in][]{2004AJ....128.1177V}, the Sculptor dSph
(DART, Hill et al. in prep.), the Sgr dSph \citep{2007A&A...465..815S,
2005A&A...441..141M, 2004A&A...414..503B} and Carina
\citep{2008AJ....135.1580K} 
\label{fig:comp2scla}} 
\end{figure}

Figure~\ref{fig:comp2scla} compares the $\alpha$-element, [Mg/Fe],
distribution in the central region of Fornax dSph, with the results of a
similar FLAMES study of the Sculptor dSph (89 stars, Hill et al., in
prep, \citealt{Tolstoy09}), the Milky Way \citep{2004AJ....128.1177V},
the Sgr dSph \citep{2007A&A...465..815S, 2005A&A...441..141M,
2004A&A...414..503B} and Carina \citep{2008AJ....135.1580K}

Sculptor is  to  date   the only  dSph   galaxy with  a  statistically
significant number  of stars observed with high  [Mg/Fe], matching the
locus of the Milky Way halo and thick disk stars. Nevertheless the few
stars gathered  at [Fe/H] $\leq -2.5$  in Carina  and Fornax, and also
the Sagittarius and Draco \citep{CohenHuang09} stars with  [Fe/H]
$\sim -1.5$ suggest that it is very  likely that all Local Group dSphs
possess $\alpha$-overabundant old stellar populations (see also
\citet{2009AJ....137...62S} for the LeoII  dSph from lower resolution
spectroscopy).  The metallicity at which [Mg/Fe] starts  to  decrease
(sometimes called ``the knee''), as a   consequence  of the  onset of
the  SNe Ia explosions, depends   on the specific  star  formation
history of each galaxy, and  in particular  on its  efficiency 
\citep[e.g.,][]{Tinsley79}.  The  Sculptor dSph is currently the only
system for  which we can identify this point ([Fe/H]=$-1.8$) with any
accuracy. Intriguingly this is roughly the same metallicity which
defines the break between two distinct kinematic and spatial components
in Sculptor \citep[][]{2004ApJ...617L.119T, 2008ApJ...681L..13B}.

All four dSphs  in Figure~\ref{fig:comp2scla} clearly have lower star
formation  efficiency than  the Milky  Way, because  SNe Ia start to
explode  at [Fe/H]  $<< -0.8$, which is at much lower values than the
``knee'' of the MW.  It is interesting to see how Sculptor, Fornax
and Sagittarius form a sequence increasing ``knee'' metallicity
position, reflecting increasing strength and duration of star formation
periods that also lead to higher mean metallicities. This is also
reflected in the metallicity luminosity relation observed for dSph
galaxies \citep[e.g.][]{1998ARA&A..36..435M, Kirby08}.
The "knee" position in the low luminosity Draco may be as low as ([Fe/H]$\leq -2.5$) \citep{CohenHuang09}.

The [Mg/Fe] values for all dSph also reach   lower values than  in the
Milky  Way, as predicted by chemical   evolution models
\citep{lanfranchi04, Revaz09}. This is  a consequence of the absence  of
a balance between SNe Ia ejecta and on-going star formation.

These three galaxies also exhibit a very small dispersion in [Mg/Fe] at
a given [Fe/H], and what there is comes mostly, if not all, from the
measurement errors.  Conversely, Carina seems to shows hints of a
larger dispersion in [Mg/Fe] \citep[][]{2008AJ....135.1580K}, which is
presumably the result of an highly episodic star formation history
\citep{smecker-hane96, hurley-keller98}.  Long periods of quiescence
between the star formation events can translate into scattered abundance
ratios \citep{Matteucci90, Gilmore91}.  \citet{Revaz09} predict that low
mass systems (initial baryonic + dark matter mass $<$ 3$\times10^8$
\msol, such as Carina, can have episodic bursts of star formation, as a
consequence of their long gas cooling time. Also additional factors
may play a role in shaping the star formation history of galaxies of
similar masses, such as the environment (galaxy orbit around the
Milky-Way for example).

\subsection{Nucleosynthesis}

\begin{figure}[!htp]
\centering
\includegraphics[width=\hsize]{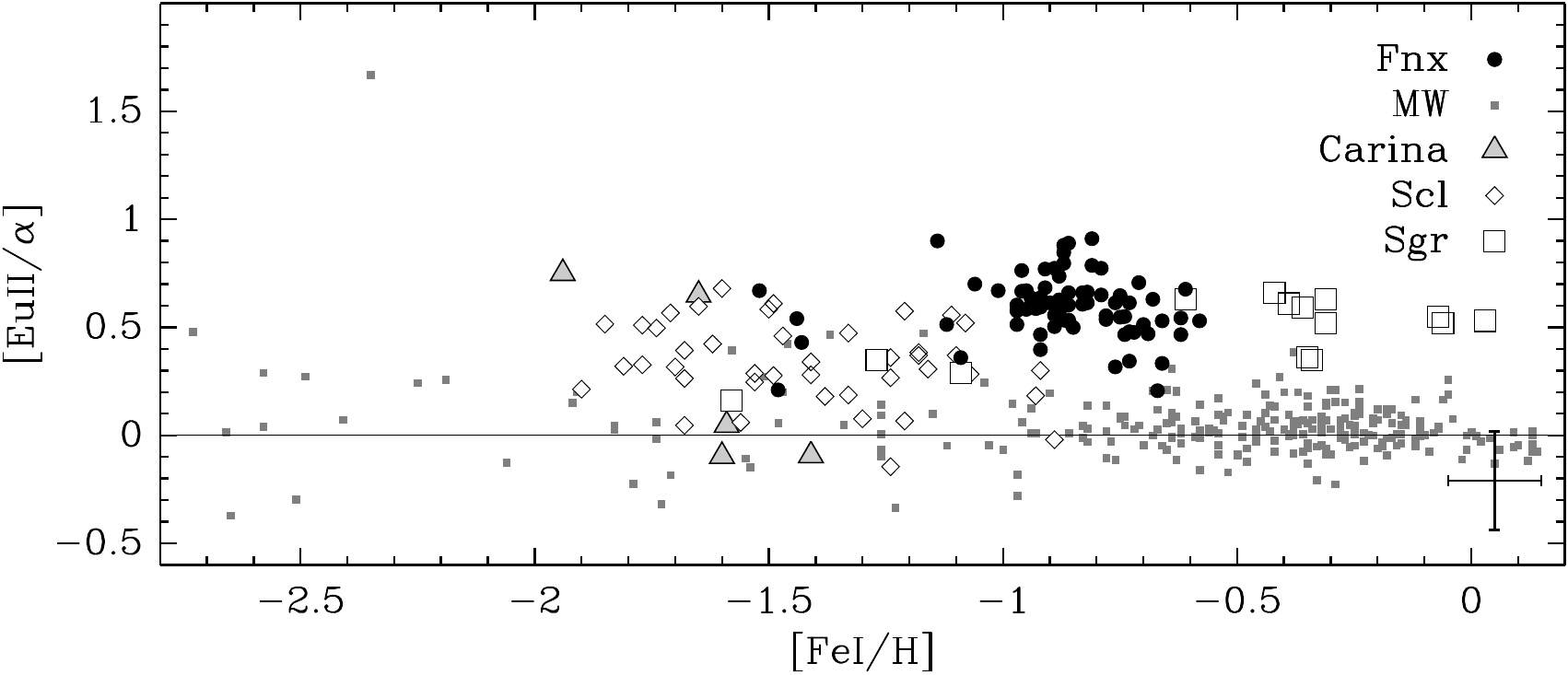}
\caption{The relation between [Eu/Mg] and [Fe/H] in our Fornax star
sample as compared to the Milky Way \citep[original references
in][]{2004AJ....128.1177V}, and three other dSph galaxies: Sagittarius
\citep[][private communication]{2005ASPC..336..221M}, Sculptor (DART
sample, Hill et al., in preparation) and Carina
\citep{2003AJ....125..684S}.  
\label{fig:heavy_alpha}} 
\end{figure}

Figure~\ref{fig:heavy_alpha} extends the comparison of Fornax with other
dSph galaxies to the case of the neutron capture elements. It supports
different sites and relative contributions for the production of
$\alpha$- and \rpr\ elements.  If Eu and $\alpha$-elements were
synthesized in the same sites, the ratio of [Eu/$\alpha$] would stay
constant and identical in all galaxies, which it clearly does not.

This is not the first evidence for a decoupling of the production of
$\alpha$ and \rpr\ elements : it is already well known among the
very metal-poor stars of the Milky Way. \citet{2008ARA&A..46..241S}
compiled different analyses of Milky Way halo metal-poor stars
demonstrating that, while there is very little dispersion in [Mg/Fe]
(taken as representative of the $\alpha$ elements), [Eu/Fe] exhibits a
very large scatter at metallicities around [Fe/H]$\sim -3.0$
\citep[e.g.,][]{1995AJ....109.2757M, 2002AJ....123..404F,
2004A&A...416.1117C, 2004ApJ...612.1107C, 2005A&A...439..129B,
2007A&A...476..935F}. Not only does this indicate that the production of
\rpr\ elements was rare while there was profusion of $\alpha$-elements
in the Milky Way at early times, but this also points towards a
different mass range of progenitor.

Several scenarios have been proposed for the origin of the \rpr\
elements, including high entropy neutrino winds
\citep[e.g.][]{1994ApJ...433..229W, 2009ApJ...694L..49F} and prompt
supernova explosions of 8-10\msol progenitors
\citep[e.g.][]{2003ApJ...593..968W}. This latter scenario is
characterized by a lack of $\alpha$ elements and only a small amount of
iron-peak nuclei. Whilst the exact low mass boundaries vary, the origin
of $\alpha$ and \rpr\ elements in respectively massive and light SNeII
reach broad agreement in  reproducing Milky Way halo metal-poor
stars
\citep[e.g.,][]{1992ApJ...391..719M,1999ApJ...521..691T,2006A&A...448..557C,
2007PhR...442..237Q}, including the downturn in the second-peak \rpr\ at
the very lowest metallicities \citep{2007A&A...476..935F,
2009A&A...494.1083A}.

The dSph galaxies assembled here provide another perspective on the
origin of Eu. Indeed, it is noteworthy that the two galaxies which have
distinctively high [Eu/$\alpha$], over-abundant by +0.7 and +0.5 dex
compared to the Milky Way, also reveal a significant influence of AGB
winds in their chemical evolution. [Ba/Fe] is in general extremely high
in most of our sample of Fornax stars, and this is also true for Sgr
 as well as the LMC \citep{2008A&A...480..379P}. Interestingly, Sgr
and Fnx dSphs are also the most massive dSphs in the Local Group,
reaching the highest metallicities of their family of galactic systems
and are the only ones to possess globular clusters, implying that higher
mass galaxies in the vicinity of the Milky Way are better able to hold
on to their gas, allowing them to reach this evolutionary stage 
where AGB contribute significantly to the enrichment. Given this strong
influence of the \spr\ in Fornax, one may question whether the s-process
can participate a significant fraction of the Eu production in this
galaxy.

\begin{figure}[!htp]
\centering
\includegraphics[width=\hsize]{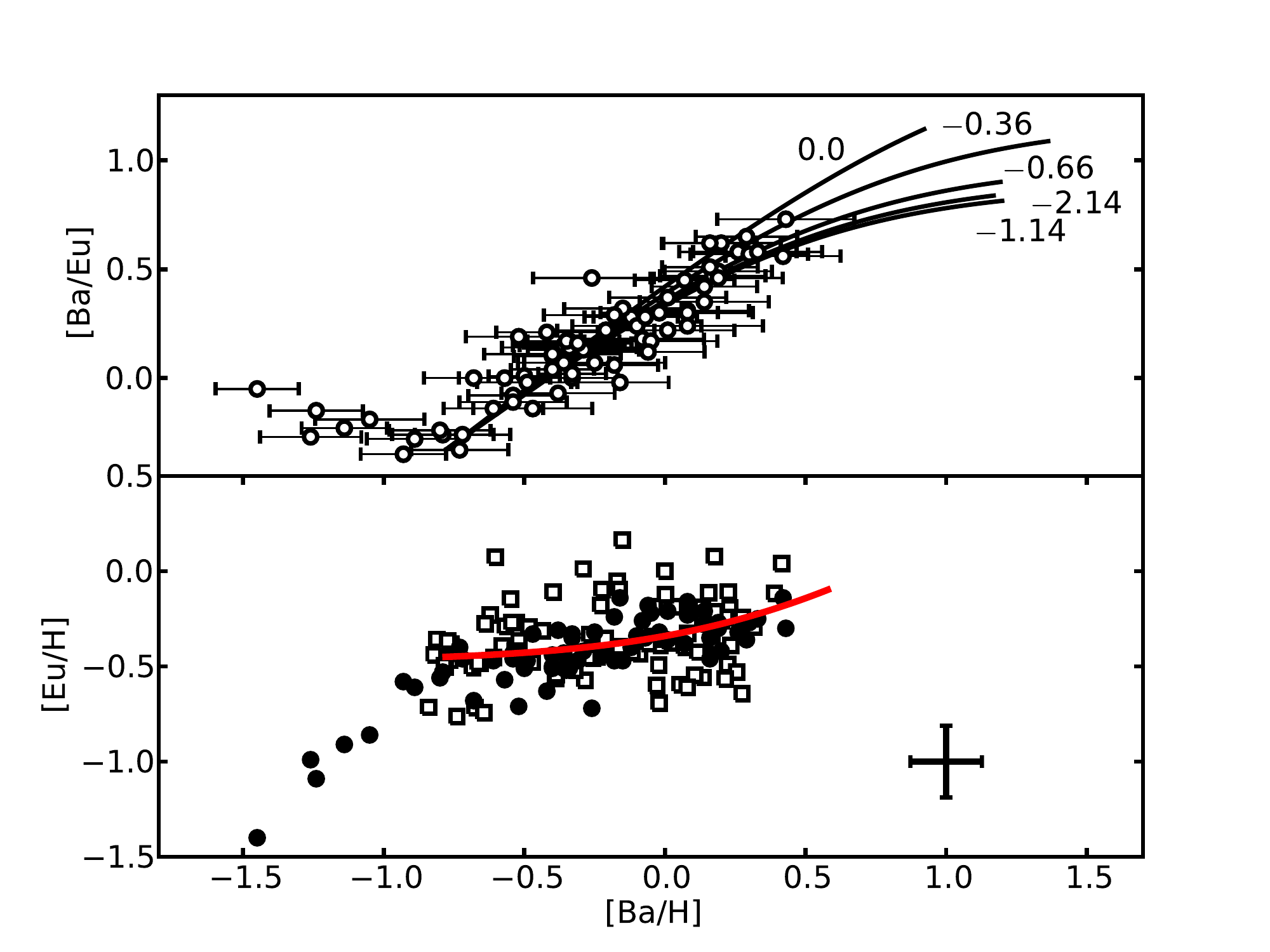}
\caption{[Eu/H] and [Ba/Eu] versus [Ba/H]. 
{\it Upper panel :} The expected evolution of [Ba/Eu] as a function of
[Ba/H] when the evolution of Ba and Eu are triggered by AGB winds only,
starting at [Ba/H]=$-0.78$ and [Ba/Eu]=$-0.33$. The five solid lines
follow the chemical evolution induced by the AGB models of
\citet{2009ApJ...696..797C} at different metallicities
([Z/Z$_{\odot}$]), which values are indicated close to the upper edges
of the model lines.  The Fornax sample stars from this work are shown in
open circles, together with the individual error bars in [Ba/H].  {\it
Lower panel :} The black points show the observations of Fornax stars
from this work. The solid line traces the locus of the evolution in Eu
and Ba induced by the yields of [Z/Z$_{\odot}$]=$-2.14$ AGBs
\citep{2009ApJ...696..797C}, starting at [Ba/H]=$-0.78$ and
[Ba/Eu]=$-0.33$. The black open squares are 85 random trials taking into
account the mean observational errors, 0.15 dex in [Ba/H] and 0.19 dex
in [Eu/H] indicated at the bottom right of the frame. They show the
scattering of the single model line induced by the observational
uncertainties only.  \label{fig:BaH_EuH}} \end{figure}

\citet{2003ApJ...592L..67A} have demonstrated the \spr\ origin of Eu in
two Milky Way halo stars strongly enriched in \spr\ elements. This is
thus a good starting point for the investigation of Fornax at [Fe/H]$\ge
-1$ where the \spr\ begins to dominate.  The effective \spr\ component
integrates successive generations of AGB stars according to the chemical
evolution of the galactic system which is considered, hence mixing the
products of different AGB yields, themselves functions of the initial
metallicity, stellar mass, 13C pocket efficiency, and other physical
properties \citep{2004ApJ...601..864T}. At [Fe/H]=$-1$, where [Ba/Fe]
rises strongly, Fornax is already a chemically complex system with
signatures of SNe~Ia and AGB products. We will try to reproduce the
behavior of Ba and Eu from there on.

Figure~\ref{fig:BaH_EuH} presents the variation of [Eu/H] and [Ba/Eu] as
a function of [Ba/H] for our sample of Fornax stars. Analogous to
\citet{2005ASPC..336..221M} in the case of the Sgr dSph, we find that a
continuous enrichment by 90\% \spr\ and 10\% \rpr\ as defined in the
solar neighborhood \citep{2004ApJ...617.1091S} and starting at
[Ba/H]$\sim -0.8$ and [Ba/Eu]$\sim -0.3$ reproduces the observed trend
between Ba and Eu.  This is equivalent to considering that, from
[Fe/H]=$-1$ on, the nucleosynthesis in AGBs is responsible for 90\% of
the neutron-capture elements via the \spr\ and only 10\% is contributed
by the \rpr .  Still, the exact \rpr\ and \spr\ fractions depend on the
prescriptions for AGB yields which are considered. Recently,
\citet{2009ApJ...696..797C} calculated that not only the yields of Ba
and Eu of a 2\msol AGB vary with [Fe/H], but also their ratio, leading
to an increased fraction of Eu relative to Ba with decreasing
metallicity. In the upper panel of Figure~\ref{fig:BaH_EuH}, the
evolution in [Ba/Eu] as a function of [Ba/H], induced by each of the 5
different metallicities ([Z/Z$_{\odot}$]) of
\citet{2009ApJ...696..797C}, are overplotted on the measurements of our
sample of Fornax stars. The [Z/Z$_{\odot}$]=0.0 and $-0.36$ AGBs tend to
produce too high [Ba/Eu], and so does the $-0.66$ model, albeit
marginally. It is nearly impossible to distinguish between the $-1.14$
and $-2.14$ models, both passing satisfactorily through the data points.
In the lower panel of Figure~\ref{fig:BaH_EuH}, we now consider
[Z/Z$_{\odot}$]= $-2.14$ AGBs only. This single model is then convolved
with the mean observed errors in [Ba/H] (0.15 dex) and [Eu/H] (0.19 dex)
resulting in the scattered squares which convincingly match the
observations. In conclusion, whilst the high [Eu/$\alpha$] of the 1 to
4~Gyr Fornax stars is primarily due to their low [$\alpha$/Fe] content,
the strong influence of the AGB nucleosynthesis increases their Eu
abundance and accentuates this feature of their abundance ratio. This
provides an explanation for the observed plateau in [Eu/Fe] at [Fe/H]$>
-1$ seen in Fornax, while it decreases in the Milky Way after the onset
the SNe Ia explosions.  An additional \rpr\ is required only if too high
metallicity AGBs are considered, such as the solar neighborhood standard
model one. A refined and more detailed chemical evolution modeling is
mandatory to evaluate whether a range in AGB metallicities (rather than
a single one as investigated here) is necessary to reproduce Fornax
properties.

\section{Conclusions}

Thanks to the  multi-fibre capability of  FLAMES we have been able  to
make detailed abundance measurements of 14 elements for a sample of 81
individual RGB  stars  in the  central  25$'$  of Fornax.  This   is a
significant, even dramatic, improvement on the previous UVES sample of 3
individual  field  stars  \citet{2003AJ....125..684S}.  We   use new
automatic procedures to measure absorption line equivalent widths, new
spherical stellar atmosphere models and new  line list, all adapted to
the  resolution  and wavelength  coverage  of  GIRAFFE.   Our thorough
investigation of  the potential  systematic errors firmly  demonstrate
that our abundances  are accurate.  We summarize  below the most
important results of this work:

$\bullet$ Although our sample was randomly chosen from the entire
breadth of the RGB (see Figure~\ref{fig:fnx-cmd-target}), it is
dominated by a relatively young (1 to 4 Gyr) and metal rich
($-1.5<$[Fe/H] $<-0.5$) stars
.

$\bullet$ These stars are significantly underabundant in [$\alpha$/Fe]
with  respect to the Milky-Way.  We observe a slightly different
behaviour among the $\alpha$-elements such that the underabundance of Ti
and Ca is more  pronounced than that of Mg and Si.  Either these two
groups of   elements  have different nucleosynthetis,  or  a large
fraction of both Ca and  Ti is  produced  in low metallicity SNIa.
However, we cannot discard the possiblity that non-LTE effects have
caused these differential trends.

$\bullet$ The dispersion in [$\alpha$/Fe], and also other abundance
ratios, is found to be very  small, essentially within the measurement
errors, which is similar to what is seen in Sculptor and Sagittarius
dSph  and also the Milky Way halo.

$\bullet$ Our sample includes one  star, at [Fe/H]=$-2.5$, which
contains above solar values of $\alpha$-elements, equivalent to a
Galactic halo  star at the same [Fe/H].  This  and a few similar stars
already found in Sculptor, Carina, and Sagittarius dSph  galaxies
suggest that an  old  stellar population with above solar
$\alpha$-element abundances is a common feature in all galaxies.

$\bullet$ [Ni/Fe] and [Cr/Fe] are  typically sub-solar in our sample.
The different behaviour seen in Fornax  stars compared to the  Galactic
trend  suggests  that the iron-peak elements are sensitive to the
evolutionary history  of the  parent galaxy and in  particular the low
SNe Ia yields are  due to the excess of low metallicity progenitors in
dSph galaxies. This  hypothesis is also supported by the Na-Ni
relationship in Fornax, which extends the  relation found in the Milky
Way to lower [Na/Fe] and [Ni/Fe].

$\bullet$ [Ba/Fe] and [La/Fe] values rise steeply with increasing
metallicity, with [Ba/Fe] reaching much higher levels than in our
Galaxy. Conversely, [Y/Fe] remains constant. The high [Ba/Y] can be
explained by a attriburing an important contribution to low metallicity
AGBs. This also implies that the high [Eu/$\alpha$] is not only a
consequence of the low abundance in $\alpha$ elements but is also linked
to the strong influence of AGBs at [Fe/H]$<$-1, which increases the
abundance of Eu.

\begin{acknowledgements}
This work was made possible by a grant from NWO (614.031.018). BL and ET
thank Paris Observatory for hospitality and the Programme National
Galaxies financial support.  We thank Marco Gullieuszik and Enrico Held
for sending us in advance of publication their IR photometry for stars
in our sample. MS thanks the National Science Foundation for support
under AST-0306884. E.T. gratefully acknowledges support from an NWO-VICI grant.
\end{acknowledgements}

\bibliographystyle{aa} 
\bibliography{bruno} 

\appendix 
\renewcommand{\thetable}{A.\arabic{table}} 
\onecolumn

\begin{table}
\begin{center}
\caption{ 
\vrad\ and associated errors for our Fornax targets.  
{\bf BOLD} is a probable foreground star, since it falls outside the
range of \vrad\ for membership in Fornax (see Figure~\ref{fig:histovrad}).  
\st{Struck-out} stars have been rejected because their September average velocity was
significantly different ($>$3 km/s) to their January velocity.  This
suggests the possibility that they are binary stars and they 
were therefore rejected from further analysis.
\label{tab:fnx-vrad}}
\resizebox{\textwidth}{!}{
\begin{tabular}{lrr|lrr|lrr}
\hline \hline
Star&     \vrad&    Stddev& Star&     \vrad&    Stddev& Star&     \vrad&    Stddev   \\
\hline
    {BL038}&       {  62.86}&     {   0.74}&     {BL147}&       {  52.39}&     {   0.61}&     {BL221}&       {  54.25}&     {   0.32}\\          
    {BL045}&       {  79.54}&     {   0.23}&     {BL148}&       {  65.48}&     {   0.65}&     {BL227}&       {  52.15}&     {   0.67}\\          
    {BL052}&       {  30.32}&     {   1.25}&     {BL149}&       {  50.70}&     {   0.40}&     {BL228}&       {  55.00}&     {   0.34}\\          
    {BL061}&       {  82.38}&     {   1.12}&     {BL150}&       {  68.70}&     {   1.12}&     {BL229}&       {  63.09}&     {   0.57}\\          
    {BL065}&       {  63.86}&     {   0.78}&     {BL151}&       {  41.41}&     {   0.58}&     {BL231}&       {  63.28}&     {   1.11}\\          
 \st{BL069}&    \st{  68.85}&  \st{   2.63}&     {BL153}&       {  50.68}&     {   0.53}&     {BL233}&       {  58.91}&     {   0.40}\\          
 \st{BL070}&    \st{  60.43}&  \st{   2.85}&     {BL155}&       {  46.18}&     {   0.21}&     {BL239}&       {  47.47}&     {   0.35}\\
\hline
\end{tabular}
}
\end{center}
\end{table}

\begin{landscape}
\begin{table}
\begin{center}
\caption{Stellar parameters of the model used for our sample of 81 stars, along with
the minimum $EW$ used for each star.
\label{tab:fnx-stelparmod}}
\begin{tabular}{lrrrrr|lrrrrr|lrrrrr|}
\hline \hline
Star & \teff& \logg& [Fe/H]& \vmic& $EW_{\mathrm{min}}$ &
Star & \teff& \logg& [Fe/H]& \vmic& $EW_{\mathrm{min}}$ &
Star & \teff& \logg& [Fe/H]& \vmic& $EW_{\mathrm{min}}$ \\
\hline
BL038& 3980&  0.69& -0.88&  2.2& 14.24& BL148& 4123&  0.72& -0.63&  2.2& 20.62& BL218& 3939&  0.67& -0.60&  2.0& 16.17\\   
BL045& 4122&  0.85& -1.09&  2.1& 12.15& BL149& 4100&  0.88& -0.91&  2.2& 14.30& BL221& 4056&  0.81& -0.86&  2.1& 13.80\\   
BL052& 3997&  0.72& -1.02&  2.4& 14.24& BL150& 4025&  0.80& -0.83&  2.2& 15.29& BL227& 4046&  0.84& -0.91&  2.0& 15.40\\   
BL065& 4330&  0.97& -1.43&  1.9& 11.33& BL151& 4024&  0.81& -0.86&  2.1& 13.37& BL228& 3992&  0.71& -0.88&  2.3& 13.31\\   
BL076& 4065&  0.83& -0.85&  2.2& 13.04& BL155& 4060&  0.90& -0.75&  2.3& 17.21& BL229& 4014&  0.80& -0.71&  2.3& 15.73\\   
BL077& 4026&  0.80& -0.79&  2.2& 12.71& BL156& 4099&  0.89& -1.13&  2.2& 13.48& BL233& 4048&  0.83& -0.68&  2.2& 14.52\\   
BL079& 4036&  0.76& -0.52&  2.1& 17.98& BL158& 4078&  0.87& -0.87&  2.0& 15.79& BL239& 4123&  0.68& -0.91&  2.1& 13.80\\
\hline
\end{tabular}
\end{center}
\end{table}
\end{landscape}

\begin{center}
\begin{longtable}{rccrrrrrrrrrrr}
\caption{Coordinates and photometric temperatures in different colours, shifted on the
\teff($V-I$) scale.  \teff\ is the average of the four different $T_{V-X}$, 
$BC_{V}$ the bolometric correction and $M_{\mathrm{Bol}}$ the bolometric magnitude.
\label{tab:fnxfi-photo}}\\
\hline \hline
Star&  RA (J2000) & DEC (J2000) & $T_{V-I}$& $T_{V-J}$& $T_{V-H}$& $T_{V-K}$& \teff& \logg&
$BC_{V}$&  $M_{\mathrm{Bol}}$\\
\hline
\endfirsthead
\hline \hline
Star&  RA (J2000) & DEC (J2000) & $T_{V-I}$& $T_{V-J}$& $T_{V-H}$& $T_{V-K}$& \teff& \logg&
$BC_{V}$&  $M_{\mathrm{Bol}}$\\
\hline
\endhead
\hline \multicolumn{14}{r}{\emph{Continued on next page}}
\endfoot
\hline
\endlastfoot
BL038 & 02 40 20.45 & --34 24 00.1 &  3984&  3976&  3996&  3964&  3980&  0.69&  --0.93&  --3.32\\  
BL045 & 02 40 07.52 & --34 23 31.8 &  4114&  4134&  4128&  4110&  4122&  0.85&  --0.77&  --3.07\\  
BL052 & 02 40 10.42 & --34 25 17.6 &  4009&  3999&  3995&  3984&  3997&  0.72&  --0.91&  --3.27\\  
BL061 & 02 39 28.59 & --34 18 38.0 &  4316&  4341&  4335&  4333&  4331&  0.84&  --0.60&  --3.30\\  
BL065 & 02 39 22.15 & --34 19 40.3 &  4315&  4340&  4334&  4332&  4330&  0.97&  --0.59&  --2.99\\  
BL069 & 02 39 29.56 & --34 25 10.6 &  4143&  3999&  4033&  3994&  4042&  0.72&  --0.86&  --3.29\\  
BL070 & 02 39 40.46 & --34 19 38.8 &  3939&  3928&  3931&  3935&  3933&  0.68&  --1.00&  --3.29\\  
BL076 & 02 39 31.49 & --34 23 05.1 &  4088&  4061&  4063&  4048&  4065&  0.83&  --0.83&  --3.05\\
\end{longtable}
\end{center}

\begin{center}
\tiny
{\sffamily
\tabcolsep 3pt
\begin{longtable}{|llrr|rrrrrrrrrrrrrrrr|}
\caption{List of all the lines, their associated parameters and their measured equivalent width ($EW$s in m\AA, measured by \texttt{DAOSPEC}) for all our Fornax stars. Part 1/5.
\label{tab:fnxfi-ewlist1}}\\
\cline{5-20}
\multicolumn{4}{c|}{} &
\multicolumn{16}{c|}{} \\
\cline{1-4}
\multicolumn{4}{|c|}{} &
\multicolumn{16}{c|}{Equivalent witdh, one star per column, BLxxx} \\
$\lambda$(\AA)& elem& $\chi$& \gf& 
038& 045& 052& 065& 076& 077& 079& 081& 084& 085& 091& 092& 096& 097& 100& 104\\
\hline
\endfirsthead
\hline \hline
$\lambda$(\AA)& elem& $\chi$& \gf& 
038& 045& 052& 065& 076& 077& 079& 081& 084& 085& 091& 092& 096& 097& 100& 104\\
\hline
\endhead
\hline \multicolumn{20}{r}{\emph{Continued on next page}}
\endfoot
\hline
\endlastfoot
6154.23& Na\,{\sc i} &    2.10&   -1.56&   38.8&   ...&   ...&   ...&  23.5&   ...&  35.7&  36.3&   ...&   ...&  25.6&  15.5&  23.4&  19.8&   ...&   ...\\
6160.75& Na\,{\sc i} &    2.10&   -1.26&   58.4&   ...&  23.3&   ...&  38.2&  38.6&  65.5&  54.5&   ...&   ...&  38.5&  21.1&  47.6&  39.7&   ...&  28.5\\
5528.41& Mg\,{\sc i} &    4.35&   -0.36&  221.1& 195.6& 212.0& 165.9& 207.0& 200.9& 253.7& 208.8& 217.3& 112.8& 203.8& 210.0& 220.0& 234.8& 198.8& 219.0\\
6318.72& Mg\,{\sc i} &    5.11&   -1.97&   43.6&   ...&   ...&   ...&  34.9&  24.6&   ...&   ...&   ...&   ...&  31.1&  38.9&  52.2&   ...&  17.7&  34.0\\
6319.24& Mg\,{\sc i} &    5.11&   -2.21&    ...&  27.0&  26.9&   ...&   ...&  23.3&  36.2&  24.9&   ...&   ...&   ...&  30.6&  42.1&   ...&  31.8&   ...\\
6319.49& Mg\,{\sc i} &    5.11&   -2.43&    ...&   ...&   ...&   ...&   ...&   ...&   ...&   ...&   ...&   ...&   ...&   ...&   ...&   ...&   ...&   ...\\
6125.03& Si\,{\sc i} &    5.62&   -1.57&   25.0&  24.7&   ...&  18.3&   ...&   ...&   ...&   ...&   ...&   ...&  22.4&   ...&   ...&   ...&   ...&  34.2\\
6142.48& Si\,{\sc i} &    5.62&   -1.51&    ...&   ...&   ...&   ...&   ...&   ...&   ...&   ...&   ...&   ...&   ...&   ...&   ...&   ...&   ...&   ...\\
\end{longtable}
}
\end{center}

\addtocounter{table}{-1}
\begin{center}
\tiny
{\sffamily
\tabcolsep 3pt
\begin{longtable}{|llrr|rrrrrrrrrrrrrrrr|}
\caption{List of all the lines, their associated parameters and their measured equivalent width ($EW$s in m\AA, measured by \texttt{DAOSPEC}) for all our Fornax stars.  Part 2/5.
\label{tab:fnxfi-ewlist2}}\\
\cline{5-20}
\multicolumn{4}{c|}{} &
\multicolumn{16}{c|}{} \\
\cline{1-4}
\multicolumn{4}{|c|}{} &
\multicolumn{16}{c|}{Equivalent witdh, one star per column, BLxxx} \\
$\lambda$(\AA)& elem& $\chi$& \gf& 
113& 115& 123& 125& 132& 135& 138& 140& 141& 146& 147& 148& 149& 150& 151& 155\\ 
\hline
\endfirsthead
\hline \hline
$\lambda$(\AA)& elem& $\chi$& \gf& 
113& 115& 123& 125& 132& 135& 138& 140& 141& 146& 147& 148& 149& 150& 151& 155\\ 
\hline
\endhead
\hline \multicolumn{20}{r}{\emph{Continued on next page}}
\endfoot
\hline
\endlastfoot
6154.23& Na\,{\sc i} &    2.10&   -1.56&   41.3&   ...&   ...&  20.5&  20.4&  20.9&   ...&   ...&  21.9&  20.2&   ...&  46.1&   ...&  33.3&  24.7&  27.3\\
6160.75& Na\,{\sc i} &    2.10&   -1.26&   59.5&   ...&  31.9&  66.9&  22.8&  49.0&   ...&  33.4&  19.6&  26.1&  18.9&  81.9&  41.0&  42.1&  43.0&  58.0\\
5528.41& Mg\,{\sc i} &    4.35&   -0.36&  229.7& 168.1& 218.5& 234.1& 214.5& 219.7& 211.1& 203.8& 201.3& 213.3& 179.4& 221.2& 203.4& 211.4& 215.0& 205.7\\
6318.72& Mg\,{\sc i} &    5.11&   -1.97&    ...&  16.2&  34.9&  49.1&  34.8&   ...&  35.6&  46.7&  39.3&  30.9&  18.4&  49.4&   ...&  48.0&  35.6&   ...\\
6319.24& Mg\,{\sc i} &    5.11&   -2.21&    ...&  16.2&   ...&   ...&   ...&   ...&  20.5&   ...&   ...&  16.8&   ...&  31.6&  29.6&  15.2&  25.0&  43.2\\
6319.49& Mg\,{\sc i} &    5.11&   -2.43&    ...&   ...&   ...&   ...&  17.4&   ...&   ...&  29.5&   ...&   ...&  30.2&   ...&   ...&   ...&   ...&   ...\\
6125.03& Si\,{\sc i} &    5.62&   -1.57&    ...&   ...&   ...&   ...&  20.1&  20.8&   ...&   ...&   ...&   ...&   ...&   ...&  31.7&   ...&   ...&   ...\\
6142.48& Si\,{\sc i} &    5.62&   -1.51&    ...&   ...&   ...&   ...&   ...&   ...&   ...&   ...&   ...&   ...&   ...&   ...&   ...&   ...&   ...&   ...\\
\end{longtable}
}
\end{center}

\addtocounter{table}{-1}
\begin{center}
\tiny
{\sffamily
\tabcolsep 3pt
\begin{longtable}{|llrr|rrrrrrrrrrrrrrrr|}
\caption{List of all the lines, their associated parameters and their measured equivalent width ($EW$s in m\AA, measured by \texttt{DAOSPEC}) for all our Fornax stars.  Part 3/5.
\label{tab:fnxfi-ewlist3}}\\
\cline{5-20}
\multicolumn{4}{c|}{} &
\multicolumn{16}{c|}{} \\
\cline{1-4}
\multicolumn{4}{|c|}{} &
\multicolumn{16}{c|}{Equivalent witdh, one star per column, BLxxx} \\
$\lambda$(\AA)& elem& $\chi$& \gf& 
156& 158& 160& 163& 166& 168& 171& 173& 180& 185& 190& 195& 196& 197& 203& 204\\
\hline
\endfirsthead
\hline \hline
$\lambda$(\AA)& elem& $\chi$& \gf& 
156& 158& 160& 163& 166& 168& 171& 173& 180& 185& 190& 195& 196& 197& 203& 204\\
\hline
\endhead
\hline \multicolumn{20}{r}{\emph{Continued on next page}}
\endfoot
\hline
\endlastfoot
6154.23& Na\,{\sc i} &    2.10&   -1.56&    ...&  22.3&   ...&  41.9&  19.6&   ...&   ...&  25.5&  19.3&  49.6&   ...&  18.2&   ...&  23.6&  24.3&   ...\\
6160.75& Na\,{\sc i} &    2.10&   -1.26&    ...&  42.3&  40.7&  53.0&  47.3&  47.9&   ...&  59.0&  45.5&  54.0&   ...&  23.7&  19.0&  42.0&  53.5&  37.6\\
5528.41& Mg\,{\sc i} &    4.35&   -0.36&  202.6& 214.5& 206.5& 221.8& 216.6& 205.1& 214.3& 225.0& 209.3& 228.0& 209.7& 196.9& 208.5& 209.9& 212.6& 217.3\\
6318.72& Mg\,{\sc i} &    5.11&   -1.97&    ...&  37.4&  29.8&  41.1&  37.8&  32.0&  38.7&  36.9&  33.1&  47.8&  29.6&  32.7&   ...&  40.3&  37.3&   ...\\
6319.24& Mg\,{\sc i} &    5.11&   -2.21&   20.3&  35.0&   ...&   ...&  25.3&   ...&  30.1&  25.5&   ...&  25.5&   ...&   ...&   ...&   ...&  28.9&   ...\\
6319.49& Mg\,{\sc i} &    5.11&   -2.43&    ...&   ...&   ...&   ...&   ...&   ...&   ...&   ...&   ...&   ...&   ...&   ...&   ...&  31.2&   ...&   ...\\
6125.03& Si\,{\sc i} &    5.62&   -1.57&    ...&   ...&   ...&   ...&   ...&   ...&   ...&   ...&   ...&   ...&   ...&   ...&   ...&   ...&   ...&   ...\\
6142.48& Si\,{\sc i} &    5.62&   -1.51&    ...&   ...&   ...&   ...&   ...&   ...&   ...&   ...&   ...&  18.4&   ...&   ...&   ...&   ...&   ...&   ...\\
\end{longtable}
}
\end{center}

\addtocounter{table}{-1}
\begin{center}
\tiny
{\sffamily
\tabcolsep 3pt
\begin{longtable}{|llrr|rrrrrrrrrrrrrrrr|}
\caption{List of all the lines, their associated parameters and their measured equivalent width ($EW$s in m\AA, measured by \texttt{DAOSPEC}) for all our Fornax stars.  Part 4/5.
\label{tab:fnxfi-ewlist4}}\\
\cline{5-20}
\multicolumn{4}{c|}{} &
\multicolumn{16}{c|}{} \\
\cline{1-4}
\multicolumn{4}{|c|}{} &
\multicolumn{16}{c|}{Equivalent witdh, one star per column, BLxxx} \\
$\lambda$(\AA)& elem& $\chi$& \gf& 
205& 208& 210& 211& 213& 216& 218& 221& 227& 228& 229& 233& 239& 242& 247& 250\\ 
\hline
\endfirsthead
\hline \hline
$\lambda$(\AA)& elem& $\chi$& \gf& 
205& 208& 210& 211& 213& 216& 218& 221& 227& 228& 229& 233& 239& 242& 247& 250\\ 
\hline
\endhead
\hline \multicolumn{20}{r}{\emph{Continued on next page}}
\endfoot
\hline
\endlastfoot
6154.23& Na\,{\sc i} &    2.10&   -1.56&   37.4&  38.0&  36.5&  40.5&   ...&  42.5&  27.2&   ...&  33.8&  16.6&  29.5&  37.6&  16.8&   ...&  24.6&  44.5\\
6160.75& Na\,{\sc i} &    2.10&   -1.26&   50.6&  51.6&  51.1&  57.3&  40.1&  65.8&  60.3&  27.5&  48.6&  29.5&  51.5&  63.1&   ...&  28.9&  54.0&  88.7\\
5528.41& Mg\,{\sc i} &    4.35&   -0.36&  217.6& 216.7& 209.4& 211.1& 200.0& 231.6& 234.9& 215.0& 205.8& 215.0& 234.1& 214.4& 204.7& 209.8& 220.6& 230.1\\
6318.72& Mg\,{\sc i} &    5.11&   -1.97&   46.5&  41.8&  55.4&  46.9&  39.1&  51.1&  47.3&   ...&   ...&  31.8&  35.8&  44.6&   ...&  41.6&   ...&  58.6\\
6319.24& Mg\,{\sc i} &    5.11&   -2.21&    ...&  36.2&   ...&  23.7&   ...&  25.3&  33.9&  19.2&  29.8&   ...&   ...&   ...&  19.4&   ...&  43.4&  34.6\\
6319.49& Mg\,{\sc i} &    5.11&   -2.43&    ...&   ...&  35.7&   ...&   ...&   ...&   ...&   ...&   ...&   ...&   ...&  24.0&   ...&  36.0&   ...&   ...\\
6125.03& Si\,{\sc i} &    5.62&   -1.57&   41.7&   ...&   ...&  48.4&   ...&   ...&   ...&   ...&  16.4&   ...&   ...&   ...&   ...&   ...&   ...&   ...\\
6142.48& Si\,{\sc i} &    5.62&   -1.51&    ...&   ...&   ...&   ...&   ...&   ...&   ...&   ...&  18.5&   ...&  21.8&   ...&   ...&   ...&   ...&   ...\\
\end{longtable}
}
\end{center}

\addtocounter{table}{-1}

\begin{center}
\tiny
{\sffamily
\tabcolsep 3pt
\begin{longtable}{|llrr|rrrrrrrrrrrrrrrrr|}
\caption{List of all the lines, their associated parameters and their measured equivalent width ($EW$s in m\AA, measured by \texttt{DAOSPEC}) for all our Fornax stars.  Part 5/5.
\label{tab:fnxfi-ewlist5}}\\
\cline{5-21}
\multicolumn{4}{c|}{} &
\multicolumn{17}{c|}{} \\
\cline{1-4}
\multicolumn{4}{|c|}{} &
\multicolumn{17}{c|}{Equivalent witdh, one star per column, BLxxx} \\
$\lambda$(\AA)& elem& $\chi$& \gf& 
253& 257& 258& 260& 261& 262& 266& 267& 269& 278& 279& 295& 300& 304& 311& 315& 323\\
\hline
\endfirsthead
\hline \hline
$\lambda$(\AA)& elem& $\chi$& \gf& 
253& 257& 258& 260& 261& 262& 266& 267& 269& 278& 279& 295& 300& 304& 311& 315& 323\\
\hline
\endhead
\hline \multicolumn{21}{r}{\emph{Continued on next page}}
\endfoot
\hline
\endlastfoot
6154.23& Na\,{\sc i} &    2.10&   -1.56&   41.9&  37.6&  45.3&   ...&  25.0&   ...&   ...&  31.7&  24.5&  52.8&  16.2&  48.8&  33.4&  17.0&  24.3&   ...&   ...\\
6160.75& Na\,{\sc i} &    2.10&   -1.26&   68.6&  62.1&  63.6&  34.0&  32.5&  25.2&   ...&  58.8&  51.6&  96.0&   ...&  70.2&  52.3&  38.9&  47.2&  26.3&  44.3\\
5528.41& Mg\,{\sc i} &    4.35&   -0.36&  227.2& 223.4& 220.8& 205.5& 219.7& 217.4& 169.5& 216.3& 221.3& 234.8& 164.9& 225.7& 240.7& 208.9& 208.1& 207.0& 220.0\\
6318.72& Mg\,{\sc i} &    5.11&   -1.97&   44.0&  64.6&  47.2&  32.4&  25.6&  50.3&  27.4&  43.6&  28.3&   ...&   ...&  49.7&  42.8&   ...&  41.7&   ...&  30.7\\
6319.24& Mg\,{\sc i} &    5.11&   -2.21&   26.0&  25.8&   ...&   ...&   ...&  20.5&   ...&   ...&  26.4&  38.8&   ...&  41.9&   ...&   ...&  25.7&   ...&   ...\\
6319.49& Mg\,{\sc i} &    5.11&   -2.43&    ...&   ...&   ...&   ...&   ...&   ...&   ...&  28.2&   ...&   ...&   ...&   ...&  31.0&  18.5&   ...&   ...&  15.2\\
6125.03& Si\,{\sc i} &    5.62&   -1.57&    ...&   ...&   ...&  31.7&   ...&   ...&   ...&  32.3&  38.8&   ...&   ...&  56.2&   ...&   ...&   ...&  17.4&   ...\\
6142.48& Si\,{\sc i} &    5.62&   -1.51&    ...&   ...&   ...&   ...&   ...&   ...&   ...&  26.0&   ...&   ...&  16.5&   ...&   ...&   ...&   ...&  16.9&   ...\\
\end{longtable}
}
\end{center}

\begin{center}
\scriptsize
\tabcolsep 3pt
\begin{longtable}{l|r|r|r|r|r|r|r|r|}
\caption{Abundance ratio of the Fornax Field stars, where we list the abundance, 
its associated error ($\sigma$) and the number of lines used (n). 
The quoted error is the error on [element/H], not [X/Fe].  Part 1
\label{tab:fnxfi-tababo1}}\\
\hline \hline
Star&
[Na\,{\sc i}/Fe]  $\sigma$ (n)& [Mg\,{\sc i}/Fe]  $\sigma$ (n)&
[Si\,{\sc i}/Fe]  $\sigma$ (n)& [Ca\,{\sc i}/Fe]  $\sigma$ (n)& 
[Ti\,{\sc i}/Fe]  $\sigma$ (n)& [Ti\,{\sc ii}/Fe] $\sigma$ (n)& 
[Cr\,{\sc i}/Fe]  $\sigma$ (n)& [Fe\,{\sc i}/H]   $\sigma$ (n)\\
\hline
\endfirsthead
\hline \hline
Star&
[Na\,{\sc i}/Fe]  $\sigma$ (n)& [Mg\,{\sc i}/Fe]  $\sigma$ (n)&
[Si\,{\sc i}/Fe]  $\sigma$ (n)& [Ca\,{\sc i}/Fe]  $\sigma$ (n)& 
[Ti\,{\sc i}/Fe]  $\sigma$ (n)& [Ti\,{\sc ii}/Fe] $\sigma$ (n)& 
[Cr\,{\sc i}/Fe]  $\sigma$ (n)& [Fe\,{\sc i}/H]   $\sigma$ (n)\\
\hline
\endhead
\hline \multicolumn{9}{r}{\emph{Continued on next page}}
\endfoot
\hline
\endlastfoot
BL038&  -0.46 $\pm$ 0.16  (2)&   0.03 $\pm$ 0.16  (2)&   0.35 $\pm$ 0.15  (4)&  -0.25 $\pm$ 0.10  (8)&  -0.13 $\pm$ 0.09  (6)&   0.18 $\pm$ 0.13  (3)&  -0.11 $\pm$ 0.23  (1)&  -0.88 $\pm$ 0.07 (43)\\
BL045&    ... $\pm$  ...  (0)&   0.02 $\pm$ 0.13  (2)&   0.25 $\pm$ 0.17  (2)&  -0.33 $\pm$ 0.06  (8)&  -0.23 $\pm$ 0.05 (10)&   0.03 $\pm$ 0.10  (4)&  -0.16 $\pm$ 0.16  (1)&  -1.09 $\pm$ 0.05 (48)\\
BL052&  -0.85 $\pm$ 0.34  (1)&   0.07 $\pm$ 0.24  (2)&   0.16 $\pm$ 0.34  (1)&  -0.42 $\pm$ 0.12  (8)&  -0.37 $\pm$ 0.13  (7)&  -0.12 $\pm$ 0.34  (2)&  -0.31 $\pm$ 0.34  (1)&  -1.02 $\pm$ 0.06 (37)\\
BL065&    ... $\pm$  ...  (0)&   0.00 $\pm$ 0.18  (1)&   0.32 $\pm$ 0.16  (2)&  -0.09 $\pm$ 0.12  (9)&  -0.18 $\pm$ 0.07  (7)&   0.28 $\pm$ 0.11  (4)&  -0.31 $\pm$ 0.18  (1)&  -1.43 $\pm$ 0.07 (42)\\
BL076&  -0.66 $\pm$ 0.13  (2)&  -0.14 $\pm$ 0.13  (2)&  -0.15 $\pm$ 0.11  (3)&  -0.19 $\pm$ 0.07  (8)&  -0.20 $\pm$ 0.06  (9)&   0.13 $\pm$ 0.16  (3)&  -0.21 $\pm$ 0.19  (1)&  -0.86 $\pm$ 0.06 (43)\\
BL077&  -0.78 $\pm$ 0.14  (1)&  -0.27 $\pm$ 0.08  (3)&  -0.09 $\pm$ 0.10  (2)&  -0.46 $\pm$ 0.06  (8)&  -0.38 $\pm$ 0.06 (10)&  -0.01 $\pm$ 0.11  (2)&  -0.42 $\pm$ 0.14  (1)&  -0.79 $\pm$ 0.06 (39)\\
BL079&  -0.65 $\pm$ 0.31  (2)&  -0.01 $\pm$ 0.31  (2)&  -0.13 $\pm$ 0.31  (2)&  -0.19 $\pm$ 0.17  (7)&   0.00 $\pm$ 0.16  (8)&  -0.09 $\pm$ 0.32  (2)&  -0.05 $\pm$ 0.44  (1)&  -0.57 $\pm$ 0.08 (39)\\
BL081&  -0.70 $\pm$ 0.13  (2)&  -0.32 $\pm$ 0.13  (2)&  -0.45 $\pm$ 0.13  (2)&  -0.34 $\pm$ 0.06  (9)&  -0.19 $\pm$ 0.07 (10)&  -0.05 $\pm$ 0.13  (3)&  -0.32 $\pm$ 0.19  (1)&  -0.62 $\pm$ 0.06 (40)\\
\end{longtable}
\end{center}

\addtocounter{table}{-1}
\begin{center}
\scriptsize
\tabcolsep 3pt
\begin{longtable}{l|r|r|r|r|r|r|r|}
\caption{Abundance ratio of the Fornax Field stars, where we list the abundance, 
its associated error ($\sigma$) and the number of lines used (n). 
The quoted error is the error on [element/H], not [X/Fe].  Part 2
\label{tab:fnxfi-tababo2}}\\
\hline \hline
Star&
[Fe\,{\sc ii}/H]  $\sigma$ (n)& [Ni\,{\sc i}/Fe]  $\sigma$ (n)&
 [Y\,{\sc ii}/Fe] $\sigma$ (n)& [Ba\,{\sc ii}/Fe] $\sigma$ (n)&
[La\,{\sc ii}/Fe] $\sigma$ (n)& [Nd\,{\sc ii}/Fe] $\sigma$ (n)&
[Eu\,{\sc ii}/Fe] $\sigma$ (n)\\
\hline
\endfirsthead
\hline \hline
Star&
[Fe\,{\sc ii}/H]  $\sigma$ (n)& [Ni\,{\sc i}/Fe]  $\sigma$ (n)&
 [Y\,{\sc ii}/Fe] $\sigma$ (n)& [Ba\,{\sc ii}/Fe] $\sigma$ (n)&
[La\,{\sc ii}/Fe] $\sigma$ (n)& [Nd\,{\sc ii}/Fe] $\sigma$ (n)&
[Eu\,{\sc ii}/Fe] $\sigma$ (n)\\
\hline
\endhead
\hline \multicolumn{8}{r}{\emph{Continued on next page}}
\endfoot
\hline
\endlastfoot
BL038&  -0.70 $\pm$ 0.08  (4)&  -0.13 $\pm$ 0.06 (15)&   0.10 $\pm$ 0.23  (1)&   0.80 $\pm$ 0.16  (2)&   0.51 $\pm$ 0.23  (1)&   0.22 $\pm$ 0.17  (3)&   0.88 $\pm$ 0.23  (1)\\
BL045&  -0.78 $\pm$ 0.07  (5)&  -0.22 $\pm$ 0.07 (11)&    ... $\pm$  ...  (0)&  -0.05 $\pm$ 0.15  (2)&   0.17 $\pm$ 0.16  (1)&  -0.06 $\pm$ 0.16  (1)&   0.22 $\pm$ 0.16  (1)\\
BL052&  -0.52 $\pm$ 0.17  (2)&  -0.28 $\pm$ 0.10 (11)&  -0.16 $\pm$ 0.34  (1)&   0.57 $\pm$ 0.28  (2)&    ... $\pm$  ...  (0)&   0.22 $\pm$ 0.24  (2)&   0.91 $\pm$ 0.34  (1)\\
BL065&  -1.06 $\pm$ 0.15  (5)&  -0.15 $\pm$ 0.07  (8)&    ... $\pm$  ...  (0)&   0.19 $\pm$ 0.13  (2)&    ... $\pm$  ...  (0)&   0.49 $\pm$ 0.18  (1)&   0.37 $\pm$ 0.18  (1)\\
BL076&  -0.43 $\pm$ 0.12  (4)&  -0.25 $\pm$ 0.08 (13)&    ... $\pm$  ...  (0)&   0.50 $\pm$ 0.13  (2)&    ... $\pm$  ...  (0)&   0.18 $\pm$ 0.12  (3)&   0.46 $\pm$ 0.19  (1)\\
BL077&  -0.38 $\pm$ 0.21  (4)&  -0.23 $\pm$ 0.06 (14)&    ... $\pm$  ...  (0)&   0.29 $\pm$ 0.11  (2)&  -0.12 $\pm$ 0.14  (1)&   0.23 $\pm$ 0.11  (3)&   0.39 $\pm$ 0.14  (1)\\
BL079&  -0.15 $\pm$ 0.30  (2)&  -0.18 $\pm$ 0.12 (14)&   0.14 $\pm$ 0.44  (1)&   1.04 $\pm$ 0.44  (1)&   0.52 $\pm$ 0.44  (1)&   0.45 $\pm$ 0.25  (3)&   0.79 $\pm$ 0.44  (1)\\
BL081&  -0.44 $\pm$ 0.17  (4)&  -0.24 $\pm$ 0.06 (14)&  -0.34 $\pm$ 0.19  (1)&   0.91 $\pm$ 0.13  (2)&   0.40 $\pm$ 0.19  (1)&   0.21 $\pm$ 0.14  (3)&   0.40 $\pm$ 0.19  (1)\\
\end{longtable}
\end{center}
\clearpage

\end{document}